\documentclass[aps,prd,reprint,superscriptaddress,nofootinbib,longbibliography,floatfix]{revtex4-2}
\usepackage[T1]{fontenc}
\usepackage[utf8]{inputenc}
\usepackage{amsmath,amssymb,mathtools,bm,mathrsfs,slashed}
\usepackage{booktabs,array,graphicx}
\usepackage{microtype}
\usepackage{hyperref}
\usepackage{units}
\hypersetup{colorlinks=true,linkcolor=blue,citecolor=blue,urlcolor=blue,
 pdftitle={Octupole parton structure of spin-3/2 targets},
 pdfauthor={Brandon B. Le and Dustin Keller}}
\newcommand{\dd}{\mathrm{d}}
\newcommand{\ii}{\mathrm{i}}
\newcommand{\e}{\mathrm{e}}
\newcommand{\Tr}{\operatorname{Tr}}
\newcommand{\tr}{\operatorname{tr}}
\newcommand{\STF}{\operatorname{STF}_{2}}
\newcommand{\Sym}{\operatorname{Sym}}
\newcommand{\diag}{\operatorname{diag}}
\newcommand{\Rea}{\operatorname{Re}}
\newcommand{\Ima}{\operatorname{Im}}
\newcommand{\M}{M_A}
\newcommand{\bk}{\bm{k}_T}
\newcommand{\bp}{\bm{p}_T}
\newcommand{\bb}{\bm{b}_T}
\newcommand{\bPh}{\bm{P}_{hT}}
\newcommand{\calF}{\mathcal F}
\newcommand{\calG}{\mathcal G}
\newcommand{\calT}{\mathcal T}
\newcommand{\calH}{\mathcal H}
\newcommand{\cC}{\mathscr C}
\newcommand{\cS}{\mathscr S}
\newcommand{\Conv}{\mathcal C}
\newcommand{\Dpol}{D_{\ell}}
\newcommand{\qq}[3]{h^{q[#3]}_{#1#2}}
\newcommand{\gh}[3]{h^{g[#3]}_{#1#2}}

\begin{document}

\title{Octupole parton structure of spin-3/2 targets}
\author{Brandon B. Le}
\email{sxh3qf@virginia.edu}
\author{Dustin Keller}
\affiliation{Department of Physics, University of Virginia, Charlottesville, Virginia 22904, USA}

\date{\today}

\begin{abstract}
In spin-$3/2$ targets, octupole polarization can reveal quark and gluon correlations beyond those probed by vector polarization and quadrupole alignment. Here, we construct the complete leading-twist gluon transverse-momentum-dependent distribution basis, with 32 independent coefficients for each specified Wilson-line and color structure. Fourteen of these describe octupole dependence in gluon number, helicity, and linear polarization. In heavy-quark-pair leptoproduction at small, nonzero pair imbalance, comparing target states with opposite octupoles and identical lower moments isolates the octupole contribution. Beam-helicity reversal and joint angular projections in target orientation and pair-imbalance azimuth can then separate all fourteen gluon octupole responses at leading-power Born order. We also derive the incoming-target quark responses in semi-inclusive deep-inelastic scattering and Drell–Yan production. Time-reversal relations between distributions remain valid under symmetry-preserving renormalization and evolution, while predictions across processes require factorization formulas connecting those distributions to scattering observables. Finally, we connect this framework to polarized $^{7}$Li through its collinear limits and nuclear spin–orbital correlations.
\end{abstract}
\maketitle

\tableofcontents

\newpage

\section{Introduction}
\label{sec:scope}

The polarization of a spin-$3/2$ target includes an octupole moment independent of its vector polarization and quadrupole alignment. This rank-three moment carries additional spin information, namely, even with the lower moments fixed, it can change gluon number, helicity, and interference between opposite gluon helicities. Interpreting these effects requires a complete operator basis connecting target polarization to parton correlations. We formulate this basis in terms of transverse-momentum-dependent (TMD) distributions, which retain the parton’s intrinsic transverse momentum and the distribution correlator’s Wilson-line structure. Our principal focus in this paper is the gluon octupole sector, which we construct a complete distribution basis for and determine how polarized scattering can isolate its responses. The quark sector provides the corresponding incoming-target distributions for semi-inclusive deep-inelastic scattering (SIDIS) and Drell--Yan (DY) production, as well as with a comparison to existing higher-spin classifications.

Inclusive deep-inelastic scattering for arbitrary spin~\cite{JaffeManohar1989}, covariant multipole expansions of local currents, and Lorentz-tensor operator bases~\cite{Cotogno2020,CosynVera2025} provide the representation theory foundations for this work. Spin-1 quark distributions and semi-inclusive extensions describe tensor-polarized structure~\cite{BacchettaMulders2000,KumanoSong2021,ZhaoSIDIS2025}, and the covariant spin-1 SIDIS decomposition of Cosyn and Weiss outlines higher-spin polarization and azimuthal structures~\cite{CosynWeiss2026}. Higher multipoles already appear in nuclear calculations, including polarized distributions of $^{7}$Li and $K=3$ collinear spin-dependent $u$- and $d$-quark multipoles of $^{11}$B~\cite{Cloet2006}. For spin $3/2$ specifically, Zhao \emph{et al.} classified 32 leading-twist quark fragmentation functions, fourteen associated with octupole polarization~\cite{Zhao2022}, and derived SIDIS responses for a polarized spin-$3/2$ detected hadron~\cite{ZhaoSIDIS2024}. Quark and gluon generalized parton distributions (GPDs), transversity GPDs, and forward-PDF positivity constraints are established as well~\cite{Fu2022,Fu2024,Fu2026}. The quark spin-tensor classification and collinear distribution sets are therefore prior results. The analysis in this paper concerns TMD distributions of the incoming target, including intrinsic parton transverse momentum and the distribution correlator's Wilson-line structure.

Our construction builds on spin-$1/2$ helicity and discrete-symmetry methods~\cite{LorcePasquini2013}, gluon parametrizations for unpolarized and vector-polarized targets~\cite{MuldersRodrigues2001}, and their tensor-polarized extension~\cite{Boer2016,Cotogno2017}. We couple the established four-component Stokes or $U(2)$ decomposition of the transverse gluon polarization matrix~\cite{LyubovitskijSchmidt2021} to rank-three target polarization. At finite transverse momentum, the allowed structures couple target-spin rank $K$, transverse target-helicity magnitude $m$, and orbital symmetric-traceless (STF) rank $n$, which describes the angular dependence carried by the transverse momentum. In the physical two-dimensional transverse plane, a nonzero-rank STF tensor has only two helicity components. Matching them to the partonic helicity channel gives
\begin{equation}
 N_q=N_g=32,\qquad N_K=4K+2\quad(K=0,1,2,3).
 \label{eq:introcount}
\end{equation}
The first two counts refer to the full quark and gluon bases, while the last one gives the number of coefficients at each target-spin rank.

Our principal result is the complete finite-$k_T$ gluon distribution basis for an arbitrary physical spin-$3/2$ density matrix and its polarization-selective hard responses. Its octupole sector contains fourteen coefficients: three for gluon number, four for helicity, and seven for linear polarization. The latter describe interference between opposite gluon helicities. This two-unit helicity change couples to target components with $m=0,1,2,3$. For $m=3$ specifically, the
orbital ranks are one and five, giving two unique structures. The lower-rank gluon sector is written in a trace/STF basis without redundant tensors. The quark helicity-flip channel carries one unit of partonic helicity change and is retained in a common convention for SIDIS and DY. Completeness follows from transverse-helicity matching and an independent check in the combined parton--target spin space.

We connect this basis to measurements in heavy-quark-pair leptoproduction using the photon--gluon-fusion amplitude~\cite{Boer2011,Pisano2013}. Physical target preparations with rank-three polarization, beam-helicity reversal, and joint target/recoil angular projections separate all fourteen gluon octupole responses at Born order under explicit nondegeneracy conditions. Longitudinal observables and fully transverse octupole ($TTT$) projections isolate the rank-three target contribution from lower-rank polarization. The transverse projections also include the rank-five orbital term. Standard SIDIS and DY hard kernels~\cite{Bacchetta2007,Arnold2009} similarly give the complete incoming-target SIDIS catalog and explicit DY octupole convolutions.

Time reversal relates the future-pointing SIDIS and past-pointing DY quark staple-shaped Wilson lines~\cite{Collins2002,CollinsMetz2004}, where the three octupole number coefficients change sign, while the eleven helicity or transverse-spin coefficients do not. These operator relations must be distinguished from angular and charge-conjugation signs in the hard scattering. Measured asymmetries also contain process-dependent hard analyzers and companion distributions or fragmentation functions. For gluons, reversal acts on the complete Wilson-path pair and color trace~\cite{Buffing2013}. Wilson-line relations therefore specify which target correlations can be compared, while the factorization formulas determine how those correlations enter the measured responses.

The framework applies to the stable $^{7}$Li ground state, which has $J^P=3/2^-$~\cite{IAEA}. For this target, the spin density matrix describes the prepared nuclear polarization, while binding, orbital motion, and rescattering enter the quark and gluon correlators. We recover the known collinear distribution sets, local-moment constraints, and positivity conditions in our normalization, providing consistency checks for nuclear calculations of these correlators. The analysis assumes parity-conserving QCD, leading power in the hard scale, and target-correlator link and rapidity prescriptions with no additional preferred transverse direction. The single-hadron SIDIS formulas describe current fragmentation at $P_{hT}/z\ll Q$, while the heavy-pair formulas describe small pair imbalance with a collinearly expanded Born hard part. The DY calculation concerns electromagnetic continuum dilepton production at $q_T\ll Q$, inclusive over the hadronic remnants. Hard responses are displayed at leading-power Born order, whereas the reversed-link identities are statements about the defined TMD operators beyond that approximation. Operator classification and renormalization are treated separately from the process-dependent factorization of measured recoil~\cite{Handbook,Rogers2010}.

The remainder of this paper is organized as follows. Sections~\ref{sec:spin} and \ref{sec:count} establish the spin, correlator, and tensor conventions and the Wilson-line relations. Section~\ref{sec:gluonSF} develops the gluon observables, and Sec.~\ref{sec:quarkobservables} presents the SIDIS and DY quark responses. Section~\ref{sec:collinear} connects their collinear limits to nuclear structure and calibrated extraction. The appendices give convention dictionaries and consistency checks, and the computational companion is available in Ref.~\cite{Li7TMDs}.

\section{Spin and parton correlators}
\label{sec:spin}

\subsection{Spin polarization}

We describe the target density matrix in terms of its spin multipoles, which specify the polarization dependence of the parton correlators. We use the basis $(3/2,1/2,-1/2,-3/2)$ and write $J_z=\diag(3/2,1/2,-1/2,-3/2)$. The raising operator is
\begin{equation}
 J_+=\begin{pmatrix}
 0&\sqrt3&0&0\\0&0&2&0\\0&0&0&\sqrt3\\0&0&0&0
 \end{pmatrix},\qquad J_-=J_+^\dagger.
 \label{eq:J}
\end{equation}
The Cartesian generators are $J_x=(J_++J_-)/2$ and $J_y=(J_+-J_-)/(2\ii)$, which satisfy $[J_a,J_b]=\ii\epsilon_{abc}J_c$ and $\bm J^2=15/4$. Under rotations, the operator space decomposes into scalar, vector, quadrupole, and octupole sectors, $0\oplus1\oplus2\oplus3$, with dimensions $1+3+5+7=16$. A normalized Hermitian density matrix therefore has fifteen independent polarization parameters.

The irreducible quadrupole and octupole operators are
\begin{align}
 Q_{ab}&=\tfrac12\{J_a,J_b\}-\tfrac54\delta_{ab}I,
 \label{eq:Q}\\
 O_{abc}&=\Sym(J_aJ_bJ_c) -\frac{41}{60}
 (\delta_{ab}J_c+\delta_{ac}J_b+\delta_{bc}J_a).
 \label{eq:QO}
\end{align}
Here, $a,b,c=x,y,z$, and $\Sym$ averages over all six ordered products, including repeated indices. The subtractions remove the Cartesian traces, and the quadratic coefficient follows from $\bm J^2=15/4$. To fix the cubic coefficient, we use $\sum_aJ_aJ_cJ_a=[J(J+1)-1]J_c$, which gives
\begin{equation}
 \sum_a\Sym(J_aJ_aJ_c)
 =\left[J(J+1)-\tfrac13\right]J_c=\tfrac{41}{12}J_c.
\end{equation}
Contracting the repeated Cartesian indices in the three delta terms of Eq.~\eqref{eq:QO} gives $5J_c$, so their coefficient must be $41/60$. Therefore, $Q_{aa}=0$ and $O_{aac}=0$ as operator identities. These normalizations agree with the spin-density convention of Refs.~\cite{Zhao2022,Fu2026}.

We define the target moments by
\begin{equation}
 S_a=\Tr(\rho J_a),\quad T_{ab}=\Tr(\rho Q_{ab}),\quad
 R_{abc}=\Tr(\rho O_{abc}).
 \label{eq:momentdefinitions}
\end{equation}
Conversely, these moments reconstruct the density matrix as
\begin{equation}
 \rho=\frac14\left[I+\frac45S_aJ_a+
 \frac23T_{ab}Q_{ab}+\frac89R_{abc}O_{abc}\right].
 \label{eq:rho}
\end{equation}
Repeated Cartesian indices are summed, so nondiagonal tensor components occur with their full multiplicities. To fix the expansion coefficients, let $\Pi_K$ denote the identity projector on rank-$K$ symmetric-traceless tensors in three spatial dimensions. Operators of different irreducible rank are orthogonal under the spin trace, allowing their coefficients to be determined independently. With the normalizations
\begin{align}
 \Tr(J_aJ_b)&=5\delta_{ab},\nonumber\\
 \Tr(Q_{ab}Q_{cd})&=6(\Pi_2)_{ab,cd},\nonumber\\
 \Tr(O_{abc}O_{def})&=\tfrac92(\Pi_3)_{abc,def},
 \label{eq:trace_metrics}
\end{align}
tracing Eq.~\eqref{eq:rho} with $J$, $Q$, or $O$ recovers the moment definitions in Eq.~\eqref{eq:momentdefinitions}, while the scalar part fixes $\Tr\rho=1$. A physical state must also satisfy $\rho\succeq0$, as arbitrary real values of the fifteen moments need not give a positive density matrix.

\begin{table}[t]
\caption{Target components $A_{Km}$, with spin rank $K$ and transverse helicity magnitude $m$. Each entry at $m>0$ denotes a transverse symmetric-traceless tensor with two independent real components.}
\label{tab:polarization}
\centering
\begin{ruledtabular}
\begin{tabular}{ccccc}
 $K$ & $m=0$ & $m=1$ & $m=2$ & $m=3$\\
 0 & $1$ & --- & --- & ---\\
 1 & $S_L$ & $S_T$ & --- & ---\\
 2 & $S_{LL}$ & $S_{LT}$ & $S_{TT}$ & ---\\
 3 & $S_{LLL}$ & $S_{LLT}$ & $S_{LTT}$ & $S_{TTT}$
\end{tabular}
\end{ruledtabular}
\end{table}

To separate longitudinal and transverse polarization, choose a spin quantization axis $z$ and use $i,j,k=x,y$ for transverse Cartesian indices. These indices have the Euclidean metric $\delta_{ij}$. For two free transverse indices, define
\begin{equation}
 \STF X^{ij}=\tfrac12(X^{ij}+X^{ji})-\tfrac12\delta^{ij}X^{aa}.
 \label{eq:stf2}
\end{equation}
The trace here is over the two transverse directions. For higher rank, $\STF$ denotes symmetrization and removal of all transverse traces. The target components are defined by
\begin{align}
 S_L&=S_z,& S_T^i&=S_i,\nonumber\\
 S_{LL}&=T_{zz},& S_{LT}^i&=2T_{zi},\nonumber\\
 S_{TT}^{ij}&=2\STF T_{ij},\nonumber\\
 S_{LLL}&=R_{zzz},& S_{LLT}^i&=R_{zzi},\nonumber\\
 S_{LTT}^{ij}&=4\STF R_{zij},&
 S_{TTT}^{ijk}&=4\STF R_{ijk},
 \label{eq:componentdefinitions}
\end{align}
where the factors of two and four fix the polarization normalization. The inverse relations recover the three-dimensional tensors:
\begin{align}
 T_{ij}&=\tfrac12S_{TT}^{ij}-\tfrac12\delta_{ij}S_{LL},&
 T_{zi}&=\tfrac12S_{LT}^i,\nonumber\\
 R_{zij}&=\tfrac14S_{LTT}^{ij}-\tfrac12\delta_{ij}S_{LLL},&
 R_{zzi}&=S_{LLT}^i,\nonumber\\
 R_{ijk}&=\tfrac14\bigl[S_{TTT}^{ijk}-\delta_{ij}S_{LLT}^k
 \nonumber\\
 &\hspace{9mm}-\delta_{ik}S_{LLT}^j-\delta_{jk}S_{LLT}^i\bigr].
 \label{eq:componentinverse}
\end{align}
For example, $R_{zxx}+R_{zyy}=-R_{zzz}$. The subtraction in $S_{LTT}$ therefore separates the longitudinal octupole from the transverse rank-two component of the same three-dimensional tensor. We denote these components by $A_{Km}$, as in Table~\ref{tab:polarization}. The label $K$ specifies the irreducible spin rank, and $m$ specifies the magnitude of its transverse helicity. For $m=0$, there is one signed real component, and for every $m>0$, there are two real components. The vector, quadrupole, and octupole sectors therefore contain $3$, $5$, and $7$ polarization parameters.

The longitudinal components follow from the spin-state populations. Let $p_\Lambda=\rho_{\Lambda\Lambda}$, with $\sum_\Lambda p_\Lambda=1$. The longitudinal moments are
\begin{align}
 S_L&=\tfrac32(p_{3/2}-p_{-3/2})
       +\tfrac12(p_{1/2}-p_{-1/2}),\nonumber\\
 S_{LL}&=p_{3/2}+p_{-3/2}-p_{1/2}-p_{-1/2},\nonumber\\
 S_{LLL}&=\tfrac3{10}
 [p_{3/2}-p_{-3/2}-3(p_{1/2}-p_{-1/2})].
 \label{eq:popmoments}
\end{align}
Conversely,
\begin{align}
 p_{\pm3/2}&=\tfrac14
 (1\pm\tfrac65S_L+S_{LL}\pm\tfrac23S_{LLL}),\nonumber\\
 p_{\pm1/2}&=\tfrac14
 (1\pm\tfrac25S_L-S_{LL}\mp2S_{LLL}).
 \label{eq:popinverse}
\end{align}
Nonnegativity of the four populations bounds the allowed longitudinal polarizations. In particular, $S_L=S_{LL}=0$ permits $|S_{LLL}|\leq1/2$, while the unrestricted longitudinal octupole range is $|S_{LLL}|\leq9/10$. The diagonal state with populations $(p_{3/2},p_{1/2},p_{-1/2},p_{-3/2})=(1/3,0,1/2,1/6)$ has $S_L=S_{LL}=0$ and $S_{LLL}=1/2$. Reversing their order reverses only its nonzero octupole moment.

Off-diagonal density-matrix elements encode transverse polarization. We resolve these coherences by spin rank with orthonormal spherical spin operators:
\begin{equation}
 \tau_{KM}=\sqrt{\frac{2K+1}{2J+1}}
 \sum_{\Lambda\Lambda'}
 \langle J\Lambda;KM\mid J\Lambda'\rangle\,
 |J\Lambda'\rangle\langle J\Lambda|,
 \label{eq:spherical}
\end{equation}
where $J=\tfrac32$ and $M=-K,\ldots,K$ is the spherical magnetic index. With real Clebsch--Gordan phases,
\begin{align}
 \Tr(\tau_{KM}^\dagger\tau_{K'M'})&=\delta_{KK'}\delta_{MM'},\nonumber\\
 \tau_{KM}^\dagger&=(-1)^M\tau_{K,-M},\qquad \tau_{00}=I/2.
\end{align}
The trace $\Tr$ in this spin construction acts in the four-dimensional target space. Below, $\Tr_D$, $\tr_c$, and $\Tr_2$ denote Dirac, color, and two-dimensional transverse polarization traces, respectively. The density matrix and its spherical moments obey
\begin{equation}
 \rho=\sum_{KM}t_{KM}\tau_{KM},\qquad
 t_{KM}=\Tr(\tau_{KM}^\dagger\rho),\quad t_{00}=\tfrac12.
 \label{eq:tomography}
\end{equation}
For example, $\tau_{10}=J_z/\sqrt5$, $\tau_{20}=Q_{zz}/2$, and $\tau_{30}=\sqrt5\,O_{zzz}/3$. The Clebsch--Gordan selection $\Lambda'=\Lambda+M$ identifies $LLT$, $LTT$, and $TTT$ with helicity differences one, two, and three. Since several tensor ranks $K$ can contribute at the same helicity difference, the coefficient $t_{3M}$ isolates the octupole coherence from lower-rank moments. Equations~\eqref{eq:componentdefinitions} and \eqref{eq:tomography} therefore provide equivalent Cartesian and spherical descriptions of any prepared or modeled target state.

\subsection{Kinematics and transverse tensors}
\label{sec:kin}

We use $g^{\mu\nu}=\diag(1,-1,-1,-1)$, $\epsilon^{0123}=+1$, and lightlike vectors
\begin{equation}
 n^\mu=\frac{(1,0,0,-1)}{\sqrt2},\qquad
 \bar n^\mu=\frac{(1,0,0,1)}{\sqrt2},\qquad n\cdot\bar n=1.
\end{equation}
The light-front components are $v^+=v\cdot n$ and $v^-=v\cdot\bar n$. With $M_A$ the target mass and $Q$ the hard scale, the target and parton momenta are
\begin{align}
 P^\mu&=P^+\bar n^\mu+\frac{\M^2}{2P^+}n^\mu,\nonumber\\
 k^\mu&=xP^+\bar n^\mu+k^-n^\mu+k_T^\mu,
 \label{eq:kin}
\end{align}
where the partonic fraction $x=k^+/P^+$ is defined relative to the whole target momentum. In quark DIS at Born order, it equals $x_{\mathrm{Bj}}=Q^2/(2P\cdot q)$, but other hard subprocesses can select a different fraction, as in Sec.~\ref{sec:gluonSF}. For a nucleus, a variable defined per average nucleon is approximately $Ax_{\mathrm{Bj}}$. The target-rest-frame definition using a reference mass $m_N$ instead gives $x_N^{\mathrm{exp}}=(M_A/m_N)x_{\mathrm{Bj}}$.

We express the rest-frame spin tensors of Sec.~\ref{sec:spin} covariantly using three orthonormal spacelike basis vectors orthogonal to $P$:
\begin{equation}
 e_L^\mu=\frac{P^+}{\M}\bar n^\mu-\frac{\M}{2P^+}n^\mu,
 \qquad e_A\cdot e_B=-\delta_{AB}.
 \label{eq:tetrad}
\end{equation}
Here, $e_x^\mu=(0,1,0,0)$, $e_y^\mu=(0,0,1,0)$, and $A,B=L,x,y$. Setting
\begin{align}
 \mathsf S^\mu&=S_Ae_A^\mu,\nonumber\\
 \mathsf T^{\mu\nu}&=T_{AB}e_A^\mu e_B^\nu,\nonumber\\
 \mathsf R^{\mu\nu\alpha}&=R_{ABC}e_A^\mu e_B^\nu e_C^\alpha
\end{align}
produces $P$-orthogonal polarization tensors with the same independent components. A spatial rotation $\mathscr R$ of the preparation axis acts as $\rho\mapsto D^{3/2}(\mathscr R)\rho D^{3/2}(\mathscr R)^\dagger$. A boost along the chosen $z$ axis introduces no additional spin rotation, while a general frame change includes the Wigner rotation of the chosen spin basis. All components entering a cross section must be expressed relative to its specified light-front and azimuthal axes.

The Lorentz transverse projector is $g_T^{\mu\nu}=g^{\mu\nu}-\bar n^\mu n^\nu-n^\mu\bar n^\nu$, and in Cartesian transverse components, $g_T^{ij}=-\delta^{ij}$. Throughout the tensor decompositions, Latin transverse indices are Euclidean and are contracted with $\delta_{ij}$. We define
\begin{equation}
 E=(\epsilon_{ij})=\begin{pmatrix}0&1\\-1&0\end{pmatrix},
 \qquad \widetilde v^i=\epsilon_{ia}v^a,
 \label{eq:euclidean}
\end{equation}
so that $\widetilde v=(v_y,-v_x)$ and $\bk^2=k_x^2+k_y^2=-k_T\cdot k_T$.  This dual acts in the Euclidean transverse plane, which is is separate from lowering a Lorentz index.

The dimensionless momentum tensors are denoted by $\mathsf k_n$, distinguishing their orbital rank $n$ from the target-spin rank $K$:
\begin{equation}
 \mathsf k_n^{i_1\cdots i_n}(\bk)=\frac1{\M^n}
 \STF(k^{i_1}\cdots k^{i_n}),\qquad \mathsf k_0=1.
 \label{eq:Kn}
\end{equation}
For the first three ranks,
\begin{align}
 \mathsf k_1^i&=k^i/\M,\nonumber\\
 \mathsf k_2^{ij}&=\frac{k^ik^j-\delta^{ij}\bk^2/2}{\M^2},\nonumber\\
 \mathsf k_3^{ijk}&=\frac1{\M^3}\bigl[k^ik^jk^k
 -\tfrac{\bk^2}{4}(\delta^{ij}k^k+\delta^{ik}k^j+\delta^{jk}k^i)\bigr].
 \label{eq:Kexamples}
\end{align}
Explicit rank-four and rank-five expressions are given in Appendix~\ref{sec:inverse}. A real rank-$n$ transverse STF tensor, with $n>0$, has two independent components. The subscript $+$ below denotes a complex transverse component, distinct from a light-front superscript $+$ or a staple label $[+]$. Define
\begin{equation}
 A_+=A^{x\cdots x}+\ii A^{yx\cdots x}=a+\ii b.
 \label{eq:complexcomponent}
\end{equation}
Tracelessness implies that replacing two $x$ indices by two $y$ indices changes the sign. A component with $j$ indices equal to $y$ is therefore $(-1)^{j/2}a$ for even $j$, and $(-1)^{(j-1)/2}b$ for odd $j$. Under a transverse rotation through $\theta$, $A_+\mapsto\e^{\ii n\theta}A_+$. In particular,
\begin{equation}
 (\mathsf k_n)_+=2^{1-n}\kappa^n,\qquad
 \kappa=\frac{k_x+\ii k_y}{\M}.
 \label{eq:Kcomplex}
\end{equation}
for $n\geq1$. The dual of a nonzero-rank tensor is
\begin{equation}
 \widetilde A^{i_1\cdots i_m}
 =\STF(\epsilon_{i_1a}A^{ai_2\cdots i_m}),\qquad
 (\widetilde A)_+=-\ii A_+.
 \label{eq:dual}
\end{equation}
For an already-STF tensor in two dimensions, the contraction on the right is itself symmetric and traceless. A transverse scalar has no such dual, so a tilde on an explicitly transverse tensor denotes this dual. On a correlator or a scalar function, it instead denotes the Fourier transform from transverse momentum in $b_T$ space, as specified in the corresponding definition.

We express angular dependence through the scalar contractions
\begin{align}
 \cC_m[A_m]&=\mathsf k_m^{i_1\cdots i_m}A_m^{i_1\cdots i_m},\nonumber\\
 \cS_m[A_m]&=-\mathsf k_m^{i_1\cdots i_m}\widetilde A_m^{i_1\cdots i_m}.
 \label{eq:CS}
\end{align}
Set $\cC_0[A_0]=A_0$ and $\cS_0=0$. If $A_{Km,+}=s_{Km}\e^{\ii m\phi_{Km}}$, direct contraction gives
\begin{align}
 \cC_m[A_{Km}]&=\Rea(A_{Km,+}^*\kappa^m)\nonumber\\
 &=s_{Km}(|\bk|/\M)^m\cos m(\phi_k-\phi_{Km}),\nonumber\\
 \cS_m[A_{Km}]&=\Ima(A_{Km,+}^*\kappa^m)\nonumber\\
 &=s_{Km}(|\bk|/\M)^m\sin m(\phi_k-\phi_{Km}).
 \label{eq:CSharmonics}
\end{align}
For example, the rank-one sine contraction is $\cS_1[S_T]=-\epsilon_{ij}k^iS_T^j/\M$. The squared tensor norm is $A_m:A_m=2^{m-1}|A_{m,+}|^2$, so $s_{Km}$ is the magnitude of the complex component, not the full Euclidean tensor norm. Equations~\eqref{eq:Kn}--\eqref{eq:CSharmonics} specify all factors and signs used in the correlators and angular weights.

\subsection{Gauge-invariant correlators and projections}
\label{sec:operators}

For a mixed target state, we use the expectation value
\begin{equation}
 \langle\mathcal O\rangle_\rho
 =\sum_{\Lambda\Lambda'}\rho_{\Lambda\Lambda'}
 \langle P,\Lambda'|\mathcal O|P,\Lambda\rangle,
 \label{eq:ensemble}
\end{equation}
with state normalization
\begin{equation}
 \langle P',\Lambda'|P,\Lambda\rangle
 =2P^0(2\pi)^3\delta^{(3)}(\bm P'-\bm P)
 \delta_{\Lambda'\Lambda}.
\end{equation}
The expectation value is linear in $\rho$ and therefore in the irreducible moments $S$, $T$, and $R$. The operator decomposition then contains no products of independent polarization moments.

A Wilson line along a path $\mathscr P$ is
\begin{equation}
 U_{\mathscr P}[a,b]=\mathcal P\exp\left[-\ii g_s
 \int_a^b\dd z^\mu A_\mu^c(z)t^c\right],
 \label{eq:wilson}
\end{equation}
where $\mathcal P$ denotes ordering along the path $\mathscr P$, and the sign is fixed by the covariant-derivative convention. Distribution correlators in SIDIS and Drell--Yan have, respectively, future- and past-pointing staple links, including their transverse connectors. The link geometry and, for gluons, its color class form part of the TMD definition~\cite{Collins2002,Boer2016,Handbook}.

For a quark flavor $q$, the unsubtracted correlator is
\begin{equation}
 \begin{split}
 \Phi^{q,[U]}_{ab,\mathrm{unsub}}(x,\bk;\rho)
 ={}&\int\frac{\dd\xi^-\dd^2\xi_T}{(2\pi)^3}
 \e^{\ii xP^+\xi^- -\ii\bk\cdot\bm\xi_T}\\
 &\times\left.\langle\bar\psi_b^q(0)U[0,\xi]
 \psi_a^q(\xi)\rangle_\rho\right|_{\xi^+=0},
 \end{split}
 \label{eq:Phiunsub}
\end{equation}
where indices $a,b$ are Dirac indices, and color is contracted through $U$. The corresponding gluon definition is
\begin{equation}
 \begin{aligned}
 &\Gamma^{g,[U,U'],ij}_{\mathrm{unsub}}(x,\bk;\rho)\\
 &\quad=\frac2{P^+}\int\frac{\dd\xi^-\dd^2\xi_T}{(2\pi)^3}
 \e^{\ii xP^+\xi^- -\ii\bk\cdot\bm\xi_T}\\
 &\qquad\times\Bigl.\Bigl\langle\tr_c\bigl[
 F^{+i}(0)U[0,\xi]F^{+j}(\xi)U'[\xi,0]\bigr]
 \Bigr\rangle_\rho\Bigr|_{\xi^+=0}.
 \end{aligned}
 \label{eq:Gammaunsub}
\end{equation}
We use $\tr_c(t^at^b)=\delta^{ab}/2$. The prefactor $2/P^+$ converts the color trace to the color-contracted field-strength normalization when the links reduce to one adjoint transporter and makes the correlator have mass dimension $-2$, as for the quark correlator. The two-link notation includes the specified Wilson-loop and color structure of the process.

The leading-twist fields are the quark component $\psi_+=\gamma^-\gamma^+\psi/2$ and the transverse field strengths $F^{+i}$, whose partonic spin content determines the leading-twist projections. Powers of transverse momentum in their decomposition label angular correlations within this operator sector, rather than additional powers of $1/Q$. Exactly lightlike unsubtracted correlators also require a rapidity regulator, a soft subtraction, and ultraviolet (UV) renormalization. We work in a fixed subtraction prescription and suppress its common scale arguments only in the tensor formulas. In impact-parameter space, with $\widetilde\Phi=\int\dd^2\bk\,\e^{\ii\bk\cdot\bb}\Phi$, the quark definition has the form
\begin{equation}
 \widetilde\Phi^q_{\mathrm{ren}}
 =\lim_{\mathrm{reg}\to0}Z_q R_q\widetilde\Phi^q_{\mathrm{unsub}},
 \label{eq:subtraction}
\end{equation}
and similarly for gluons. Here, $Z_q$ is the UV counterterm, and $R_q$ includes the rapidity counterterm and the appropriate vacuum soft-factor combination. An isotropic subtraction preserves the physical tensor count. The full coefficient arguments are $(x,\bk^2;\mu,\zeta;[U])$ for quarks and $(x,\bk^2;\mu,\zeta;[U,U'])$ for gluons. Evanescent tensors used in dimensional regularization require a specified renormalization prescription but do not add physical two-dimensional coefficients~\cite{Handbook}.

With $\sigma^{\mu\nu}=\ii[\gamma^\mu,\gamma^\nu]/2$ and $\gamma_5=\ii\gamma^0\gamma^1\gamma^2\gamma^3$, define the quark number, helicity, and transverse-spin projections by
\begin{align}
 \calF^q&=\tfrac12\Tr_D(\Phi^q\gamma^+),\nonumber\\
 \calG^q&=\tfrac12\Tr_D(\Phi^q\gamma^+\gamma_5),\nonumber\\
 \calT^{q,i}&=\tfrac12\Tr_D(\Phi^q\ii\sigma^{i+}\gamma_5).
 \label{eq:qproject}
\end{align}
These projections reconstruct the leading-twist quark correlator as
\begin{equation}
 \Phi^q_{(2)}=\tfrac12\bigl[
 \calF^q\gamma^-+\calG^q\gamma_5\gamma^-
 +\calT^{q,i}\ii\sigma^{i-}\gamma_5\bigr],
 \label{eq:qreconstruct}
\end{equation}
and traces in Eq.~\eqref{eq:qproject} invert this expression. The leading-twist quark correlator therefore has four real partonic components: two scalars and the two components of $\calT^q$.

For gluons, the corresponding four-component decomposition is
\begin{equation}
 \Gamma^{g,ij}=\frac{x}{2}
 \bigl[\delta^{ij}\calF^g+\ii\epsilon_{ij}\calG^g+\calH^{g,ij}\bigr],
 \label{eq:gammaproj}
\end{equation}
where $\calH^{g,ij}=\calH^{g,ji}$ and $\calH^{g,ii}=0$. The inverse projections are
\begin{align}
 \calF^g&=\delta_{ij}\Gamma^{g,ij}/x,\nonumber\\
 \calG^g&=-\ii\epsilon_{ij}\Gamma^{g,ij}/x,\nonumber\\
 \calH^{g,ij}&=(2/x)\STF\Gamma^{g,ij}.
 \label{eq:gproject}
\end{align}
Hermiticity makes $\calF^g$, $\calG^g$, and the two components of $\calH^g$ real. The trace and antisymmetric parts describe unpolarized and circularly polarized gluons. The STF part describes interference between gluon helicities $+1$ and $-1$ and therefore carries two units of helicity change.

Each scalar TMD coefficient specifies a partonic projection, a target multipole, and a transverse harmonic. For example, the rank-three, $M=0$ projection of $\calG^q$ selects the $S_{LLL}$ coefficient, but the full helicity response of a state with $S_L\ne0$ may also contain a lower-rank term. Appendix~\ref{sec:inverse} gives the two helicity-flip inversions. The extraction of these operator coefficients from measured convolutions is discussed in Sec.~\ref{sec:nuclear}.

\section{Tensor basis and Wilson-line relations}
\label{sec:count}

\begin{table*}[!t]
\caption{Left: orbital STF ranks $n$ for each target transverse rank $m$ and parton channel. Right: numbers of independent scalar coefficients at target multipole rank $K$, valid separately for quarks and gluons. The helicity-flip projection is $\calT^q$ for quarks and $\calH^g$ for gluons.}
\label{tab:orbital}\label{tab:count}
\centering
\begin{ruledtabular}
\begin{tabular}{cccc@{\qquad\qquad}crrrr}
 \multicolumn{4}{c}{Orbital ranks}&\multicolumn{5}{c}{Independent coefficients}\\
 $m$&$\calF,\calG$ ($r=0$)&$\calT^q$ ($r=1$)&$\calH^g$ ($r=2$)
 &$K$&$\calF$&$\calG$&Helicity flip&Total\\
 0&0&1&2&0&1&0&1&2\\
 1&1&0,2&1,3&1&1&2&3&6\\
 2&2&1,3&0,4&2&3&2&5&10\\
 3&3&2,4&1,5&3&3&4&7&14\\
 \multicolumn{4}{c}{}&All ranks&8&8&16&32
\end{tabular}
\end{ruledtabular}
\end{table*}

\subsection{Complete leading-twist basis}

We construct the basis by matching target and orbital helicities to the parton projection, then imposing parity. This follows the helicity and discrete-symmetry organization of generalized correlators~\cite{LorcePasquini2013}. The parton projection has transverse rank $r=0$ for number and helicity, $r=1$ for quark transverse spin, and $r=2$ for gluon linear polarization. A target component of transverse rank $m>0$ has helicities $\pm m$, while an orbital STF tensor of rank $n>0$ has helicities $\pm n$. Their product then has helicities $\pm m\pm n$. Matching to output helicity $+r$ gives, for $r>0$,
\begin{equation}
 n=\begin{cases}
 r,&m=0,\\
 |m-r|\ \text{or}\ m+r,&m>0.
 \end{cases}
 \label{eq:rankrule}
\end{equation}
For $r=0$, the only orbital rank is $n=m$, while parity distinguishes the scalar and pseudoscalar contractions. When $m,r>0$, the two solutions of Eq.~\eqref{eq:rankrule} have different orbital harmonics and are independent at generic nonzero $k_T$. Additional powers of $\bk^2$ are absorbed into the scalar coefficients, so the rule exhausts the transverse covariants linear in target polarization. Table~\ref{tab:orbital} lists the allowed ranks. In two dimensions, STF products have only sum and difference harmonics. The intermediate ranks familiar from three-dimensional angular-momentum coupling do not occur, since in gluon linear polarization, there is no $n=2$ term for $m=2$ and no $n=3$ term for $m=3$.

Parity next fixes which covariants require a transverse dual. To keep the large target momentum along $+z$, combine spatial parity with a rotation by $\pi$ about $y$, which reflects the transverse coordinates as $(x,y)\mapsto(x,-y)$. Because spin is axial, a component of target-spin rank $K$ transforms as
\begin{equation}
 A_{Km,+}\mapsto(-1)^K A_{Km,+}^*,\qquad
 \kappa\mapsto\kappa^*.
 \label{eq:reflection}
\end{equation}
$\cC_m$ and $\cS_m$ then have reflection signs $(-1)^K$ and $(-1)^{K+1}$, respectively. The number projection is a scalar, the helicity projection a pseudoscalar, quark transverse spin an axial vector, and gluon linear polarization an ordinary rank-two tensor. The allowed scalar projections are then
\begin{equation}
 \begin{array}{c|cc}
 &K\text{ even}&K\text{ odd}\\ \hline
 \calF^{q,g}&\cC_m&\cS_m\\
 \calG^{q,g}&\cS_m&\cC_m
 \end{array}.
 \label{eq:parityscalar}
\end{equation}
For quark helicity flip, a dual is required for even $K$ and absent for odd $K$, while gluon helicity flip has the opposite assignment. Since $\cS_0=0$, the scalar projections have no odd-$K$, $m=0$ number term and no even-$K$, $m=0$ helicity term.

At fixed $K$, the number and helicity channels together contain $2K+1$ coefficients: one at $m=0$ and two at every $m>0$. Equation~\eqref{eq:rankrule} supplies another $2K+1$ in the helicity-flip channel for either parton species. The total is therefore
\begin{equation}
 N_K=4K+2,\qquad \sum_{K=0}^{3}N_K=32.
 \label{eq:count32}
\end{equation}
The right block of Table~\ref{tab:count} separates these coefficients by parton projection.

The number and helicity coefficients are denoted by $f^a_{Km}$ and $g^a_{Km}$, where $a=q,g$ identifies the parton species and, when specified, the quark flavor. Antiquarks carry the label $\bar q$, and the notation $q/A$ or $g/A$ also specifies the parent hadron. Helicity-flip coefficients are written $h^{a[n]}_{Km}$, where the bracket $[n]$ gives the explicit orbital rank, and a final square bracket in the superscript specifies the Wilson link, as in $h_{Km}^{q[n],[-]}$. Upright labels such as $\mathrm{SIDIS}$ and $\mathrm{DY}$ identify processes. All TMD coefficients have mass dimension $-2$ and the scale and link arguments specified in Sec.~\ref{sec:operators}. Appendix~\ref{sec:inverse} relates these labels to conventional names.

The complete scalar projections, common in form to quarks and gluons, are
\begin{align}
 \calF^a={}&f^a_{00}+\cS_1[S_T]f^a_{11}+S_{LL}f^a_{20}
 \nonumber\\
 &+\cC_1[S_{LT}]f^a_{21}+\cC_2[S_{TT}]f^a_{22}
 \nonumber\\
 &+\cS_1[S_{LLT}]f^a_{31}+\cS_2[S_{LTT}]f^a_{32}
 \nonumber\\
 &+\cS_3[S_{TTT}]f^a_{33},
 \label{eq:Ffull}\\
 \calG^a={}&S_Lg^a_{10}+\cC_1[S_T]g^a_{11}
 \nonumber\\
 &+\cS_1[S_{LT}]g^a_{21}+\cS_2[S_{TT}]g^a_{22}
 \nonumber\\
 &+S_{LLL}g^a_{30}+\cC_1[S_{LLT}]g^a_{31}
 \nonumber\\
 &+\cC_2[S_{LTT}]g^a_{32}+\cC_3[S_{TTT}]g^a_{33}.
 \label{eq:Gfull}
\end{align}
While these forms are shared by quarks and gluons, their coefficient functions are independent. Eqs.~\eqref{eq:CS}--\eqref{eq:CSharmonics} fix the tensor normalization. For example, $\calF^q_{33}=-\mathsf k_3^{abc}\epsilon_{ad}S_{TTT}^{dbc}f^q_{33}$, whereas $\calG^q_{33}=\mathsf k_3^{abc}S_{TTT}^{abc}g^q_{33}$. The octupole number coefficients $f^a_{3m}$ describe momentum asymmetries of unpolarized partons, while $g^a_{3m}$ resolve parton helicity in the same target-spin sector. Each is a linear response to the corresponding component of $\rho$.

With the gluon trace and helicity sectors specified by Eqs.~\eqref{eq:Ffull} and \eqref{eq:Gfull}, the remaining sector is linear polarization. Define ordinary STF tensors $\mathcal B_{(K)}^{ij}$ and impose the parity assignment by writing
\begin{equation}
 \calH^{g,ij}_{(K)}=
 \begin{cases}
 \mathcal B_{(K)}^{ij},&K\ \text{even},\\
 \widetilde{\mathcal B}_{(K)}^{ij},&K\ \text{odd}.
 \end{cases}
 \label{eq:Hparity}
\end{equation}
Their explicit forms are
\begin{align}
 \mathcal B_{(0)}^{ij}={}&\mathsf k_2^{ij}\gh{0}{0}{2},
 \label{eq:H0}\\
 \mathcal B_{(1)}^{ij}={}&S_L\mathsf k_2^{ij}\gh{1}{0}{2}
 +\STF(S_T^i\mathsf k_1^j)\gh{1}{1}{1}
 \nonumber\\
 &+\mathsf k_3^{ija}S_T^a\gh{1}{1}{3},
 \label{eq:H1}\\
 \mathcal B_{(2)}^{ij}={}&S_{LL}\mathsf k_2^{ij}\gh{2}{0}{2}
 +\STF(S_{LT}^i\mathsf k_1^j)\gh{2}{1}{1}
 \nonumber\\
 &+\mathsf k_3^{ija}S_{LT}^a\gh{2}{1}{3}
 +S_{TT}^{ij}\gh{2}{2}{0}
 \nonumber\\
 &+\mathsf k_4^{ijab}S_{TT}^{ab}\gh{2}{2}{4},
 \label{eq:H2}\\
 \mathcal B_{(3)}^{ij}={}&S_{LLL}\mathsf k_2^{ij}\gh{3}{0}{2}
 +\STF(S_{LLT}^i\mathsf k_1^j)\gh{3}{1}{1}
 \nonumber\\
 &+\mathsf k_3^{ija}S_{LLT}^a\gh{3}{1}{3}
 +S_{LTT}^{ij}\gh{3}{2}{0}
 \nonumber\\
 &+\mathsf k_4^{ijab}S_{LTT}^{ab}\gh{3}{2}{4}
 +S_{TTT}^{ija}\mathsf k_1^a\gh{3}{3}{1}
 \nonumber\\
 &+\mathsf k_5^{ijabc}S_{TTT}^{abc}\gh{3}{3}{5}.
 \label{eq:B3}
\end{align}
The dual acts on one free index as in Eq.~\eqref{eq:dual}. Substitution of $\calH^g=\sum_K\calH^g_{(K)}$ into Eq.~\eqref{eq:gammaproj} completes the gluon correlator. The seven terms in $\mathcal B_{(3)}$, together with the three number and four helicity terms, give fourteen octupole gluon coefficients. The seven linear-polarization coefficients describe interference between gluon helicities differing by two, balanced by target coherence and orbital momentum. Quark transversity instead involves a one-unit parton helicity change, although its octupole sector also has seven helicity-flip coefficients. At finite $k_T$, the gluon octupole thus extends the lower-spin parametrizations with two orbital branches in each nonzero-$m$ sector.

For $m=2$, the orbital ranks are zero and four: $h^{g[0]}_{32}$ couples to $LTT$ without an explicit momentum tensor, while $h^{g[4]}_{32}$ carries the fourth harmonic. For $m=3$, the ranks are one and five, and the low-rank $TTT$ term is $S_{TTT}^{ija}k^a/\M$. The $TTT$ pair $h^{g[1]}_{33},h^{g[5]}_{33}$ couples the same three-unit target coherence to different orbital transfers. The straight-link collinear limit imposes further constraints (see Sec.~\ref{sec:collinear}), and Sec.~\ref{sec:gluonSF} develops their separation by a specified hard analyzer.

Writing $\calT^q=\sum_{K=0}^{3}\calT^q_{(K)}$, the sixteen quark transverse-spin terms are
\begin{align}
 \calT^{q,i}_{(0)}={}&\epsilon_{ia}\mathsf k_1^a\qq{0}{0}{1},
 \label{eq:T0}\\
 \calT^{q,i}_{(1)}={}&S_L\mathsf k_1^i\qq{1}{0}{1}
 +S_T^i\qq{1}{1}{0}+\mathsf k_2^{ia}S_T^a\qq{1}{1}{2},
 \label{eq:T1}\\
 \calT^{q,i}_{(2)}={}&\epsilon_{ia}\bigl[
 S_{LL}\mathsf k_1^a\qq{2}{0}{1}+S_{LT}^a\qq{2}{1}{0}
 \nonumber\\
 &\quad+\mathsf k_2^{ab}S_{LT}^b\qq{2}{1}{2}
 +S_{TT}^{ab}\mathsf k_1^b\qq{2}{2}{1}
 \nonumber\\
 &\quad+\mathsf k_3^{abc}S_{TT}^{bc}\qq{2}{2}{3}\bigr],
 \label{eq:T2}\\
 \calT^{q,i}_{(3)}={}&S_{LLL}\mathsf k_1^i\qq{3}{0}{1}
 +S_{LLT}^i\qq{3}{1}{0}
 \nonumber\\
 &+\mathsf k_2^{ia}S_{LLT}^a\qq{3}{1}{2}
 +S_{LTT}^{ia}\mathsf k_1^a\qq{3}{2}{1}
 \nonumber\\
 &+\mathsf k_3^{iab}S_{LTT}^{ab}\qq{3}{2}{3}
 +S_{TTT}^{iab}\mathsf k_2^{ab}\qq{3}{3}{2}
 \nonumber\\
 &+\mathsf k_4^{iabc}S_{TTT}^{abc}\qq{3}{3}{4}.
 \label{eq:T3}
\end{align}
The displayed signs define the coefficients in our Euclidean and dual conventions, with parity requiring the quadrupole dual. Absence of an explicit transverse momentum in an orbital-rank-zero term does not alone ensure its collinear survival, as time reversal also constrains it. The seven terms in Eq.~\eqref{eq:T3}, together with three number and four helicity coefficients, give the fourteen-function quark octupole used in Sec.~\ref{sec:sidis}. Its channels match the quark fragmentation classification~\cite{Zhao2022}, but the matrix elements here describe an incoming target and have distribution-side link properties. Appendix~\ref{app:literature} gives the channel and projection conversion.

\subsection{Independence and relation to lower-spin conventions}
\label{sec:redundancy}

The helicity construction above gives a spanning, independent basis at generic transverse momentum. A separate upper bound follows from the combined parton--target spin matrix. Two parton polarizations and four target states form an $8\times8$ Hermitian matrix with 64 real parameters. Taking $\bk$ along $x$, the parity-preserving reflection exchanges opposite helicities. After irrelevant helicity phases are removed, its two eigenspaces each have dimension four. Reflection invariance makes the Hermitian matrix block diagonal in these spaces, leaving $4^2+4^2=32$ real parameters. The displayed basis attains this bound, and exact coefficient-to-correlator maps provide a further finite-dimensional check~\cite{Li7TMDs}.

Comparison with lower-spin parametrizations requires some care because terms that appear distinct before the STF projection can become redundant afterward. In the physical two-dimensional transverse plane, two STF matrices,
\begin{equation}
 A=\begin{pmatrix}a&b\\b&-a\end{pmatrix},\qquad
 B=\begin{pmatrix}c&d\\d&-c\end{pmatrix},
\end{equation}
satisfy
\begin{equation}
 AB=(ac+bd)I+(ad-bc)E.
 \label{eq:AB}
\end{equation}
Hence, $\STF(AB)=\STF(EAB)=0$. Likewise, for rank-three STF tensors with complex components $a+\ii b$ and $c+\ii d$,
\begin{equation}
 A_3^{iab}B_3^{jab}
 =2\bigl[(ac+bd)\delta^{ij}+(ad-bc)\epsilon_{ij}\bigr].
 \label{eq:A3B3}
\end{equation}
After STF projection on the free indices, these identities eliminate the proposed intermediate-rank tensors built from $S_{TT}\mathsf k_2$, its octupole analogue $S_{LTT}\mathsf k_2$, and the rank-three contraction $S_{TTT}\mathsf k_3$, which vanish identically in the physical transverse plane.

In this subsection and Appendix~\ref{app:literature}, the upright superscript $\mathrm{B}$ labels the convention of Boer \emph{et al.}~\cite{Boer2016}, also used in Ref.~\cite{Cotogno2017}. Their spin-1 gluon parametrization includes $S_{TT}^{\{i}{}_{\alpha}k_T^{j\}\alpha}h_{1TT}^{\perp g,\mathrm{B}}$, with braces denoting the unnormalized symmetric sum. The key point is that the two-index momentum is a rank-two STF tensor, rather than the dyadic $k_T^j k_T^\alpha$:
\begin{equation}
 k_T^{j\alpha}\equiv k_T^j k_T^\alpha
 +\frac{\bk^2}{2}g_T^{j\alpha}
 =M_A^2\mathsf k_2^{j\alpha}.
 \label{eq:referenceSTF}
\end{equation}
This is the definition in Eq.~(95) of Ref.~\cite{Boer2016}. For transverse Cartesian components, $S_{TT}^{i}{}_{\alpha}=-S_{TT}^{i\alpha}$, so the symmetric contraction is purely trace. Keeping the reference's polarization normalization fixed and converting its Lorentz index contractions consistently, denote the resulting independent trace coefficient by $\mathrm{min}$. Then,
\begin{equation}
 f_{1TT}^{g,\mathrm{min}}
 =f_{1TT}^{g,\mathrm{B}}-h_{1TT}^{\perp g,\mathrm{B}}.
 \label{eq:litcombination}
\end{equation}
The same difference appears in the corresponding helicity matrix and positivity relation~\cite{Cotogno2017}. A symmetric $LT$ term also contains both trace and STF parts,
\begin{equation}
 S^ik^j+S^jk^i=2\STF(S^ik^j)+\delta^{ij}S\cdot k.
\end{equation}
Separating these parts shifts the scalar $LT$ coefficient and fixes the normalization of its linear-polarization partner. The eighteen independent coefficients with $K\leq2$ therefore reorganize the lower-spin structures without redundancy. Comparisons with published model functions require these coefficient transformations~\cite{Xie2026}, with the full conversion dictionary for all three gluon polarization projections given in Appendix~\ref{app:literature}.

\subsection{Wilson-line reversal}
\label{sec:universality}

The tensor basis fixes the allowed polarization dependence, while operator universality relates matrix elements with identical or symmetry-related field, link, and subtraction definitions. Using these relations to make cross-process predictions also requires factorization formulas that identify the relevant operators in each measurement~\cite{Collins2002,CollinsMetz2004}. We first derive the target-spin transformation, which needs no nuclear model. Under $\Theta=PT$, the intrinsic target parity cancels between the bra and ket. In the ordered spin basis, its spin action is $\mathsf U_T\mathsf K$, where $\mathsf K$ denotes complex conjugation, and
\begin{equation}
 (\mathsf U_T)_{\Lambda'\Lambda}
 =(-1)^{J-\Lambda}\delta_{\Lambda',-\Lambda}.
 \label{eq:PTspin}
\end{equation}
The transformed density matrix is $\rho^\Theta=\mathsf U_T\rho^*\mathsf U_T^\dagger$. Since $\mathsf U_TJ_a^*\mathsf U_T^\dagger=-J_a$, the operators in Eqs.~\eqref{eq:Q} and \eqref{eq:QO} give
\begin{equation}
 S_a^\Theta=-S_a,\qquad T_{ab}^\Theta=T_{ab},\qquad
 R_{abc}^\Theta=-R_{abc}.
 \label{eq:PTmoments}
\end{equation}
Thus, every real Cartesian component of $A_{Km}$ acquires $(-1)^K$, including off-diagonal coherences as well as longitudinal populations. The transformation preserves positivity and normalization of $\rho$.

Combined $PT$ leaves the momentum labels unchanged and reverses the Wilson paths. Denote the complete transformed link and color-trace specification by $\mathcal U^\Theta$. The field transformation also conjugates the explicit Dirac and transverse factors because it is antiunitary. For a real projected component $\mathcal X=\calF,\calG,\calT^i$, or $\calH^{ij}$, it gives
\begin{equation}
 \mathcal X^{a,[\mathcal U^\Theta]}(x,\bk;\rho^\Theta)
 =\eta_{\mathcal X}\mathcal X^{a,[\mathcal U]}(x,\bk;\rho),
 \label{eq:PTprojection}
\end{equation}
with $\eta_{\calF}=\eta_{\calH}=+1$ and $\eta_{\calG}=\eta_{\calT}=-1$ in the conventions of Sec.~\ref{sec:operators}. For example, the explicit $\ii$ multiplying the antisymmetric gluon tensor must be conjugated, as its helicity coefficient cannot be classified from the target tensor alone. Resolving Eq.~\eqref{eq:PTprojection} on the independent real covariants gives
\begin{align}
 f_{Km}^{a,[\mathcal U^\Theta]}&=(-1)^K f_{Km}^{a,[\mathcal U]},\nonumber\\
 g_{Km}^{a,[\mathcal U^\Theta]}&=(-1)^{K+1}g_{Km}^{a,[\mathcal U]},\nonumber\\
 h_{Km}^{q[n],[\mathcal U^\Theta]}&=(-1)^{K+1}h_{Km}^{q[n],[\mathcal U]},\nonumber\\
 h_{Km}^{g[n],[\mathcal U^\Theta]}&=(-1)^K h_{Km}^{g[n],[\mathcal U]}.
 \label{eq:Tparity}
\end{align}
There is no additional orbital factor $(-1)^n$, since combined $PT$ leaves the momentum unchanged, and the real covariants already include the required duals. The lower-spin limits reproduce the standard classifications~\cite{Bacchetta2007,Boer2016}. At spin $3/2$, the quark basis has twenty link-even and twelve link-odd coefficients, and the gluon basis has sixteen of each. Link-odd functions can be nonzero for a fixed staple direction, and reversal relates their signs to those of the reversed operator.

For the standard quark TMD definitions, current-fragmentation SIDIS selects a future-pointing staple $[+]$, whereas color-singlet continuum Drell--Yan selects a past-pointing staple $[-]$. This identification follows from the eikonal color flow and the factorization analysis~\cite{Collins2002,CollinsMetz2004}. Applying Eq.~\eqref{eq:Tparity} to $K=3$ yields
\begin{align}
 f_{3m}^{q,[-]}&=-f_{3m}^{q,[+]},&&m=1,2,3,\nonumber\\
 g_{3m}^{q,[-]}&=+g_{3m}^{q,[+]},&&m=0,1,2,3,\nonumber\\
 h_{3m}^{q[n],[-]}&=+h_{3m}^{q[n],[+]},&&(m,n)\in\mathscr I_q,
 \label{eq:octSIDISDY}
\end{align}
where the seven helicity-flip labels are
\begin{equation}
 \mathscr I_q=\{(0,1),(1,0),(1,2),(2,1),(2,3),(3,2),(3,4)\}.
 \label{eq:qoctindices}
\end{equation}
Each equality compares the same target and flavor at the same $x$, $\bk^2$, $\mu$, and $\zeta$, in the same polarization convention. The relations hold separately for antiquarks because time reversal does not exchange quarks and antiquarks. They give three sign-reversing number-channel correlations and eleven sign-preserving helicity or transverse-spin correlations. The three $f_{3m}$ are octupole analogues of Sivers-type momentum asymmetries. Their reversal follows from operator symmetry whether or not a nuclear model predicts a nonzero value. The seven quark $h$ functions are link even in this sector, although the unpolarized-target Boer--Mulders function is link odd. The full operator and target structure determine naive time-reversal behavior, not just the function's letter or transverse-momentum power alone. Table~\ref{tab:octuniversality} collects these assignments. These laws apply to the complete nuclear matrix element, including binding, spin recoupling, and rescattering in a $^{7}$Li TMD. A reduction to nucleon distributions must reproduce the same transformation within its stated approximation. Rotating a calibrated target state or changing the experimental azimuth convention is a separate operation that must be done before applying Eq.~\eqref{eq:octSIDISDY}.

\begin{table}[tb]
\caption{Octupole coefficients under complete Wilson-line reversal. The number of coefficients is per indicated parton species and link class. For quarks, the reversal relates the standard SIDIS and DY staples. For gluons, it relates a specified pair of reversed link/color operators.}
\label{tab:octuniversality}
\centering
\begin{ruledtabular}
\begin{tabular}{lccc}
 Parton channel & Number & Quark sign & Gluon sign\\
 $f_{3m}$, $m=1,2,3$ & 3 & $-$ & $-$\\
 $g_{3m}$, $m=0,1,2,3$ & 4 & $+$ & $+$\\
 $h_{3m}^{[n]}$, allowed $(m,n)$ & 7 & $+$ & $-$\\
 Link even & & 11 & 4\\
 Link odd & & 3 & 10
\end{tabular}
\end{ruledtabular}
\end{table}

For gluons, reversal must specify both Wilson paths and the color trace. In the shorthand $[\eta_1,\eta_2]$, the path from $0$ to $\xi$ is $U^{[\eta_1]}$ and the return path is $(U^{[\eta_2]})^\dagger$, with $\eta_i=+,-$. Complete reversal maps
\begin{equation}
 [+,+]\longleftrightarrow[-,-],\qquad
 [+,-]\longleftrightarrow[-,+].
 \label{eq:gluonlinkpairs}
\end{equation}
For either such pair, or a fully specified reversed traced-loop operator, Eq.~\eqref{eq:Tparity} gives
\begin{align}
 f_{3m}^{g,[\mathcal U^\Theta]}&=-f_{3m}^{g,[\mathcal U]},\nonumber\\
 g_{3m}^{g,[\mathcal U^\Theta]}&=+g_{3m}^{g,[\mathcal U]},\nonumber\\
 h_{3m}^{g[n],[\mathcal U^\Theta]}&=-h_{3m}^{g[n],[\mathcal U]}.
 \label{eq:octgluonreverse}
\end{align}
Here, the allowed helicity-flip labels are
\begin{equation}
 \mathscr I_g=\{(0,2),(1,1),(1,3),(2,0),(2,4),(3,1),(3,5)\}.
 \label{eq:goctindices}
\end{equation}
In particular, octupole linear gluon polarization is link odd, unlike linear gluon polarization in an unpolarized target. Different paths or traced loops can define independent matrix elements, including several color structures at the same transverse rank~\cite{Buffing2013,Boer2016}. The 32-dimensional basis applies separately to each specified operator, rather than counting all
functions across all link geometries. Definite-rank generalized-universality constructions organize this color dependence through additional operator matrix elements and process-dependent factors~\cite{Buffing2013}. Our basis resolves their target dependence, but it doesn't supply the complete rank-five color-operator expansion.

The heavy-pair leptoproduction calculation below uses the $[+,+]$ class. Comparison with a reaction selecting $[-,-]$ can use Eq.~\eqref{eq:octgluonreverse} once both responses are factorized into the corresponding operators, although a mixed-direction class requires additional nonperturbative input. Ordinary electromagnetic DY at Born order annihilates a quark and an antiquark and therefore does not directly analyze these gluon TMDs. Gluon contributions from perturbative matching or other hard channels require their own factorization formula. A gauge-invariant single-target correlator does not suffice for a measured hadron--hadron reaction, where color entanglement can obstruct factorization for observed colored final-state systems~\cite{Rogers2010}. This limitation leaves the reversal identity of a defined operator intact and does not rule out colored final states in lepton--hadron scattering. Their soft factors and recoil measurement still require explicit treatment.

\subsection{Renormalization, evolution, and observable comparisons}
\label{sec:renormalization}

The operator relation in Eq.~\eqref{eq:octSIDISDY} extends beyond the Born hard approximation when the reversed two-line quark staples use a common $PT$-symmetric regulator, UV counterterms, and rapidity subtraction~\cite{CollinsMetz2004}. Comparing between different subtraction schemes requires a finite conversion. For general gluon color operators, reversal acts on the full subtraction prescription, without assuming a universal scalar soft factor. For standard two-line TMDs at nonzero $b_T$, evolution is linear and independent of target spin within a fixed parton representation~\cite{Echevarria2014,Echevarria2015}. To apply this statement to the basis, let $\widetilde C_\nu$ be the coefficient of a fixed impact-parameter tensor, with the Fourier phase kept in the tensor. The evolution operator $\mathscr E_a$ then gives, for quarks,
\begin{equation}
 \begin{split}
 \widetilde C_\nu^{[-]}(\mu,\zeta)
 &=\mathscr E_q\,\widetilde C_\nu^{[-]}(\mu_0,\zeta_0)\\
 &=s_\nu\mathscr E_q\,\widetilde C_\nu^{[+]}(\mu_0,\zeta_0)
 =s_\nu\widetilde C_\nu^{[+]}(\mu,\zeta),
 \end{split}
 \label{eq:universalityevolution}
\end{equation}
where $s_\nu$ is the quark sign in Table~\ref{tab:octuniversality}. The reversal sign is therefore preserved when the two operators are compared at matched scales. This does not require momentum-space TMD ratios to remain fixed across scales, and matching is still channel dependent (Sec.~\ref{sec:evolution}). Higher orbital ranks need their own matching calculations, and sharing a link sign does not identify a high-rank coefficient with a single ordinary Sivers or twist-three matrix element.

To connect these operator relations to measurements, the two quark processes must also include their different companion operators:
\begin{align}
 W_{\mathrm{SIDIS}}&\sim\Phi_A^{q,[+]}\otimes H_{\mathrm{SIDIS}}\otimes\Delta_h^q,
 \nonumber\\
 W_{\mathrm{DY}}&\sim\Phi_A^{q,[-]}\otimes H_{\mathrm{DY}}
                 \otimes\bar\Phi_B^{q,[-]}+(q\leftrightarrow\bar q).
 \label{eq:processoperators}
\end{align}
Secs.~\ref{sec:sidis} and \ref{sec:DY} give the normalization and convolutions for these schematic products. SIDIS pairs the target distribution with a fragmentation function, whereas DY pairs it with a second-hadron distribution. Distribution-link reversal supplies no sign rule for the Collins fragmentation function, whose out-state matrix element requires a different symmetry analysis~\cite{CollinsMetz2004}. A universality test compares calibrated structure functions using common target coefficients constrained by Eq.~\eqref{eq:octSIDISDY}, consistent evolution, and the respective process convolutions. A target-distribution sign relation does not generally imply the same sign relation for measured asymmetries. Beyond the displayed Born kernels, predictions also require perturbative hard factors, recoil resummation, and assessment of power corrections.

\section{Gluon octupole observables}
\label{sec:gluonSF}

\subsection{Hard matrix and Born analyzing powers}

To turn the gluon basis into observables, we combine a hard polarization analyzer with target preparations that isolate rank three. Heavy-quark-pair and dijet production in lepton--hadron scattering provide suitable channels~\cite{Boer2011,Pisano2013}. Here, we calculate the heavy-pair Born analyzer, then combine opposite octupole preparations, beam-helicity reversal, and angular projections to separate the responses. The process selects a future-pointing $[+,+]$ link pair through its final-state color flow~\cite{Pisano2013,Cotogno2017}, with the second link conjugated as required by the field ordering in Eq.~\eqref{eq:Gammaunsub}. This operator choice is fixed throughout the section.

We use a right-handed frame with the target and incoming collinear gluon along $+z$, the virtual photon along $-z$, and $+x$ in the lepton plane. Before specifying detailed hard kinematics, write the target's transverse gluon polarization matrix as
\begin{equation}
 D_g=\frac{\Gamma^g}{x}
 =\frac12\begin{pmatrix}
 \calF^g+\calH^{g,xx}&\calH^{g,xy}+\ii\calG^g\\
 \calH^{g,xy}-\ii\calG^g&\calF^g-\calH^{g,xx}
 \end{pmatrix}.
 \label{eq:gD}
\end{equation}
The hard scattering is described by a separate polarization matrix. If $\mathcal M_i$ is the hard amplitude for incoming Cartesian gluon polarization $i$, define $B_{ij}=\sum_X\mathcal M_i\mathcal M_j^*$, summing over unobserved final-state quantum numbers. The field ordering in Eq.~\eqref{eq:Gammaunsub} puts the conjugate-amplitude index first in the correlator, so the polarization response is
\begin{equation}
 \mathcal W_g=\Tr_2(BD_g)=\sum_{ij}B_{ij}(D_g)_{ji}.
 \label{eq:hardcontraction}
\end{equation}
For a pure incoming polarization $e_i$, the Gram matrix $e_i^*e_j$ gives the squared amplitude $|\sum_i\mathcal M_ie_i|^2$ through Eq.~\eqref{eq:hardcontraction}. This checks the index order and fixes the circular-polarization sign.

The Stokes decomposition resolves the hard matrix into its unpolarized, circular, and two linear-polarization components~\cite{LyubovitskijSchmidt2021}:
\begin{equation}
 B=b_UI+b_G\ii E+b_C\sigma_3+b_S\sigma_1,
 \label{eq:Bhard}
\end{equation}
where $\sigma_3=\diag(1,-1)$ and $\sigma_1$ has unit off-diagonal entries. Its real coefficients are
\begin{align}
 b_U&=\tfrac12(B_{xx}+B_{yy}),&b_G&=\Ima B_{xy},\nonumber\\
 b_C&=\tfrac12(B_{xx}-B_{yy}),&b_S&=\Rea B_{xy}.
 \label{eq:bproject}
\end{align}
Their contraction with the target matrix gives
\begin{align}
 \mathcal W_g&=b_U\calF^g+b_G\calG^g
              +b_C\calH^{g,xx}+b_S\calH^{g,xy}\nonumber\\
 &=b_U\calF^g+b_G\calG^g+\Rea(b_+^*\calH_+^g),
 \label{eq:gmaster}
\end{align}
with $b_+=b_C+\ii b_S$ and $\calH_+^g=\calH^{g,xx}+\ii\calH^{g,xy}$.

\begin{table}[t]
\caption{Gluon linear-polarization kernels in Eq.~\eqref{eq:gkernelrule}. For each row, $m\leq K\leq3$. The coefficients $c_{mn}$ follow from the STF normalization.}
\label{tab:gkernels}
\centering
\begin{ruledtabular}
\begin{tabular}{cccl}
 $m$&$n$&$c_{mn}$&$\Psi$\\
 0&2&$1/2$&$2\phi_k-2\phi_B$\\
 1&1&$1/2$&$\phi_k+\phi_{K1}-2\phi_B$\\
 1&3&$1/4$&$3\phi_k-\phi_{K1}-2\phi_B$\\
 2&0&$1$&$2\phi_{K2}-2\phi_B$\\
 2&4&$1/4$&$4\phi_k-2\phi_{K2}-2\phi_B$\\
 3&1&$1$&$3\phi_{K3}-\phi_k-2\phi_B$\\
 3&5&$1/4$&$5\phi_k-3\phi_{K3}-2\phi_B$
\end{tabular}
\end{ruledtabular}
\end{table}

The hard coefficients determine which parton polarizations the measurement can analyze, while the target structure remains in $D_g$. For example, $b_G=0$ removes sensitivity to all eight gluon helicity coefficients. For the linear-polarization harmonics, write $b_+=b_{\mathrm{lin}}\exp(2\ii\phi_B)$, with $b_{\mathrm{lin}}=|b_+|$, and set $t=|\bk|/M_A$. Each coefficient contributes
\begin{equation}
 b_{\mathrm{lin}}\,s_{Km}\,c_{mn}\,t^n h^{g[n]}_{Km}
 \begin{cases}
 \cos\Psi,&K\text{ even},\\
 \sin\Psi,&K\text{ odd},
 \end{cases}
 \label{eq:gkernelrule}
\end{equation}
where $s_{00}=1$ and $s_{K0}=A_{K0}$ are signed scalar amplitudes. For $m>0$, $s_{Km}=|A_{Km,+}|$ is the complex-component magnitude defined in Eq.~\eqref{eq:CSharmonics}. Table~\ref{tab:gkernels} gives the normalization and phase of each harmonic. The sine for odd $K$ comes from the dual factor $-\ii$. Applying each row to all $K\geq m$ up to three generates the sixteen helicity-flip terms. Multiplying Eq.~\eqref{eq:Ffull} by $b_U$ gives the eight number terms, and multiplying Eq.~\eqref{eq:Gfull} by $b_G$ gives the eight helicity terms. These rules give the complete response for any calculated $B$. The hard matrix fixes $\phi_B$, which need not be a final-state momentum azimuth. When $b_{\mathrm{lin}}=0$, the linear-polarization response vanishes and the angle is immaterial.

We now calculate the hard matrix for $\ell g\to\ell'Q\bar Q$. Let $q$ and $k$ be the virtual-photon and incoming collinear-gluon momenta and $p_1,p_2$ be the final heavy-quark momenta, with mass $m_Q$. The two photon--gluon-fusion diagrams give the vertex with coupling and color factors stripped off~\cite{Boer2011,Pisano2013}:
\begin{equation}
 V^{\mu\alpha}=
 \gamma^\mu\frac{\slashed p_1-\slashed q+m_Q}
 {(p_1-q)^2-m_Q^2}\gamma^\alpha +\gamma^\alpha\frac{\slashed p_1-\slashed k+m_Q}
 {(p_1-k)^2-m_Q^2}\gamma^\mu,
 \label{eq:bornvertex}
\end{equation}
with $\overline V=\gamma^0V^\dagger\gamma^0$. The two terms correspond to the two orderings of the photon and gluon vertices on the heavy-quark line. The amplitude contains $\bar u(p_1)t^aV^{\mu\alpha}v(p_2)$ contracted with the lepton current and incoming gluon polarization. The cancellation between the diagrams ensures gauge invariance. Define $r_t=p_1-q$ and $r_u=p_1-k$. For the photon contraction, the on-shell spinors obey $\bar u\slashed q=\bar u(m_Q-\slashed r_t)$ and $\slashed qv=(\slashed r_u-m_Q)v$. These factors cancel the propagator denominators, leaving $-\bar u\gamma^\alpha v$ and $+\bar u\gamma^\alpha v$ from the two diagrams. The gluon contraction likewise gives $+\bar u\gamma^\mu v$ and $-\bar u\gamma^\mu v$. With $p_1+p_2=q+k$, the resulting Ward identities are
\begin{equation}
 \bar u(p_1)q_\mu V^{\mu\alpha}v(p_2)=0,\qquad
 \bar u(p_1)k_\alpha V^{\mu\alpha}v(p_2)=0.
 \label{eq:wards}
\end{equation}
These identities apply to the full amplitude between on-shell external states, rather than to each diagram separately.

For a massless positive-energy incoming electron, the physical helicity label $\lambda=\pm1$ is the eigenvalue of $\operatorname{diag}(\bm\sigma\cdot\hat{\bm l},\bm\sigma\cdot\hat{\bm l})$. With $j_\mu=\bar u(l')\gamma_\mu u_\lambda(l)$, sum the final-electron helicities to define $L_{\mu\nu}=\sum_{h'}j_\mu^*j_\nu$. The conjugate current comes first in this definition, whereas the Born hard trace below has the amplitude first and therefore contracts with $L_{\nu\mu}=\sum_{h'}j_\mu j_\nu^*$. Explicitly,
\begin{equation}
 L_{\mu\nu}=2\bigl(l_\mu l'_\nu+l_\nu l'_\mu-g_{\mu\nu}l\cdot l'
 +\ii\lambda\epsilon_{\mu\nu\alpha\beta}l^\alpha l'^\beta\bigr).
 \label{eq:leptonic}
\end{equation}
Summing final spins and colors and averaging the incoming color gives
\begin{equation}
 \begin{split}
 B_{ij}={}&\frac{e^4e_Q^2g_s^2T_F}{Q^4}\,L_{\nu\mu}\\
 &\times\Tr_D\bigl[(\slashed p_1+m_Q)V^{\mu i}
 (\slashed p_2-m_Q)\overline V^{\nu j}\bigr],
 \end{split}
 \label{eq:BornB}
\end{equation}
The interchange matches the amplitude-first heavy trace to the conjugate-current-first $L_{\mu\nu}$. The SIDIS hard trace uses the reverse order, already matched by $L_{\mu\nu}$. The color trace $\sum_a\tr_c(t^at^a)=T_F(N_c^2-1)$, with $T_F=1/2$, gives the factor in Eq.~\eqref{eq:BornB} after averaging over incoming gluon colors. The hard matrix is not averaged over gluon spin, an unpolarized incoming gluon instead being represented by $D_{\mathrm{in}}=I/2$. The on-shell partonic cross section is
\begin{align}
 \dd\widehat\sigma[D_{\mathrm{in}}]&=
 \frac{\dd\Pi_3}{2\widehat s_{\ell g}}\Tr_2(BD_{\mathrm{in}}), \nonumber\\
 \dd\Pi_3&=(2\pi)^4\delta^{(4)}(l+k-l'-p_1-p_2)\nonumber\\
 &\quad\times\prod_{r=l',p_1,p_2}
 \frac{\dd^3\bm r}{(2\pi)^3\,2r^0},
 \label{eq:Bornphase}
\end{align}
with $\widehat s_{\ell g}=2l\cdot k$. Equations~\eqref{eq:leptonic}, \eqref{eq:bornvertex}, \eqref{eq:BornB}, and \eqref{eq:bproject} now specify the complete Born analyzer, including its normalization. Since $B$ is an amplitude Gram matrix, it is positive semidefinite, with eigenvalues
\begin{equation}
 \lambda_\pm(B)=b_U\pm\sqrt{b_G^2+b_C^2+b_S^2}.
\end{equation}
For $b_U>0$, the dimensionless analyzing vector is
\begin{equation}
 \bm{\mathcal A}_g=\frac{(b_G,b_C,b_S)}{b_U},\qquad |\bm{\mathcal A}_g|\leq1.
 \label{eq:analyzingbound}
\end{equation}
The components measure the hard process's sensitivity to circular and linear gluon polarization at the chosen kinematics. The bound follows from the perturbative hard matrix, requiring no assumption of pointwise positivity for a subtracted target TMD.

\begin{figure}[t]
 \centering
 \includegraphics[width=\columnwidth]{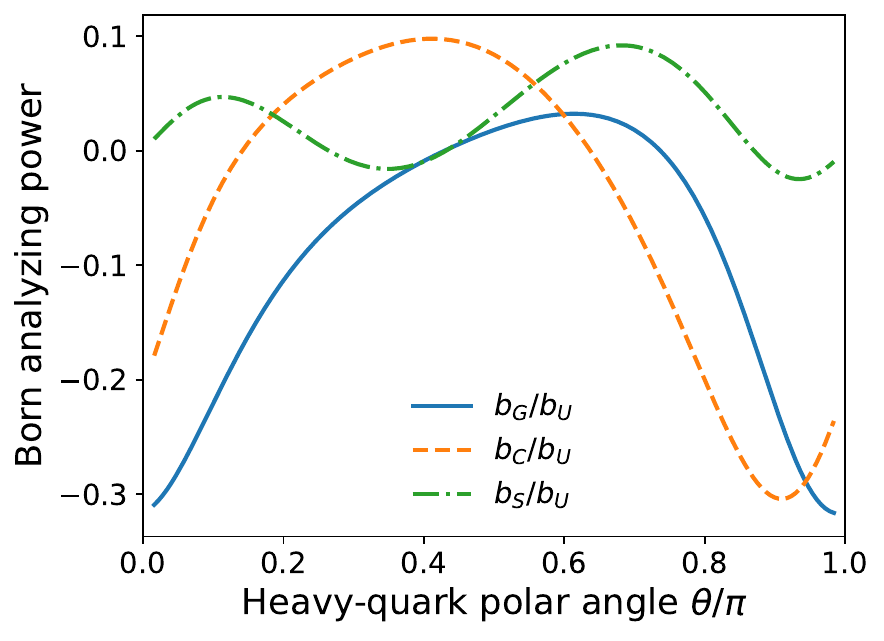}
 \caption{Born hard polarization analyzing powers for $\ell g\to\ell'Q\bar Q$ as functions of the heavy-quark polar angle, with the kinematics specified in the text. The ratios $b_G/b_U$, $b_C/b_U$, and $b_S/b_U$ quantify circular and linear gluon-polarization sensitivity at physical electron helicity $\lambda=+1$ and satisfy Eq.~\eqref{eq:analyzingbound}.}
 \label{fig:born}
\end{figure}

Figure~\ref{fig:born} illustrates the analyzing powers at $\sqrt{(q+k)^2}=\unit[5]{GeV}$, $Q^2=\unit[4]{GeV^2}$, $m_Q=\unit[1.5]{GeV}$, and lepton energy $\unit[10]{GeV}$ in the $\gamma^*g$ center-of-mass frame, with heavy-quark azimuth $\phi=0.4$ and electron helicity $+1$. The heavy-quark polar angle $\theta$ is measured from the incoming gluon. Common couplings and the color factor cancel in $\bm{\mathcal A}_g$. Beam helicity separates the target channels: at Born order, $B(-\lambda)=B(\lambda)^*$ in these conventions, so $b_U,b_C,b_S$ are helicity even, while $b_G$ is helicity odd and vanishes for an unpolarized lepton. The antisymmetric leptonic tensor therefore supplies octupole gluon-helicity sensitivity.

For reference, let $b^{\mathrm{red}}$ denote coefficients computed with the common prefactor $e^4e_Q^2g_s^2T_F/Q^4$ omitted from Eq.~\eqref{eq:BornB}. At $\theta=0.8$,
\begin{equation}
 (b_U^{\mathrm{red}},b_G^{\mathrm{red}},b_C^{\mathrm{red}},b_S^{\mathrm{red}})
 \simeq(1684,+123.6,110.2,5.977)\,\mathrm{GeV}^2.
 \label{eq:Bornbenchmark}
\end{equation}
The reduced eigenvalues are approximately $\unit[1.518\times10^3]{GeV^2}$ and $\unit[1.850\times10^3]{GeV^2}$. At this point, the normalized Ward residuals are at the $10^{-16}$ level. Appendix~\ref{app:validation} describes further diagnostics, and Ref.~\cite{Li7TMDs} supplies them and the data underlying Fig.~\ref{fig:born}.

\subsection{Small-imbalance observables}

Nearly back-to-back heavy quarks allow the hard analyzer to probe the gluon TMD. Define their imbalance and relative transverse momentum:
\begin{equation}
 \bm q_{\mathrm{pair},T}=\bm p_{1T}+\bm p_{2T},\qquad
 \bm p_{\mathrm{rel},T}=\tfrac12(\bm p_{1T}-\bm p_{2T}).
 \label{eq:pairkin}
\end{equation}
The TMD regime has $|\bm q_{\mathrm{pair},T}|$ small compared with $Q$, $|\bm p_{\mathrm{rel},T}|$, and $\sqrt{m_Q^2+p_{\mathrm{rel},T}^2}$. Neglecting target-mass power corrections, Born kinematics selects
\begin{equation}
 x_g=x_{\mathrm{Bj}}\left(1+\frac{M_{Q\bar Q}^2}{Q^2}\right),\qquad
 M_{Q\bar Q}^2=(p_1+p_2)^2.
 \label{eq:chi}
\end{equation}
The pair mass fixes the correlator's argument to $x=x_g$, generally different from $x_{\mathrm{Bj}}$. To retain the small imbalance, first extract the transverse-momentum conservation factor and then expand the hard part about a collinear incoming gluon. At Born order, the factor is $\delta^{(2)}(\bm q_{\mathrm{pair},T}-\bk)$, as the measured imbalance equals the gluon's intrinsic transverse momentum. Before including flux and measurement Jacobians, the polarization response is
\begin{equation}
 \begin{split}
 \mathcal W_g^{\mathrm{Born}}(x_g,\bm q_{\mathrm{pair},T})
 =\int\dd^2\bk\,&\delta^{(2)}(\bm q_{\mathrm{pair},T}-\bk)\\
 &\times\Tr_2[B D_g(x_g,\bk)].
 \end{split}
 \label{eq:BornTMDresponse}
\end{equation}
The hard matrix is evaluated at collinear Born kinematics, while the target correlator retains the measured imbalance. Substitution in Eq.~\eqref{eq:gmaster}, using Table~\ref{tab:gkernels}, gives the individual target-spin responses.

Beyond Born order, radiation and soft recoil contribute to the measured imbalance, so it no longer isolates the gluon’s intrinsic transverse momentum. Where TMD factorization applies, the response combines the gluon correlator with process-specific soft and jet or heavy-hadron factors, whose Wilson lines, color channel, subtraction, and measurement must be specified~\cite{Handbook}. The Born analyzer supplies the hard polarization input, but the target tensor classification alone does not establish factorization~\cite{Boer2011}. Color entanglement can also obstruct TMD factorization in hadron--hadron production~\cite{Rogers2010}.

\subsection{Target preparation and response tomography}
\label{sec:octobservables}

Separating vector and octupole responses also requires controlled target moments. We construct two positive density matrices with opposite octupoles and identical lower moments by rotating the normalized operator $\tau_{30}=\sqrt5\,O_{zzz}/3$. For
$\hat n=(\sin\theta\cos\phi,\sin\theta\sin\phi,\cos\theta)$, where the angles label the preparation axis, define
\begin{align}
 X_3(\hat n)&=D^{3/2}(\phi,\theta,0)\tau_{30}
                  D^{3/2}(\phi,\theta,0)^\dagger\nonumber\\
 &=\frac{\sqrt5}{3}\left[(\hat n\cdot\bm J)^3
                    -\frac{41}{20}\hat n\cdot\bm J\right],\nonumber\\
 \rho_\pm(\hat n)&=\tfrac14I\pm\beta_3 X_3(\hat n),\qquad
 0<\beta_3\leq\frac{\sqrt5}{6}.
 \label{eq:rotatedrho}
\end{align}
The eigenvalues of $\tau_{30}$ are $\frac{\sqrt5}{10}(1,-3,3,-1)$, so the bound $1/4-\beta_3\|\tau_{30}\|\geq0$ on the dimensionless preparation strength $\beta_3$ ensures positivity in every direction. Trace orthogonality makes the rank-one and rank-two moments vanish, and rotations preserve their absence. Along $\hat n=\hat z$, the plus preparation has $S_{LLL}=O=3\beta_3/\sqrt5$, with $0<O\leq1/2$, while the minus preparation has the opposite octupole and the same lower moments.

\begin{figure}[t]
 \centering
 \includegraphics[width=\columnwidth]{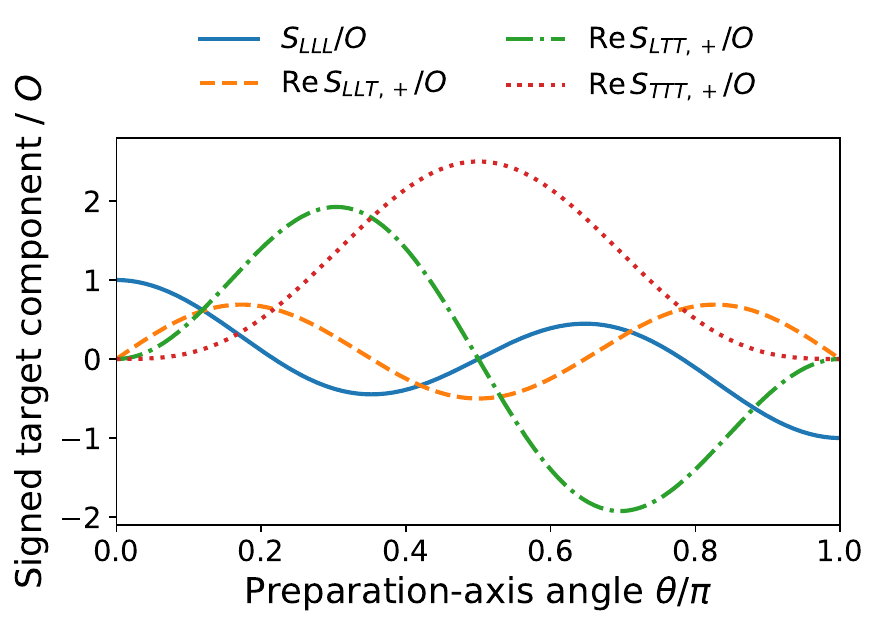}
 \caption{Components of a rotated rank-three target state, divided by its longitudinal amplitude $O$, for $\phi=0$. The curves evaluate Eq.~\eqref{eq:rotatedcomponents}. Vector and quadrupole moments vanish for every preparation-axis angle.}
 \label{fig:rotation}
\end{figure}

The Cartesian octupole tensor of the plus preparation is
\begin{equation}
 R_{abc}=\frac{5O}{2}\left[n_an_bn_c
 -\frac{n_a\delta_{bc}+n_b\delta_{ac}+n_c\delta_{ab}}5\right].
 \label{eq:Rrotated}
\end{equation}
This expression is symmetric and traceless and has $R_{zzz}=O$ for $\hat n=\hat z$. Applying Eq.~\eqref{eq:componentdefinitions} gives
\begin{align}
 S_{LLL}&=O\,P_3(\cos\theta),\nonumber\\
 S_{LLT,+}&=\tfrac O2(5\cos^2\theta-1)\sin\theta\,\e^{\ii\phi},\nonumber\\
 S_{LTT,+}&=5O\cos\theta\sin^2\theta\,\e^{2\ii\phi},\nonumber\\
 S_{TTT,+}&=\tfrac{5O}{2}\sin^3\theta\,\e^{3\ii\phi},
 \label{eq:rotatedcomponents}
\end{align}
where $P_3(c)=(5c^3-3c)/2$. Figure~\ref{fig:rotation} shows the signed complex-component amplitudes at $\phi=0$. Rotation changes the components without introducing lower spin ranks. A transverse preparation therefore contains both $LLT$ and $TTT$, whose different magnitudes follow from Eq.~\eqref{eq:componentdefinitions}. These states are mathematically allowed, though experimental use requires calibration of the prepared moments. The usual vector-polarization fraction $P_z=2S_L/3$, in particular, does not determine $S_{LL}$ or $S_{LLL}$. Appendix~\ref{app:NMR} relates the longitudinal moments to calibrated spin populations and the ideal three-transition NMR response.

Response tomography reconstructs the rank-three target response by varying the preparation direction. At fixed kinematics and beam or hard-analyzer setting, write $\mathcal W[\rho]=\Tr(\rho\widehat W)$, where the Hermitian target-spin operator $\widehat W$ describes a differential cross section or a specified angular projection. Its rank-three components are $w_{3M}=\Tr(\tau_{3M}^\dagger\widehat W)$. The difference of opposite preparations gives
\begin{equation}
 \Delta\mathcal W(\Omega)
 =\frac{\mathcal W[\rho_+(\Omega)]-
           \mathcal W[\rho_-(\Omega)]}{2\beta_3} =\Tr[X_3(\Omega)\widehat W],
 \label{eq:rotationdifference}
\end{equation}
which isolates rank three. Expanding the rotated operator in spherical components gives
\begin{equation}
 X_3(\Omega)=\sum_{M=-3}^{3}D^3_{M0}(\phi,\theta,0)\tau_{3M}.
\end{equation}
Hermiticity and Wigner-matrix orthogonality then imply
\begin{align}
 \Delta\mathcal W(\Omega)&=
 \sum_{M=-3}^{3}D^{3*}_{M0}(\Omega)w_{3M},\nonumber\\
 w_{3M}&=\frac7{4\pi}\int\dd\Omega\,
 D^3_{M0}(\Omega)\Delta\mathcal W(\Omega).
 \label{eq:rotationinverse}
\end{align}
The relation $w_{3,-M}=(-1)^Mw_{3M}^*$ leaves seven independent real components. Measurements along a finite set of generic directions can replace the angular integral by a linear inversion or likelihood fit, with acceptance included in the response matrix. Appendix~\ref{app:validation}
gives an explicit rank-seven set, which are components of the target response for a chosen measurement, rather than fourteen separate TMDs. TMD separation also requires parton polarization analysis, momentum dependence, and the appropriate fragmentation or gluon-sensitive channel. Below, we combine beam helicity with preparation/recoil projections for heavy-pair leptoproduction.

\subsection{Joint target and recoil projections}
\label{sec:jointobservables}

We now combine the prepared states with the heavy-pair analyzer above to separate the gluon octupole coefficients in a single hard channel. Keeping the hard variables fixed ($Q^2$, the pair invariant mass, the heavy-quark direction, and the lepton kinematics), we vary the pair-imbalance azimuth $\varphi=\phi_{q_{\mathrm{pair},T}}$ at fixed $t=|\bm q_{\mathrm{pair},T}|/M_A>0$, together with the preparation azimuth. Label the preparation polar angle and azimuth by $\vartheta$ and $\psi$, reserving $\theta,\phi$ for the heavy-quark direction. At leading-power Born order, the hard matrix is independent of $\varphi$, while the TMD is evaluated at $(x_g,\bk)=(x_g,\bm q_{\mathrm{pair},T})$.

Write the moments of the two preparations as
\begin{equation}
 A_{30}=\eta O\alpha_0,\qquad
 A_{3m,+}=\eta O\alpha_m\e^{\ii m\psi}\quad(m>0),
 \label{eq:cone_moments}
\end{equation}
Here, $\eta=\pm1$ selects the preparation, $0<O\leq1/2$ is its longitudinal octupole amplitude before rotation, and the signed real angular factors are
\begin{align}
 \alpha_0&=P_3(\cos\vartheta),\nonumber\\
 \alpha_1&=\tfrac12(5\cos^2\vartheta-1)\sin\vartheta,\nonumber\\
 \alpha_2&=5\cos\vartheta\sin^2\vartheta,\nonumber\\
 \alpha_3&=\tfrac52\sin^3\vartheta.
 \label{eq:cone_alpha}
\end{align}
Keeping these factors signed allows a common preparation azimuth for all components. Eq.~\eqref{eq:rotatedrho} ensures positivity, and reversing $\eta$ changes only rank three. All gluon functions below have the same future-pointing $[+,+]$ link class and scales as the heavy-pair correlator above.

\begin{table*}[t]
\caption{Joint Fourier projections of the octupole gluon response. Rows containing $m$ apply to $m=1,2,3$. The beam weight is $\lambda^\epsilon$ in the target-sign difference. The entries in the last column multiply the displayed angular function in $\mathcal W_U$ ($\epsilon=0$) or $\mathcal W_L$ ($\epsilon=1$). All functions are evaluated at the same $(x_g,k_T^2)$ and $[+,+]$ link class.}
\label{tab:oct_fourier}
\centering
\begin{ruledtabular}
\begin{tabular}{cccll}
 Coefficient & $\epsilon$ & $(p,q)$ & Angular function $\mathscr Y_\nu$
 & Response amplitude $C_\nu$\\
 $g^g_{30}$ & 1 & $(0,0)$ & $1$ & $\alpha_0 b_G(1)g^g_{30}$\\
 $h^{g[2]}_{30}$ & 0 & $(2,0)$ & $\sin(2\varphi-2\phi_B)$
 & $\alpha_0 b_{\mathrm{lin}}t^2h^{g[2]}_{30}/2$\\
 $f^g_{3m}$ & 0 & $(m,-m)$ & $\sin m(\varphi-\psi)$
 & $\alpha_m b_Ut^mf^g_{3m}$\\
 $g^g_{3m}$ & 1 & $(m,-m)$ & $\cos m(\varphi-\psi)$
 & $\alpha_m b_G(1)t^mg^g_{3m}$\\
 $h^{g[|m-2|]}_{3m}$ & 0 & $(2-m,m)$ & $\sin\Psi_m^-$
 & $\alpha_m b_{\mathrm{lin}}c_m t^{|m-2|}h^{g[|m-2|]}_{3m}$\\
 $h^{g[m+2]}_{3m}$ & 0 & $(m+2,-m)$ & $\sin\Psi_m^+$
 & $\alpha_m b_{\mathrm{lin}}t^{m+2}h^{g[m+2]}_{3m}/4$
\end{tabular}
\end{ruledtabular}
\end{table*}

Let $\sigma_{\lambda,\eta}$ be the differential cross section in the same hard variables and $\dd^2\bm q_{\mathrm{pair},T}$ for each spin setting, corrected for luminosity and acceptance. At Born order, it separates into an unpolarized-target reference and two octupole responses:
\begin{equation}
 \sigma_{\lambda,\eta}
 =\mathcal J_g\bigl[\mathcal W_0^g+
       \eta O(\mathcal W_U+\lambda\mathcal W_L)\bigr].
 \label{eq:gluon_spin_cross}
\end{equation}
The common spin-independent factor $\mathcal J_g$ contains the flux and measurement Jacobian in the phase-space convention of Eq.~\eqref{eq:Bornphase}, and it is independent of $\psi$ and $\varphi$ for the fixed Born variables. The hard coefficients retain the couplings in Eq.~\eqref{eq:BornB}. Under beam-helicity reversal, $b_G$ changes sign, whereas $b_U$ and $b_+$ are unchanged. We therefore write $b_G(\lambda)=\lambda b_G(1)$, where $b_G(1)$ is evaluated at $\lambda=+1$ with the same hard kinematics. For an unpolarized target, the reference response is
\begin{equation}
 \mathcal W_0^g=b_U f^g_{00}
  +\frac{b_{\mathrm{lin}}t^2}{2}h^{g[2]}_{00}
         \cos(2\varphi-2\phi_B),
 \label{eq:gluon_unpol_reference}
\end{equation}
which cancels in the target-octupole difference. Beam averaging or helicity weighting then selects the two reduced responses:
\begin{align}
 \mathcal W_U&=\frac1{4O\mathcal J_g}
       \sum_{\lambda,\eta=\pm1}\eta\,\sigma_{\lambda,\eta},\nonumber\\
 \mathcal W_L&=\frac1{4O\mathcal J_g}
       \sum_{\lambda,\eta=\pm1}\lambda\eta\,
              \sigma_{\lambda,\eta}.
 \label{eq:gluon_spin_differences}
\end{align}
Thus, $\mathcal W_U$ is the beam-even response containing the link-odd number and linear-polarization terms, while $\mathcal W_L$ is the beam-odd response containing the link-even gluon helicity terms. Both belong to the octupole target sector, and the subscripts label the beam operation. The full correlator gives the helicity response
\begin{equation}
 \mathcal W_L=b_G(1)\left[
 \alpha_0g^g_{30}+\sum_{m=1}^3\alpha_m t^m g^g_{3m}
                     \cos m(\varphi-\psi)\right].
 \label{eq:gluon_oct_helicity}
\end{equation}
For the beam-even response, define the lower-orbital-rank kernel factors from Table~\ref{tab:gkernels} by $c_m\equiv c_{m,|m-2|}$, so $c_1=1/2$ and $c_2=c_3=1$. The phases of the two orbital branches are
\begin{equation}
 \Psi_m^-=(2-m)\varphi+m\psi-2\phi_B,\quad
 \Psi_m^+=(m+2)\varphi-m\psi-2\phi_B.
 \label{eq:gluon_oct_angles}
\end{equation}
Then, the number and linear-polarization terms are
\begin{align}
 \mathcal W_U={}&b_U\sum_{m=1}^3\alpha_m t^m f^g_{3m}
                         \sin m(\varphi-\psi)\nonumber\\
 &+\frac{b_{\mathrm{lin}}\alpha_0t^2}{2}h^{g[2]}_{30}
                         \sin(2\varphi-2\phi_B)\nonumber\\
 &+b_{\mathrm{lin}}\sum_{m=1}^3\alpha_m\biggl[
 c_m t^{|m-2|}h^{g[|m-2|]}_{3m}\sin\Psi_m^-\nonumber\\
 &\hspace{35mm}
 +\frac{t^{m+2}}4h^{g[m+2]}_{3m}\sin\Psi_m^+\biggr].
 \label{eq:gluon_oct_unpolarized}
\end{align}
Equations~\eqref{eq:gluon_oct_helicity} and \eqref{eq:gluon_oct_unpolarized} contain all fourteen gluon octupole coefficients for this physical target family at Born order. Both azimuths are needed to separate the complete basis. For example, at $m=1$ the lower-rank linear-polarization branch and the number term share a first-order recoil harmonic in $\varphi$, but have opposite preparation phases. Varying $\psi$ distinguishes their joint harmonics. A recoil-only scan at one fixed preparation cannot resolve the complete fourteen-coefficient response.

For each response $\nu$, let $\mathscr Y_\nu$ be its angular function. Table~\ref{tab:oct_fourier} lists the Fourier pair $(p,q)$ in its phase $p\varphi+q\psi+\delta$ and the amplitude, including target and hard factors. Scan the preparation azimuth at a fixed polar angle, choosing a cone on which every $\alpha_m$ is nonzero. For example, $\cos\vartheta=3/5$ gives
\begin{equation}
 (\alpha_0,\alpha_1,\alpha_2,\alpha_3)
 =(-9/25,8/25,48/25,32/25).
 \label{eq:generic_cone}
\end{equation}
Within each beam channel, no two Fourier pairs coincide up to an overall sign, so their angular projections are independent. With the beam channels resolved, $b_Ub_G(1)b_{\mathrm{lin}}\ne0$, and $t>0$, all fourteen coefficients can therefore be separated in the ideal Born response. This establishes identifiability at fixed $(x_g,k_T^2)$ without demonstrating statistical sensitivity or experimental preparation.

Normalized moments cancel $\mathcal J_g$ but still require calibrated target strength. With $\int\dd\varphi\dd\psi$ covering both full azimuthal cycles, define
\begin{equation}
 \mathcal A_\nu=
 \frac{\mathcal Z_\nu}{O}
 \frac{\displaystyle\int\dd\varphi\,\dd\psi\,\mathscr Y_\nu
       \sum_{\lambda,\eta}\eta\lambda^{\epsilon_\nu}
          \sigma_{\lambda,\eta}}
      {\displaystyle\int\dd\varphi\,\dd\psi
       \sum_{\lambda,\eta}\sigma_{\lambda,\eta}},
 \label{eq:gluon_measured_moment}
\end{equation}
where $\mathcal Z_\nu=1$ for the constant row and $2$ otherwise. For a nonzero reference rate, the angular average of $\mathcal W_0^g$ is $b_Uf^g_{00}$, and
\begin{equation}
 \mathcal A_\nu=\frac{C_\nu}{b_Uf^g_{00}}
 \label{eq:gluon_moment_ratio}
\end{equation}
at leading-power Born order. The normalization accounts for $\langle\sin^2\rangle=\langle\cos^2\rangle=1/2$. Dividing the extracted amplitude by its known, nonzero target factors and hard analyzing powers gives a TMD ratio to $f^g_{00}$. A vanishing analyzing power leaves the moment defined at a nonzero reference rate, but prevents determination of that coefficient. Absolute TMD normalization also requires an independently normalized unpolarized response or equivalent input for $f^g_{00}$.

\subsection{Longitudinal and transverse octupole observables}
\label{sec:explicitgluon}

For a longitudinal preparation, $\alpha_0=1$ and $\alpha_{1,2,3}=0$, so the preparation azimuth drops out. The two moments of Eq.~\eqref{eq:gluon_measured_moment} require only the recoil azimuth and reduce to
\begin{equation}
 \mathcal A_L^{(0)}=\frac{b_G(1)}{b_U}
                       \frac{g^g_{30}}{f^g_{00}},\qquad
 \mathcal A_U^{(2)}=\frac{b_{\mathrm{lin}}}{b_U}\frac{t^2}{2}
                       \frac{h^{g[2]}_{30}}{f^g_{00}}.
 \label{eq:longitudinal_gluon_observables}
\end{equation}
The first moment uses a beam--target double difference to probe the finite-$k_T$ octupole helicity response, and the second averages over the beam and weights the target difference by $\sin(2\varphi-2\phi_B)$, probing linear gluon polarization in a longitudinal octupole target. Lower-rank target terms cancel because the two density matrices have identical lower moments. The unpolarized linear-gluon term cancels in the target difference and averages out in the denominator. Although leading twist, the second observable does not correspond to an additional collinear octupole PDF.

For a transverse preparation, $\alpha_1=-1/2$ and $\alpha_3=5/2$, with $\alpha_0=\alpha_2=0$. Its $LLT$ and $TTT$ components are separated by projecting the third preparation harmonic, which isolates the fully transverse octupole:
\begin{equation}
 (\mathcal P_3\mathcal W)(\psi)
 =\frac1\pi\int_0^{2\pi}\dd\psi'\,
       \cos[3(\psi-\psi')]\mathcal W(\psi').
 \label{eq:TTTprep_projector}
\end{equation}
In the ideal rank-three response, eight equally spaced preparation angles $\psi_j=2\pi j/8$ give the same projection, as one can replace the integral with $\frac14\sum_{j=0}^{7}\cos[3(\psi-\psi_j)]\mathcal W(\psi_j)$. The preparation harmonics have magnitude at most three, so this discrete scan separates the third harmonic without contamination from the $LLT$ harmonic.

After this target-angle projection, the beam-even and beam-odd responses are
\begin{align}
 \mathcal P_3\mathcal W_U&=\frac52\biggl[
 b_Ut^3f^g_{33}\sin(3\varphi-3\psi)\nonumber\\
 &\qquad+b_{\mathrm{lin}}t h^{g[1]}_{33}\sin(3\psi-\varphi-2\phi_B)\nonumber\\
 &\qquad+\frac{b_{\mathrm{lin}}t^5}{4}h^{g[5]}_{33}
                         \sin(5\varphi-3\psi-2\phi_B)\biggr],\nonumber\\
 \mathcal P_3\mathcal W_L&=\frac52 b_G(1)t^3g^g_{33}
                         \cos(3\varphi-3\psi).
 \label{eq:TTTgluon_response}
\end{align}
Next, project the recoil harmonics one, three, and five at fixed $\psi$. With $\Psi_1=3\psi-\varphi-2\phi_B$, $\Psi_3=3\varphi-3\psi$, and $\Psi_5=5\varphi-3\psi-2\phi_B$, define
\begin{align}
 M_{U,n}&=\frac{2}{5\pi}\int_0^{2\pi}\dd\varphi\,
              \sin\Psi_n\,\mathcal P_3\mathcal W_U,\nonumber\\
 M_{L,3}&=\frac{2}{5\pi}\int_0^{2\pi}\dd\varphi\,
              \cos\Psi_3\,\mathcal P_3\mathcal W_L.
 \label{eq:TTTgluon_moments}
\end{align}
with $n=1,3,5$. Fourier orthogonality yields
\begin{align}
 M_{U,1}&=b_{\mathrm{lin}}t h^{g[1]}_{33},&
 M_{U,3}&=b_Ut^3f^g_{33},\nonumber\\
 M_{U,5}&=\tfrac14b_{\mathrm{lin}}t^5h^{g[5]}_{33},&
 M_{L,3}&=b_G(1)t^3g^g_{33}.
 \label{eq:TTTgluon_isolation}
\end{align}
Then, dividing by the azimuthally averaged reference response $b_Uf^g_{00}$ gives normalized observables. The two $TTT$ linear-polarization coefficients are separated by their first and fifth recoil harmonics, while the third harmonic probes number and helicity through the different beam channels. The benchmark above has nonzero circular and linear analyzing powers and satisfies the hard nondegeneracy condition.

Experimental use requires calibrated target moments, equalized luminosities, and common acceptance treatment. Residual vector and quadrupole polarization must enter the response matrix, and incomplete angular coverage requires a fit using the angular Gram matrix for that coverage. Partial lepton polarization rescales the beam-odd response. Beyond Born order, recoil convolutions and final-state soft factors modify the pointwise relations in Eqs.~\eqref{eq:gluon_moment_ratio} and \eqref{eq:TTTgluon_isolation}. Linearity in $\rho$ preserves spin-rank projection, but the angular kernels acquire process-dependent convolutions. The Born results therefore supply a baseline for resummed extraction, without assuming that soft recoil cancels to all orders.

\section{Quark observables in SIDIS and Drell--Yan}
\label{sec:quarkobservables}

\subsection{SIDIS kinematics and cross section}
\label{sec:sidis}

We now connect the quark basis to SIDIS and Drell--Yan measurements. SIDIS pairs the target distribution with a fragmentation function, while DY pairs it with a distribution in the second incoming hadron. Together with the hard kernel, these partners analyze quark number, helicity, and transverse spin. We derive the complete SIDIS catalog and the DY octupole responses in the spin and transverse conventions of the gluon construction. The SIDIS calculation complements the established spin-$3/2$ fragmentation analyses~\cite{Zhao2022,ZhaoSIDIS2024}. Here, the octupole belongs to the incoming target, and the detected hadron is unpolarized. We consider one-photon exchange in
\begin{equation}
 \ell(l,\lambda)+A(P,\rho)\longrightarrow
 \ell(l')+h(P_h)+X,
\end{equation}
where $\lambda=\pm1$ is the incident lepton helicity. We use $q=l-l'$, $Q^2=-q^2$, and
\begin{equation}
 x_{\mathrm{Bj}}=\frac{Q^2}{2P\cdot q},\qquad
 y=\frac{P\cdot q}{P\cdot l},\qquad
 z=\frac{P\cdot P_h}{P\cdot q}.
\end{equation}
At Born order, the active quark fraction is $x=x_{\mathrm{Bj}}$. We work in the current-fragmentation TMD region specified in Sec.~\ref{sec:scope}. Choose a right-handed frame with the large target momentum along $+z$, the virtual photon along $-z$, and $+x$ along the lepton's transverse momentum. Azimuths increase from $+x$ toward $+y$. The measured hadron direction is $\hat h=\bPh/|\bPh|=(\cos\phi_h,\sin\phi_h)$, with $\hat h_\perp=(-\sin\phi_h,\cos\phi_h)$. Choosing $+z$ along the photon while keeping the same $+x$ reverses the $y$ axis. Comparison with that convention requires transforming both the azimuths and the longitudinal/transverse spin components. Our phase-space normalization follows Ref.~\cite{Bacchetta2007}.

The SIDIS convolution uses two transverse momenta. Let $\bp$ denote the outgoing quark transverse momentum relative to the detected-hadron direction. At leading power,
\begin{equation}
 \bk-\bp=\bPh/z.
 \label{eq:SIDISdelta}
\end{equation}
The hadron momentum relative to the outgoing quark is instead $\bm j_T=-z\bp$. For hadron mass $M_h$, the unsubtracted fragmentation correlator is
\begin{equation}
 \begin{split}
 \Delta_{ab,\mathrm{unsub}}(z,\bp)={}&
 \frac1{2z}\sum_X\int\frac{\dd\xi^+\dd^2\xi_T}{(2\pi)^3}
 \e^{\ii p^-\xi^+-\ii\bp\cdot\bm\xi_T}\\
 &\times\langle0|U[\infty,\xi]\psi_a(\xi)|hX;\mathrm{out}\rangle\\
 &\times\left.\langle hX;\mathrm{out}|\bar\psi_b(0)
 U[0,\infty]|0\rangle\right|_{\xi^-=0},
 \end{split}
 \label{eq:Delta}
\end{equation}
where $p^-=P_h^-/z$ and color is contracted through the Wilson lines. For the following projections and convolutions, the fragmentation correlator uses a regulator and UV/rapidity subtraction consistent with the distribution correlator in Sec.~\ref{sec:operators}. Its two leading-twist projections are
\begin{align}
 \tfrac12\Tr_D(\Delta\gamma^-)&=D_1,\nonumber\\
 \mathcal D^i\equiv\tfrac12\Tr_D(\Delta\ii\sigma^{i-}\gamma_5)
 &=-\frac{\epsilon_{ia}p^a}{M_h}H_1^\perp,
 \label{eq:FFproj}
\end{align}
and the corresponding reconstruction is
\begin{equation}
 \Delta_{(2)}=\tfrac12
 \bigl[D_1\gamma^++\mathcal D^i\ii\sigma^{i+}\gamma_5\bigr].
 \label{eq:FFrecon}
\end{equation}
Equation~\eqref{eq:FFproj} fixes the Collins fragmentation sign convention~\cite{Collins1993}. The choice of fragmentation momentum also fixes the collinear normalization. In a regulated tree-level treatment, $D_1(z)=z^2\int\dd^2\bp\,D_1(z,\bp^2)$, where the Jacobian $z^2$ converts $\bp$ to $\bm j_T$. The renormalized relation beyond this formal integral requires UV matching.

We combine the target and fragmentation coefficients in the flavor-weighted convolution
\begin{equation}
 \begin{aligned}
 &\Conv[w f D]=x\sum_{a=q,\bar q}e_a^2
 \int\dd^2\bk\,\dd^2\bp\\
 &\qquad\times\delta^{(2)}\!\left(\bk-\bp-\frac{\bPh}{z}\right)
 w\,f^aD^a.
 \end{aligned}
 \label{eq:convolution}
\end{equation}
Here, $w=w(\bk,\bp)$, $f^a=f^a(x,\bk^2)$, and $D^a=D^a(z,\bp^2)$. A dimensionless weight produces a structure function of mass dimension $-2$. Target amplitudes and leptonic depolarization factors are kept outside this definition.

The hard contraction determines how each quark projection enters the cross section. At leading-power Born order, the hadronic tensor follows from $\Tr_D(\Phi\gamma^\mu\Delta\gamma^\nu)$. Using the dual reconstructions in Eqs.~\eqref{eq:qreconstruct} and \eqref{eq:FFrecon}, its transverse photon components are
\begin{equation}
 \begin{split}
 \Tr_D(\Phi\gamma^i\Delta\gamma^j)
 ={}&\delta^{ij}\calF^qD_1+\ii\epsilon_{ij}\calG^qD_1\\
 &-\calT^{q,i}\mathcal D^j-\calT^{q,j}\mathcal D^i
 +\delta^{ij}\calT^q\cdot\mathcal D.
 \end{split}
 \label{eq:traceSIDIS}
\end{equation}
The number and helicity terms multiply $\delta^{ij}$ and $\ii\epsilon_{ij}$, respectively. The $\calT_x\mathcal D_x$ term has the tensor $\diag(-1,1)$. These trace components fix the relative signs of the helicity and Collins responses.

Let $w^{ij}$ denote the convolution of Eq.~\eqref{eq:traceSIDIS}. With $\mathbb W^{\mu\nu}=2M_AW^{\mu\nu}$, the leading transverse hadronic tensor is $\mathbb W^{ij}=(2z/x)w^{ij}$. The massless leptonic tensor in Eq.~\eqref{eq:leptonic} supplies the beam-helicity and depolarization factors. Its normalized transverse projection in our frame is
\begin{equation}
 \mathscr L_{ij}=\tfrac12\bigl[
 \delta_{ij}+\varepsilon\diag(1,-1)_{ij}
 -\ii\lambda\Dpol\epsilon_{ij}\bigr],
 \label{eq:Lnorm}
\end{equation}
where
\begin{equation}
 \varepsilon=\frac{1-y}{1-y+y^2/2},\qquad
 \Dpol=\sqrt{1-\varepsilon^2}.
 \label{eq:depolarization}
\end{equation}
Explicitly, $L_{xx}=4l_T^2+Q^2$, $L_{yy}=Q^2$, $l_T^2=Q^2(1-y)/y^2$, and $L_{xy}=-\ii\lambda Q^2(2-y)/y$. The transverse trace is $4Q^2(1-y+y^2/2)/y^2$. Contracting the two tensors in $\dd\sigma/\dd\mathcal P=
\alpha_{\mathrm{em}}^2y\,\mathbb W^{\mu\nu}L_{\mu\nu}/(8zQ^4)$ gives
\begin{align}
 \frac{\dd\sigma}{\dd\mathcal P}&=\mathcal N\mathscr L_{ij}w^{ij},\nonumber\\
 \mathcal N&=\frac{\alpha_{\mathrm{em}}^2}{xyQ^2}(1-y+y^2/2),\nonumber\\
 \dd\mathcal P&=\dd x\,\dd y\,\dd\psi_\ell\,\dd z\,\dd\phi_h\,\dd P_{hT}^2.
 \label{eq:prefactor}
\end{align}
The angle $\psi_\ell$ specifies the overall lepton-plane azimuth, and the target moments are evaluated relative to that plane. The target-mass expansion is controlled by $\gamma=2xM_A/Q$, approximately $2x_Nm_N/Q$ for $x_N\simeq Ax$ and $M_A\simeq Am_N$. We retain the leading-power Born structure functions, and target-mass kinematic factors can be restored consistently afterward~\cite{Bacchetta2007}.

Substituting $\mathcal D=-E\bp\,H_1^\perp/M_h$ separates the cross section into number, helicity, and transverse-spin responses:
\begin{equation}
 \begin{split}
 \frac1{\mathcal N}\frac{\dd\sigma}{\dd\mathcal P}
 ={}&\Conv[\calF^qD_1]+\lambda\Dpol\Conv[\calG^qD_1]\\
 &+\varepsilon\Conv\!\left[
 \frac{\calT^{q,x}p_y+\calT^{q,y}p_x}{M_h}H_1^\perp\right].
 \end{split}
 \label{eq:SIDISmaster}
\end{equation}
The fragmentation function $D_1$ accompanies quark number and helicity, with the latter selected by lepton helicity. Transverse spin requires the chiral-odd Collins function $H_1^\perp$, whose its numerator is $\Ima(\calT_+^qp_+)$. At this order $F_{UU,L}=0$. Gluon distributions do not enter this Born single-hadron kernel as independent initial partons, as their effects arise through matching, evolution, or a separately specified gluon-sensitive subprocess.

\subsection{Harmonic weights and the complete response basis}

To express the master equation in measured azimuths, rotate the integration momenta into the observed-hadron direction. This separates explicit angular phases from real convolution weights:
\begin{align}
 u&=\frac{k_x+\ii k_y}{\M}\e^{-\ii\phi_h}=a+\ii b,\nonumber\\
 v&=\frac{p_x+\ii p_y}{M_h}\e^{-\ii\phi_h}=c+\ii d.
 \label{eq:uv}
\end{align}
Here, $a=\hat h\cdot\bk/\M$, $b=\hat h_\perp\cdot\bk/\M$, $c=\hat h\cdot\bp/M_h$, and $d=\hat h_\perp\cdot\bp/M_h$. Reflection about that direction sends $(b,d)\mapsto(-b,-d)$. The delta function and radial coefficients are unchanged, whereas imaginary parts of the complex weights reverse sign and integrate to zero. The surviving real weights are
\begin{align}
 w_m&=\Rea(u^m),&&m=0,1,2,3,\nonumber\\
 b_0&=\Rea(uv),\nonumber\\
 \ell_m&=\Rea[(u^*)^{m-1}v],&&m=1,2,3,\nonumber\\
 r_m&=\tfrac12\Rea[u^{m+1}v],&&m=1,2,3.
 \label{eq:weights}
\end{align}
The factor $1/2$ in $r_m$ follows from the high-rank STF contraction in Eq.~\eqref{eq:Vcomplex}. Explicitly,
\begin{equation}
 \begin{aligned}
 &w_0=1,\quad w_1=a,\quad w_2=a^2-b^2,\quad w_3=a^3-3ab^2,\\
 &b_0=ac-bd,\quad \ell_1=c,\quad \ell_2=ac+bd,\\
 &\ell_3=(a^2-b^2)c+2abd,\\
 &r_1=\tfrac12[(a^2-b^2)c-2abd],\\
 &r_2=\tfrac12[(a^3-3ab^2)c-(3a^2b-b^3)d],\\
 &r_3=\tfrac12[(a^4-6a^2b^2+b^4)c-(4a^3b-4ab^3)d].
 \end{aligned}
 \label{eq:weightpolys}
\end{equation}
For example, $b_0=[2(\hat h\cdot\bk)(\hat h\cdot\bp)-\bk\cdot\bp]/(M_AM_h)$, whereas $\ell_2=\bk\cdot\bp/(M_AM_h)$.

For a target component $(K,m)$, the phases are
\begin{align}
 \delta_m&=m(\phi_h-\phi_{Km}),\nonumber\\
 \theta_m^-&=(2-m)\phi_h+m\phi_{Km},\nonumber\\
 \theta_m^+&=(m+2)\phi_h-m\phi_{Km}.
 \label{eq:anglesSF}
\end{align}
The lower and higher orbital-rank branches of the quark transverse-spin projection give $\ell_m$ with $\theta_m^-$ and $r_m$ with $\theta_m^+$, respectively. For $m=0$, the Collins phase is $2\phi_h$ with weight $b_0$. Odd $K$ gives sine factors through the imaginary part in Eq.~\eqref{eq:SIDISmaster}, while for even $K$, the transverse dual contributes an additional $-\ii$, giving minus cosine factors. The number and helicity phases follow from Eq.~\eqref{eq:CSharmonics}.

\begin{table*}[!t]
\caption{All 32 independent leading-twist quark SIDIS responses in the axes specified in Sec.~\ref{sec:sidis}. Target amplitudes multiply rows according to Eq.~\eqref{eq:SFrow}. Every convolution includes flavor charges and the momentum delta function in Eq.~\eqref{eq:convolution}.}
\label{tab:SF}
\centering\small
\begin{ruledtabular}
\begin{tabular*}{\textwidth}{@{\extracolsep{\fill}}cccll@{}}
$(K,m)$ & Response & $d(y,\lambda)$ & Angular factor & Structure function\\
\midrule
\multicolumn{5}{l}{$K=0$: unpolarized target (2 responses)}\\
$(0,0)$&$F^{\mathrm{U}}_{00}$&$1$&$1$&$\Conv[f^q_{00}D_1]$\\
$(0,0)$&$F^{\mathrm{T}[1]}_{00}$&$\varepsilon$&$\cos2\phi_h$&$-\Conv[b_0\qq{0}{0}{1}H]$\\
\multicolumn{5}{l}{$K=1$: vector target (6 responses)}\\
$(1,0)$&$F^{\mathrm{L}}_{10}$&$\lambda\Dpol$&$1$&$\Conv[g^q_{10}D_1]$\\
$(1,0)$&$F^{\mathrm{T}[1]}_{10}$&$\varepsilon$&$\sin2\phi_h$&$\Conv[b_0\qq{1}{0}{1}H]$\\
$(1,1)$&$F^{\mathrm{U}}_{11}$&$1$&$\sin\delta_1$&$\Conv[w_1f^q_{11}D_1]$\\
$(1,1)$&$F^{\mathrm{L}}_{11}$&$\lambda\Dpol$&$\cos\delta_1$&$\Conv[w_1g^q_{11}D_1]$\\
$(1,1)$&$F^{\mathrm{T}[0]}_{11}$&$\varepsilon$&$\sin\theta_1^-$&$\Conv[\ell_1\qq{1}{1}{0}H]$\\
$(1,1)$&$F^{\mathrm{T}[2]}_{11}$&$\varepsilon$&$\sin\theta_1^+$&$\Conv[r_1\qq{1}{1}{2}H]$\\
\multicolumn{5}{l}{$K=2$: quadrupole target (10 responses)}\\
$(2,0)$&$F^{\mathrm{U}}_{20}$&$1$&$1$&$\Conv[f^q_{20}D_1]$\\
$(2,0)$&$F^{\mathrm{T}[1]}_{20}$&$\varepsilon$&$\cos2\phi_h$&$-\Conv[b_0\qq{2}{0}{1}H]$\\
$(2,1)$&$F^{\mathrm{U}}_{21}$&$1$&$\cos\delta_1$&$\Conv[w_1f^q_{21}D_1]$\\
$(2,1)$&$F^{\mathrm{L}}_{21}$&$\lambda\Dpol$&$\sin\delta_1$&$\Conv[w_1g^q_{21}D_1]$\\
$(2,1)$&$F^{\mathrm{T}[0]}_{21}$&$\varepsilon$&$\cos\theta_1^-$&$-\Conv[\ell_1\qq{2}{1}{0}H]$\\
$(2,1)$&$F^{\mathrm{T}[2]}_{21}$&$\varepsilon$&$\cos\theta_1^+$&$-\Conv[r_1\qq{2}{1}{2}H]$\\
$(2,2)$&$F^{\mathrm{U}}_{22}$&$1$&$\cos\delta_2$&$\Conv[w_2f^q_{22}D_1]$\\
$(2,2)$&$F^{\mathrm{L}}_{22}$&$\lambda\Dpol$&$\sin\delta_2$&$\Conv[w_2g^q_{22}D_1]$\\
$(2,2)$&$F^{\mathrm{T}[1]}_{22}$&$\varepsilon$&$\cos\theta_2^-$&$-\Conv[\ell_2\qq{2}{2}{1}H]$\\
$(2,2)$&$F^{\mathrm{T}[3]}_{22}$&$\varepsilon$&$\cos\theta_2^+$&$-\Conv[r_2\qq{2}{2}{3}H]$\\
\multicolumn{5}{l}{$K=3$: octupole target (14 responses)}\\
$(3,0)$&$F^{\mathrm{L}}_{30}$&$\lambda\Dpol$&$1$&$\Conv[g^q_{30}D_1]$\\
$(3,0)$&$F^{\mathrm{T}[1]}_{30}$&$\varepsilon$&$\sin2\phi_h$&$\Conv[b_0\qq{3}{0}{1}H]$\\
$(3,1)$&$F^{\mathrm{U}}_{31}$&$1$&$\sin\delta_1$&$\Conv[w_1f^q_{31}D_1]$\\
$(3,1)$&$F^{\mathrm{L}}_{31}$&$\lambda\Dpol$&$\cos\delta_1$&$\Conv[w_1g^q_{31}D_1]$\\
$(3,1)$&$F^{\mathrm{T}[0]}_{31}$&$\varepsilon$&$\sin\theta_1^-$&$\Conv[\ell_1\qq{3}{1}{0}H]$\\
$(3,1)$&$F^{\mathrm{T}[2]}_{31}$&$\varepsilon$&$\sin\theta_1^+$&$\Conv[r_1\qq{3}{1}{2}H]$\\
$(3,2)$&$F^{\mathrm{U}}_{32}$&$1$&$\sin\delta_2$&$\Conv[w_2f^q_{32}D_1]$\\
$(3,2)$&$F^{\mathrm{L}}_{32}$&$\lambda\Dpol$&$\cos\delta_2$&$\Conv[w_2g^q_{32}D_1]$\\
$(3,2)$&$F^{\mathrm{T}[1]}_{32}$&$\varepsilon$&$\sin\theta_2^-$&$\Conv[\ell_2\qq{3}{2}{1}H]$\\
$(3,2)$&$F^{\mathrm{T}[3]}_{32}$&$\varepsilon$&$\sin\theta_2^+$&$\Conv[r_2\qq{3}{2}{3}H]$\\
$(3,3)$&$F^{\mathrm{U}}_{33}$&$1$&$\sin\delta_3$&$\Conv[w_3f^q_{33}D_1]$\\
$(3,3)$&$F^{\mathrm{L}}_{33}$&$\lambda\Dpol$&$\cos\delta_3$&$\Conv[w_3g^q_{33}D_1]$\\
$(3,3)$&$F^{\mathrm{T}[2]}_{33}$&$\varepsilon$&$\sin\theta_3^-$&$\Conv[\ell_3\qq{3}{3}{2}H]$\\
$(3,3)$&$F^{\mathrm{T}[4]}_{33}$&$\varepsilon$&$\sin\theta_3^+$&$\Conv[r_3\qq{3}{3}{4}H]$\\
\end{tabular*}
\end{ruledtabular}
\end{table*}

Table~\ref{tab:SF} assembles all thirty-two responses of the leading-power electromagnetic Born kernel in the TMD region. A full hadronic-tensor decomposition beyond leading power or at arbitrary transverse momentum has a broader scope~\cite{CosynWeiss2026}. Each row contributes a target amplitude, a leptonic factor, an angular harmonic, and a structure function:
\begin{equation}
 s_{Km}\,d(y,\lambda)\,\mathcal A(\phi_h,\phi_{Km})\,F
 \quad\text{to}\quad
 \frac1{\mathcal N}\frac{\dd\sigma}{\dd\mathcal P}.
 \label{eq:SFrow}
\end{equation}
The target amplitudes are $s_{00}=1$, $s_{K0}=A_{K0}$, and $s_{Km}=|A_{Km,+}|$ for $m>0$, as in Eq.~\eqref{eq:gkernelrule}. The structure function $F$ excludes this amplitude, the angular factor $\mathcal A$, and the leptonic factor $d$. The upright labels $\mathrm{U},\mathrm{L},\mathrm{T}$ on $F_{Km}^{\mathrm{U}}$, $F_{Km}^{\mathrm{L}}$, and $F_{Km}^{\mathrm{T}[n]}$ identify quark number, helicity, and transverse spin. Target polarization is specified separately by $(K,m)$, and $[n]$ gives the target TMD's orbital rank. Uppercase functions are therefore measurable structure functions, while lowercase coefficients inside their convolutions are TMDs. In the table only, $H_1^\perp$ is abbreviated by $H$.

Two octupole examples connect the tensor structures to a fifth harmonic and a response with no explicit hadron-azimuth dependence. The higher orbital-rank $TTT$ branch gives $\calT_+=\tfrac12 A_{33,+}^*\kappa^4h^{q[4]}_{33}$. Multiplication by $p_+/M_h$ in the hard trace produces
\begin{equation}
 \frac{s_{33}}2\Ima\!\left[
 \e^{\ii(5\phi_h-3\phi_{33})}u^4v\right]h^{q[4]}_{33}.
\end{equation}
Reflection removes the imaginary reduced weight, leaving the cross-section contribution
\begin{equation}
 \varepsilon s_{33}\sin(5\phi_h-3\phi_{33})
 \Conv[r_3h^{q[4]}_{33}H_1^\perp].
 \label{eq:fifthharmonic}
\end{equation}
The fifth harmonic combines orbital rank four in the target distribution with rank one in Collins fragmentation. Both belong to the leading-power TMD sector. The final-state spin-$3/2$ SIDIS analysis of Ref.~\cite{ZhaoSIDIS2024} contains a corresponding rank-four fragmentation channel, where the distribution and fragmentation roles are interchanged, and their nonperturbative functions are independent.

The lower orbital-rank $LTT$ branch instead has $\calT_+=A_{32,+}\kappa^*h^{q[1]}_{32}$ and produces
\begin{equation}
 \varepsilon s_{32}\sin(2\phi_{32})
 \Conv[\ell_2h^{q[1]}_{32}H_1^\perp].
 \label{eq:LTTsidis}
\end{equation}
Although the hadron azimuth cancels from this trigonometric factor, the response still depends on target orientation relative to the lepton plane and on $P_{hT}^2$ through the convolution. The thirty-two responses are independent in the combined space of target multipoles, beam helicity, and azimuths. A single preparation can combine several responses in the same hadron harmonic. For example, $g^q_{10}$ and $g^q_{30}$ both give azimuth-independent helicity responses, and several $m=1$ vector and octupole terms share phases. Separating them requires target-state variation. A third hadron harmonic also occurs in vector-target pretzelosity~\cite{Bacchetta2007}, so that harmonic alone cannot identify an octupole. The preparations in Sec.~\ref{sec:octobservables} isolate rank three from the lower moments.

\subsection{Drell--Yan kinematics and number responses}
\label{sec:DY}

In continuum Drell--Yan production, the same target-spin decomposition enters a distribution correlator with a past-pointing staple. The hard process pairs it with a second-hadron distribution rather than fragmentation. At small transverse momentum, consider
\begin{equation}
 A(P_A,\rho)+B(P_B)\longrightarrow
 \ell^-(l_-)+\ell^+(l_+)+X,
 \label{eq:DYprocess}
\end{equation}
with $q=l_-+l_+$, $Q^2=q^2>0$, and $q_T\ll Q$. We neglect lepton masses, use one-photon exchange, and sum the final lepton spins. The hard part is the standard quark--antiquark annihilation kernel. Tensor-polarized DY has a lower-spin precedent~\cite{HinoKumano1999}, and here, we use the polarized parton-model hard-contraction conventions of Ref.~\cite{Arnold2009} and supply the complete rank-three distribution in the incoming target.

Take $A$ along $+z$ and $B$ along $-z$ in the hadron center-of-mass frame. Neglecting hadron masses in the hard kinematics, but retaining $M_A,M_B$ in the TMD normalizations, put
\begin{equation}
 x_A=\frac{Q}{\sqrt{s}}\e^{\mathcal Y},\qquad
 x_B=\frac{Q}{\sqrt{s}}\e^{-\mathcal Y},\qquad
 s=(P_A+P_B)^2,
 \label{eq:DYfractions}
\end{equation}
at leading power, where $\mathcal Y$ is the dilepton rapidity. The subscripts $A,B$ label the incoming hadrons: $x_B$ is distinct from the SIDIS Bjorken variable $x_{\mathrm{Bj}}$. Replacing $Q$ by $\sqrt{Q^2+q_T^2}$ changes the hard kinematics only at subleading power. Write $\bm q_T=q_T(\cos\varphi,\sin\varphi)$ and $\hat h=\bm q_T/q_T$. Target moments use the SIDIS correlator's $+z$ convention, with azimuths measured in the hadron frame. The negative lepton's angles $(\theta_\ell,\phi_\ell)$ instead refer to the Collins--Soper dilepton rest frame. For $q_T>0$ we choose its $Z$ axis along $\hat P_A^*-\hat P_B^*$ and $X$ axis opposite $\hat P_A^*+\hat P_B^*$, with $Y=Z\times X$ and stars denoting unit hadron directions in that frame. At leading power, the transverse $X,Y$ axes correspond to $\hat h,\hat h_\perp$ in the hadron frame. Beyond this approximation, spin-basis rotations and frame corrections must be included consistently.

Let $\bar\Phi_B^q$ denote the annihilation antiquark correlator. Its leading projections are $\bar\calF_B=\tfrac12\Tr_D(\bar\Phi_B\gamma^-)$, $\bar\calG_B=\tfrac12\Tr_D(\bar\Phi_B\gamma^-\gamma_5)$, and $\bar\calT_B^i=\tfrac12\Tr_D(\bar\Phi_B\ii\sigma^{i-}\gamma_5)$. It is reconstructed with $\gamma^+$ and $\ii\sigma^{i+}\gamma_5$ in place of their minus-index counterparts in Eq.~\eqref{eq:qreconstruct}. Both incoming distributions have past-pointing staples, which remain part of their nonperturbative definitions even with a Born hard coefficient. The leading-power Born transverse hadronic tensor is
\begin{equation}
 \begin{aligned}
 W_{\mathrm{DY}}^{ij}={}&\frac1{N_c}\sum_q e_q^2
 \int\dd^2\bm k_{AT}\,\dd^2\bm k_{BT}\\
 &\times\delta^{(2)}(\bm q_T-\bm k_{AT}-\bm k_{BT})\\
 &\times\left\{\Tr_D[\Phi_A^q\gamma^i\bar\Phi_B^q\gamma^j]
                  +(q\leftrightarrow\bar q)\right\},
 \end{aligned}
 \label{eq:DYtensor}
\end{equation}
where factor $1/N_c$ applies to color-summed distribution correlators. The exchanged term uses an antiquark from $A$ and a quark from $B$, at the same target-moment and momentum arguments.

Summing the final lepton spins selects the symmetric part of the hard trace. The Dirac reconstructions give, for the first annihilation channel,
\begin{equation}
 \begin{split}
 \mathcal H_{\mathrm{DY}}^{(ij)}={}&
 \delta^{ij}(\calF_A\bar\calF_B+\calG_A\bar\calG_B)\\
 &-\calT_A^i\bar\calT_B^j-\calT_A^j\bar\calT_B^i
       +\delta^{ij}\calT_A\cdot\bar\calT_B.
 \end{split}
 \label{eq:DYtrace}
\end{equation}
The rest-frame leptonic tensor satisfies
\begin{equation}
 \frac{L_{ij}}{2Q^2}=\delta_{ij}
       -\sin^2\theta_\ell\,e_i e_j,\qquad
 \bm e=(\cos\phi_\ell,\sin\phi_\ell).
 \label{eq:DYlepton}
\end{equation}
Its contraction with Eq.~\eqref{eq:DYtrace} is
\begin{equation}
 \begin{split}
 &(1+\cos^2\theta_\ell)
       (\calF_A\bar\calF_B+\calG_A\bar\calG_B)\\
 &\quad+\sin^2\theta_\ell\,
       \Rea[\e^{-2\ii\phi_\ell}\calT_{A,+}\bar\calT_{B,+}],
 \end{split}
 \label{eq:DYangulartrace}
\end{equation}
The transverse complex components here use the recoil-aligned axes. The trace pairs number with number, helicity with helicity, and transverse spin with transverse spin. A number--helicity product gives no symmetric-photon contribution. The transverse-spin term has a second dilepton-azimuth harmonic and vanishes under a full $\phi_\ell$ average. These selection rules determine the required partner and angular projections for every target spin rank.

Superscripts $A,B$ identify the parent hadrons, and the flavor sum supplies quark and antiquark labels. For physical scalar functions, define the convolution
\begin{equation}
 \begin{aligned}
 &\mathcal C_{\mathrm{DY}}[w a_A b_B]=\frac1{N_c}\sum_q e_q^2
 \int\dd^2\bm k_{AT}\,\dd^2\bm k_{BT}\\
 &\quad\times\delta^{(2)}(\bm q_T-\bm k_{AT}-\bm k_{BT})\\
 &\quad\times w\left[a_A^{q,[-]}b_B^{\bar q,[-]}
                     +a_A^{\bar q,[-]}b_B^{q,[-]}\right].
 \end{aligned}
 \label{eq:DYconv}
\end{equation}
The two distributions have arguments $(x_A,\bm k_{AT}^2)$ and $(x_B,\bm k_{BT}^2)$, with a common UV convention and rapidity scales paired so that $\zeta_A\zeta_B=Q^4$ in the standard two-line factorization scheme. The DY convolution contains two incoming distributions and its color factor, no SIDIS factor $x_A$ or fragmentation Jacobian. The phase space and cross-section prefactor are
\begin{equation}
 \begin{split}
 \dd\mathcal P_{\mathrm{DY}}&=\dd Q^2\,\dd\mathcal Y\,\dd^2\bm q_T
                    \,\dd\cos\theta_\ell\,\dd\phi_\ell,\\
 \mathcal N_{\mathrm{DY}}&=\frac{\alpha_{\mathrm{em}}^2}{4sQ^2}.
 \end{split}
 \label{eq:DYnormalization}
\end{equation}
The differential cross section is $\mathcal N_{\mathrm{DY}}$ times the flavor convolution of Eq.~\eqref{eq:DYangulartrace}. For an unpolarized reference target, the dilepton-azimuth average is $\mathcal N_{\mathrm{DY}}(1+\cos^2\theta_\ell)\mathcal C_{\mathrm{DY}}[f_{00}^{A}f_1^B]$. The solid-angle integral of $1+\cos^2\theta_\ell$ gives $16\pi/3$, recovering the usual $4\pi\alpha_{\mathrm{em}}^2/(9sQ^2)$ normalization for $N_c=3$ and collinear Born distributions. This checks the Born operator limit, while relating a TMD integral to a renormalized PDF still requires the UV treatment discussed in Sec.~\ref{sec:collinear}.

We first isolate the number channel with an unpolarized hadron $B$, for which $\bar\calF_B=f_1^{\bar q/B}$ and $\bar\calG_B=0$. Its Boer--Mulders term may be nonzero, but averaging over $\phi_\ell$ removes the transverse-spin product. Define the averaged rate
\begin{equation}
 \bar\sigma_\eta=\frac1{2\pi}\int_0^{2\pi}\dd\phi_\ell\,
       \frac{\dd\sigma_\eta}{\dd\mathcal P_{\mathrm{DY}}}.
 \label{eq:DYaverage}
\end{equation}
The label $\eta=\pm1$ selects the opposite octupole preparations in Eq.~\eqref{eq:rotatedrho}, both of which have positive density matrices and zero vector and quadrupole moments. At a fixed preparation polar angle, their moments are $A_{3m,+}=\eta O\alpha_m\e^{\ii m\psi}$, using the cone geometry and signed $\alpha_m$ of Eqs.~\eqref{eq:cone_moments} and \eqref{eq:cone_alpha}. The lepton average is an acceptance-corrected angular projection.

Rotate both integration momenta into $\hat h$ and define
\begin{equation}
 u_A=\frac{k_{Ax}+\ii k_{Ay}}{M_A}\e^{-\ii\varphi},\qquad
 v_B=\frac{k_{Bx}+\ii k_{By}}{M_B}\e^{-\ii\varphi}.
 \label{eq:DYuv}
\end{equation}
The scalar weights are $w_m^{\mathrm{DY}}=\Rea(u_A^m)$. For $u_A=a+\ii b$, they are $a$, $a^2-b^2$, and $a^3-3ab^2$ for $m=1,2,3$, respectively. Reflecting both integration momenta about $\hat h$ leaves the DY delta function and radial coefficients unchanged. The odd weight $\Ima(u_A^m)$ integrates to zero, so $\cS_m[A_{3m}]$ contributes $\eta O\alpha_m w_m^{\mathrm{DY}}\sin m(\varphi-\psi)$. Define
\begin{align}
 F_{UU}^{\mathrm{DY}}&=\mathcal C_{\mathrm{DY}}[f_{00}^{A}f_1^B],\nonumber\\
 F_{3m}^{\mathrm{DY}}&=\mathcal C_{\mathrm{DY}}[w_m^{\mathrm{DY}}f_{3m}^{A}f_1^B],
 \label{eq:DYnumberSF}
\end{align}
with $m=1,2,3$ to write the averaged rate as
\begin{equation}
 \begin{split}
 \bar\sigma_\eta={}&\mathcal N_{\mathrm{DY}}(1+\cos^2\theta_\ell)
 \biggl[F_{UU}^{\mathrm{DY}}\\
 &\quad+\eta O\sum_{m=1}^3\alpha_m
       \sin m(\varphi-\psi)F_{3m}^{\mathrm{DY}}\biggr].
 \end{split}
 \label{eq:DYnumberrate}
\end{equation}
These preparations contain no $K=1,2$ contributions, and their difference also removes the unpolarized reference, leaving the three octupole number responses. Parity forbids $f_{30}$, so this channel has no longitudinal octupole term.

A joint scan of recoil and preparation azimuths separates these three responses. For equalized luminosities and full angular coverage in $0\leq\varphi,\psi<2\pi$, define the normalized moment
\begin{equation}
 \mathcal A_m^{\mathrm{DY}}=
 \frac2O\,
 \frac{\displaystyle\int\dd\varphi\,\dd\psi\,
       \sin m(\varphi-\psi)(\bar\sigma_+-\bar\sigma_-)}
      {\displaystyle\int\dd\varphi\,\dd\psi\,
                         (\bar\sigma_++\bar\sigma_-)}
 =\alpha_m\frac{F_{3m}^{\mathrm{DY}}}{F_{UU}^{\mathrm{DY}}},
 \label{eq:DYnumbermoment}
\end{equation}
where the factor of two compensates the squared-sine average. Interpreting the moment requires the calibrated octupole strength and a nonzero unpolarized reference. Choosing $\cos\vartheta=3/5$ gives the nonzero cone coefficients in Eq.~\eqref{eq:generic_cone}. On a transverse cone, $\alpha_1=-1/2$, $\alpha_3=5/2$, and $\alpha_2=0$. The $m=3$ projection gives $\mathcal A_3^{\mathrm{DY}} = 5F_{33}^{\mathrm{DY}}/(2F_{UU}^{\mathrm{DY}})$ and removes the accompanying $LLT$ response. It therefore isolates the $TTT$ contribution even though the preparation itself contains both components. Imperfect preparation is handled through its measured density matrix, as in Sec.~\ref{sec:nuclear}.

To express this DY response in terms of SIDIS target inputs, substitute the operator reversal relation in Eq.~\eqref{eq:octSIDISDY}:
\begin{equation}
 \begin{aligned}
 F_{3m}^{\mathrm{DY}}={}&-\frac1{N_c}\sum_q e_q^2
 \int\dd^2\bm k_{AT}\,\dd^2\bm k_{BT}\\
 &\times\delta^{(2)}(\bm q_T-\bm k_{AT}-\bm k_{BT})w_m^{\mathrm{DY}}\\
 &\times\bigl[f_{3m}^{q/A,[+]}f_1^{\bar q/B,[-]}
             +f_{3m}^{\bar q/A,[+]}f_1^{q/B,[-]}\bigr].
 \end{aligned}
 \label{eq:DYfromSIDIS}
\end{equation}
The minus sign acts on the target distributions constrained in SIDIS, after evolution to the DY scales. The second-hadron distributions remain independent inputs. Comparing this expression with the SIDIS structure function $\Conv[w_m f_{3m}^{q/A,[+]}D_1]$ requires each process's companion functions, flavor weights, transverse momentum constraints, and kinematics.

\subsection{Spin analyzers and cross-process comparisons}

To access the target helicity channel in Eq.~\eqref{eq:DYangulartrace}, use a longitudinally polarized spin-$1/2$ partner. Let $\lambda_B\in[-1,1]$ be its helicity polarization along $P_B$. In the annihilation antiquark convention, $\bar\calG_B=-\lambda_B g_{1L}^{\bar q/B}$, while the quark from $A$ has $\calG_A$ as in Eq.~\eqref{eq:Gfull}. In the exchanged channel the antiquark axial sign is on $A$, giving the same physical sign in $-\lambda_B\calG_A^{\bar q}g_{1L}^{q/B}$. This charge-conjugation sign belongs to the hard spin contraction, distinct from Wilson-line reversal~\cite{Arnold2009}.

With $w_0^{\mathrm{DY}}=1$, define
\begin{equation}
 G_{3m}^{\mathrm{DY}}=\mathcal C_{\mathrm{DY}}
       [w_m^{\mathrm{DY}}g_{3m}^{A}g_{1L}^B],\qquad m=0,1,2,3.
 \label{eq:DYhelicitySF}
\end{equation}
After the $\phi_\ell$ average, the target-preparation difference contains number and helicity responses. Averaging over opposite partner polarizations selects number, while their difference selects helicity:
\begin{equation}
 \begin{aligned}
 &\frac{\bar\sigma_{\lambda_B,\eta}-
       \bar\sigma_{\lambda_B,-\eta}}
      {2\eta O\mathcal N_{\mathrm{DY}}(1+\cos^2\theta_\ell)}\\
 &\quad=\sum_{m=1}^3\alpha_m\sin m(\varphi-\psi)F_{3m}^{\mathrm{DY}}
       -\lambda_B\alpha_0G_{30}^{\mathrm{DY}}\\
 &\qquad-\lambda_B\sum_{m=1}^3\alpha_m
                  \cos m(\varphi-\psi)G_{3m}^{\mathrm{DY}}.
 \end{aligned}
 \label{eq:DYhelicityrate}
\end{equation}
For a longitudinal octupole target ($\alpha_0=1$) and fully polarized partner ($\lambda_B=\pm1$), the beam--target double difference selects the longitudinal helicity response:
\begin{equation}
 \frac14\sum_{\lambda_B,\eta=\pm1}\lambda_B\eta
       \bar\sigma_{\lambda_B,\eta}
 =-O\mathcal N_{\mathrm{DY}}(1+\cos^2\theta_\ell)G_{30}^{\mathrm{DY}}.
 \label{eq:DYlongdouble}
\end{equation}
The minus sign comes from the longitudinal analyzing power of vector-current annihilation. The target function is unchanged under link reversal: $g_{30}^{q,[-]}$ equals $g_{30}^{q,[+]}$. Thus, a negative hard analyzing power can accompany a link-even target distribution.

Retaining the dilepton azimuth gives access to the seven octupole quark transverse-spin coefficients. An unpolarized hadron $B$ can supply their chiral-odd partner through its Boer--Mulders function, which correlates transverse quark spin with intrinsic momentum. Its transverse-spin projection in the common transverse axes is $\bar{\calT}_B^i=-\epsilon_{ij}k_{BT}^j h_1^{\perp\bar q/B}/M_B$, or $\bar{\calT}_{B,+}=\ii v_Bh_1^{\perp\bar q/B}$ in the recoil-aligned axes. The epsilon term has the opposite sign to that for a plus-moving hadron because the light-front directions are interchanged~\cite{Arnold2009}. Combining it with the target branches in Eq.~\eqref{eq:Vcomplex}, define
\begin{align}
 H_{30}^{[1],\mathrm{DY}}&=\mathcal C_{\mathrm{DY}}
  [\Rea(u_Av_B)h_{30}^{A[1]}h_1^{\perp B}],\nonumber\\
 H_{3m}^{[m-1],\mathrm{DY}}&=\mathcal C_{\mathrm{DY}}
  [\Rea((u_A^*)^{m-1}v_B)h_{3m}^{A[m-1]}h_1^{\perp B}],\nonumber\\
 H_{3m}^{[m+1],\mathrm{DY}}&=\mathcal C_{\mathrm{DY}}
  [\tfrac12\Rea(u_A^{m+1}v_B)h_{3m}^{A[m+1]}h_1^{\perp B}],
 \label{eq:DYchiralSF}
\end{align}
for $m=1,2,3$. The bracketed superscripts give the target TMD's orbital rank, while the separate staple label is $[-]$ for both incoming distributions, as specified in Eq.~\eqref{eq:DYconv}. With $\Delta_m=m(\varphi-\phi_{3m})$, the chiral-odd rank-three contribution is
\begin{equation}
 \begin{aligned}
 &\frac1{\mathcal N_{\mathrm{DY}}}
 \left.\frac{\dd\sigma}{\dd\mathcal P_{\mathrm{DY}}}
       \right|_{K=3,\,h}\\
 &\quad=\sin^2\theta_\ell\biggl\{S_{LLL}\sin2\phi_\ell H_{30}^{[1],\mathrm{DY}}\\
 &\qquad+\sum_{m=1}^3s_{3m}\bigl[
       \sin(2\phi_\ell+\Delta_m)H_{3m}^{[m-1],\mathrm{DY}}\\
 &\hspace{24mm}+\sin(2\phi_\ell-\Delta_m)H_{3m}^{[m+1],\mathrm{DY}}\bigr]\biggr\}.
 \end{aligned}
 \label{eq:DYchiralrate}
\end{equation}
To obtain the two angular phases, multiply the recoil-frame target branches $s_{3m}\e^{-\ii\Delta_m}(u_A^*)^{m-1}$ and $\tfrac12s_{3m}\e^{\ii\Delta_m}u_A^{m+1}$ by $\ii v_B$ in Eq.~\eqref{eq:DYangulartrace}. Reflection removes the imaginary reduced weights, leaving Eq.~\eqref{eq:DYchiralrate}. A SIDIS/DY comparison must therefore account for both the partner and the lepton-frame conventions.

The target $h$ functions are link even, whereas their Boer--Mulders partner is link odd. This partner is a distribution, distinct from the Collins fragmentation function in SIDIS. Interpreting the DY responses requires independent information about it, or complementary polarized-beam observables with a different chiral-odd analyzer. Together, Eqs.~\eqref{eq:DYnumberrate}, \eqref{eq:DYhelicityrate}, and \eqref{eq:DYchiralrate} display all three number, four helicity, and seven transverse-spin octupole channels in the specified beam settings. Extracting the fourteen target TMDs from those responses still requires flavor convolutions, partner distributions, acceptance corrections, and target calibration.

A joint SIDIS/DY analysis therefore starts from one set of target functions, $f_{3m}^{q,[+]}$, $g_{3m}^{q,[+]}$, and $h_{3m}^{q[n],[+]}$, at a common reference scale, together with independent fragmentation and second-hadron distributions. Table~\ref{tab:SF} supplies the SIDIS kernels, and the DY kernels use the past-staple substitutions in Eq.~\eqref{eq:octSIDISDY}. The three number channels test sign-reversing target functions, while the helicity and transverse-spin channels test unchanged target functions with different analyzers. No equality between flavors or between target and partner antiquark distributions is assumed. The operator-level null condition $C_\nu^{q,[-]}-s_\nu C_\nu^{q,[+]}=0$ at matched arguments is tested through these process convolutions. Evolution and higher-order comparisons follow Secs.~\ref{sec:universality} and \ref{sec:renormalization}. Larger-$q_T$ matching and power corrections require separate assessment, as do factorization changes from nuclear breakup tags, noninclusive remnant restrictions, or a different colored-final-state measurement.

\section{Collinear and nuclear interpretation}
\label{sec:collinear}

\subsection{Collinear distributions and spin contrasts}

We now connect the scattering responses to collinear spin contrasts, local-moment constraints, and nuclear structure. The collinear sets established by spin-$3/2$ PDF and GPD analyses~\cite{Fu2022,Fu2024,Fu2026} constrain the finite-$k_T$ extension and clarify its lowest orbital ranks. With a rotationally invariant transverse regulator, $\int\dd\phi_k\,\mathsf k_n=0$ for $n>0$, selecting orbital-rank-zero structures. The remaining radial integral is generally UV divergent in QCD because of the perturbative transverse-momentum tail, and relating it to a collinear PDF requires a UV prescription and the appropriate matching, not an unregulated integral of the soft-subtracted TMD.

We therefore define the collinear quark correlator directly through the renormalized light-ray operator,
\begin{equation}
 \begin{split}
 \Phi^q_{\mathrm{col}}(x;\mu)=\Bigl[\int\frac{\dd\xi^-}{2\pi}
 &\e^{\ii xP^+\xi^-}
 \langle\bar\psi(0)\\
 &\times U_{\mathrm{straight}}[0,\xi^-]\psi(\xi^-)\rangle_\rho
 \Bigr]_{\mathrm{UV\,ren}},
 \end{split}
 \label{eq:colop}
\end{equation}
with an analogous field-strength definition for gluons. The small-$b_T$ expansion and matching relate this operator to the subtracted TMD. To identify the allowed collinear distributions, we combine orbital-rank selection with the symmetries of the straight-link operator.

Orbital-rank selection retains the quark scalar terms $f^q_{00}$, $f^q_{20}$, $g^q_{10}$, and $g^q_{30}$, and the transverse-spin candidates $h^{q[0]}_{11}$, $h^{q[0]}_{21}$, and $h^{q[0]}_{31}$. Time reversal excludes the link-odd $K=2$ transversity coefficient from the straight-link collinear operator. The six surviving distributions give
\begin{align}
 \calF^q_{\mathrm{col}}&=f_1^q+S_{LL}f_{1LL}^q,\nonumber\\
 \calG^q_{\mathrm{col}}&=S_Lg_1^q+S_{LLL}g_{1LLL}^q,\nonumber\\
 \calT^{q,i}_{\mathrm{col}}&=S_T^ih_1^q+S_{LLT}^ih_{1LLT}^q,
 \label{eq:qcolproj}
\end{align}
which are the conventional collinear names. Their correspondence to the rank-indexed TMD coefficients is fixed by the tensor products in Sec.~\ref{sec:count}. The gluon candidates similarly consist of the four number and helicity terms and $h^{g[0]}_{22},h^{g[0]}_{32}$. Time reversal excludes the latter, which is link odd. The five collinear functions enter as
\begin{align}
 \calF^g_{\mathrm{col}}&=f_1^g+S_{LL}f_{1LL}^g,\nonumber\\
 \calG^g_{\mathrm{col}}&=S_Lg_1^g+S_{LLL}g_{1LLL}^g,\nonumber\\
 \calH^{g,ij}_{\mathrm{col}}&=S_{TT}^{ij}h_{1TT}^g.
 \label{eq:gcolproj}
\end{align}
These sets agree with the independent spin-$3/2$ forward PDF and GPD analyses~\cite{Fu2024,Fu2026}. Time-reversal invariance excludes the collinear octupole gluon-transversity PDF, but the target can still support a rank-three $LTT$ helicity-two coherence. At finite transverse momentum, this coherence can contribute to a link-odd TMD.

The longitudinal distributions compare pure target-spin states. For target helicity $\Lambda=3/2$ or $1/2$, denote the number responses by $U_{3/2},U_{1/2}$ and the helicity responses by $G_{3/2},G_{1/2}$. The target-spin eigenvalues give, for either parton species,
\begin{align}
 U_{3/2}&=f_1+f_{1LL},&U_{1/2}&=f_1-f_{1LL},\nonumber\\
 G_{3/2}&=\tfrac32g_1+\tfrac3{10}g_{1LLL},&
 G_{1/2}&=\tfrac12g_1-\tfrac9{10}g_{1LLL}.
 \label{eq:populationresponses}
\end{align}
Reversing target helicity leaves the number response unchanged and reverses the helicity response. In a positive spectral prescription, the two parton-helicity densities are $(U_\Lambda\pm G_\Lambda)/2$. The operator relations remain valid in renormalization prescriptions that do not preserve pointwise probabilities, and Appendix~\ref{sec:positivity} gives the positivity conditions. Inverting Eq.~\eqref{eq:populationresponses} gives
\begin{align}
 f_1&=\tfrac12(U_{3/2}+U_{1/2}),&
 f_{1LL}&=\tfrac12(U_{3/2}-U_{1/2}),\nonumber\\
 g_1&=\tfrac15(3G_{3/2}+G_{1/2}),&
 g_{1LLL}&=\tfrac13(G_{3/2}-3G_{1/2}).
 \label{eq:populationinverse}
\end{align}
The quadrupole function $f_{1LL}$ compares the number responses of the $|\Lambda|=3/2$ and $1/2$ sectors. Octupole helicity measures the departure from proportionality to target helicity, which would require $G_{3/2}=3G_{1/2}$. It can therefore have independent $x$ dependence even when its lowest local moments vanish, as discussed below. The same interpretation applies to $g_{30}(x,\bk^2)$ once the target multipole and orbital harmonic have been separated.

Quark transversity describes interference between opposite quark helicities. At orbital rank zero, both $S_T$ and $S_{LLT}$ supply the required one-unit target coherence. To distinguish their vector and octupole spin ranks, consider the raising block in target spin space:
\begin{equation}
 h_1J_++h_{1LLT}(O_{zzx}+\ii O_{zzy}).
 \label{eq:coherenceblock}
\end{equation}
In this block, let $a$ denote either outer transition amplitude ($-3/2\to-1/2$ or $1/2\to3/2$), and $b$ the middle transition ($-1/2\to1/2$). The spin matrices give
\begin{align}
 a&=\sqrt3\left(h_1+\tfrac25h_{1LLT}\right),\nonumber\\
 b&=2h_1-\tfrac65h_{1LLT},
 \label{eq:coherenceamplitudes}\\
 h_{1LLT}&=a/\sqrt3-b/2,\qquad
 h_1=\frac{3a}{5\sqrt3}+\frac b5.
 \label{eq:coherenceinverse}
\end{align}
Octupole transversity therefore compares the outer and middle coherences. A single off-diagonal density-matrix element can receive contributions from multiple spin ranks, so projection onto rank three isolates the octupole from rank-one transversity. Gluon linear polarization requires a two-unit partonic helicity change, with orbital-rank-zero target partners $S_{TT}$ and $S_{LTT}$ in Eqs.~\eqref{eq:H2} and \eqref{eq:B3}.

\subsection{Inclusive structure functions and local moments}
\label{sec:mellin}

In inclusive DIS, no final hadron is identified. For the diagonal collinear target multipoles, define the Born structure functions by
\begin{align}
 F_{1,A}^{(0)}&=\tfrac12\sum_q e_q^2(f_1^q+f_1^{\bar q}),\nonumber\\
 F_{1,A}^{(2)}&=\tfrac12\sum_q e_q^2(f_{1LL}^q+f_{1LL}^{\bar q}),\nonumber\\
 g_{1,A}^{(1)}&=\tfrac12\sum_q e_q^2(g_1^q+g_1^{\bar q}),\nonumber\\
 g_{1,A}^{(3)}&=\tfrac12\sum_q e_q^2(g_{1LLL}^q+g_{1LLL}^{\bar q}).
 \label{eq:inclusiveSF}
\end{align}
At massless-quark Born order, the transverse number and helicity responses are $2x[F_{1,A}^{(0)}+S_{LL}F_{1,A}^{(2)}]$ and $2x[S_Lg_{1,A}^{(1)}+S_{LLL}g_{1,A}^{(3)}]$, respectively. The number sector obeys $F_2^{(K)}=2xF_1^{(K)}$ for $K=0,2$. Equation~\eqref{eq:inclusiveSF} fixes our multipole normalization, and comparison with a $b_1$-type convention requires converting the polarization normalization. Chiral-odd quark PDFs need an additional chiral-odd analyzer and do not enter this Born inclusive electromagnetic kernel. Gluon coefficient functions and further photon-helicity structures require a higher-order hard calculation. To recover the inclusive number response from SIDIS, use an energy-weighted sum over hadron species and the fragmentation momentum sum rule, which cannot be replaced by setting an identified-hadron fragmentation function to unity.

The moments of these collinear functions are constrained by local operator spin rank. Taking the $N$th Mellin moment converts the light-ray operator into a local bilinear with $N-1$ covariant derivatives. In the number and helicity channels, the operators are
\begin{align}
 \mathcal O_{V,N}^{\mu_1\cdots\mu_N}
 &=\bar q\gamma^{\{\mu_1}\ii D^{\mu_2}\cdots
 \ii D^{\mu_N\}}q-\text{traces},\nonumber\\
 \mathcal O_{A,N}^{\mu_1\cdots\mu_N}
 &=\bar q\gamma^{\{\mu_1}\gamma_5\ii D^{\mu_2}\cdots
 \ii D^{\mu_N\}}q-\text{traces}.
 \label{eq:localoperators}
\end{align}
The transversity operator starts with $\ii\sigma^{\alpha\mu_1}\gamma_5$ and has the same $N-1$ derivatives with its mixed index symmetry. Total derivatives vanish in the forward matrix element. In the target rest frame, vector and axial-vector bilinears have rotational rank at most one, and the antisymmetric tensor bilinear also decomposes into two rotational vectors. Each derivative adds at most one unit of rotational rank, so these local quark operators have maximum rank $N$. By the Wigner--Eckart theorem, an operator cannot resolve a higher-rank target multipole. Hence,
\begin{equation}
 \text{rank-}K\text{ quark moment}=0\qquad(N<K).
 \label{eq:momentselection}
\end{equation}
This multipole constraint underlies arbitrary-spin DIS and spin-$3/2$ GPD moments~\cite{JaffeManohar1989,Fu2022,Fu2024}.

The constraints apply to specific quark--antiquark combinations. With positive-$x$ antiquark conventions, vector and transversity moments contain $f^q+(-1)^Nf^{\bar q}$ and $h^q+(-1)^Nh^{\bar q}$, while axial moments contain $g^q+(-1)^{N-1}g^{\bar q}$. The first-moment constraints are
\begin{align}
 \int_0^1\dd x\,[f_{1LL}^q-f_{1LL}^{\bar q}]&=0,\nonumber\\
 \int_0^1\dd x\,[g_{1LLL}^q+g_{1LLL}^{\bar q}]&=0,\nonumber\\
 \int_0^1\dd x\,[h_{1LLT}^q-h_{1LLT}^{\bar q}]&=0.
 \label{eq:localmoments}
\end{align}
For the rank-three functions, the second moments also vanish:
\begin{align}
 \int_0^1\dd x\,x[g_{1LLL}^q-g_{1LLL}^{\bar q}]&=0,\nonumber\\
 \int_0^1\dd x\,x[h_{1LLT}^q+h_{1LLT}^{\bar q}]&=0.
 \label{eq:secondmoments}
\end{align}
The first locally allowed rank-three moments therefore have $N=3$, with an $x^2$ weight. Because the antiquark sign changes with $N$, the constraints refer to the displayed combinations, not to an isolated positive-$x$ quark distribution.

Equations~\eqref{eq:localmoments} and \eqref{eq:secondmoments} define moments through renormalized local operators. Treating them as ordinary integrals also requires small-$x$ convergence and the appropriate treatment of distributional contributions. A continuous, integrable, nonzero combination with a vanishing unweighted integral must take both signs, but the constraint fixes no node position. These relations test a model's spin structure without predicting its shape. The argument uses local quark currents; gluon helicity has different gauge-invariant operator content and does not inherit a first-moment sum rule from it.

\subsection{Evolution and transverse-distance matching}
\label{sec:evolution}

Scale evolution acts on the coefficient of each fixed
impact-parameter tensor. For a two-line, soft-subtracted TMD with scalar UV and rapidity counterterms, the evolution at nonzero transverse separation is the same for all target multipoles within a fixed quark or gluon representation. For the coefficient $\widetilde C_\nu^a$, with $\nu$ labeling the parton channel and fixed target/orbital tensor,
\begin{align}
 \frac{\partial\widetilde C_\nu^a}{\partial\ln\sqrt\zeta}
 &=\mathcal K_a(b_T;\mu)\widetilde C_\nu^a,\nonumber\\
 \frac{\partial\widetilde C_\nu^a}{\partial\ln\mu}
 &=\gamma_F^a(\alpha_s,\zeta/\mu^2)\widetilde C_\nu^a,\nonumber\\
 \frac{\dd \mathcal K_a}{\dd\ln\mu}&=-\gamma_K^a.
 \label{eq:evolution}
\end{align}
Here, $\mathcal K_a$ is the Collins--Soper kernel and $a$ labels the parton representation. The evolution kernel has no additional target-spin label $K$. Quark and gluon kernels differ, while short-distance matching also depends on the partonic channel~\cite{Handbook}.

To obtain the impact-parameter tensor coefficient, Fourier transform the momentum tensor together with its scalar coefficient. For a radial coefficient $c(x,\bk^2)$, let $\widetilde c(x,b_T^2)$ be its scalar Fourier transform. A rank-$n$ momentum tensor transforms by differentiation:
\begin{equation}
 \begin{split}
 &\int\dd^2\bk\,\e^{\ii\bk\cdot\bb}
 \mathsf k_n^{i_1\cdots i_n}c(x,\bk^2)\\
 &\quad=\frac{(-\ii)^n}{\M^n}\STF
 \bigl(\partial_{b^{i_1}}\cdots\partial_{b^{i_n}}\bigr)
 \widetilde c(x,b_T^2).
 \end{split}
 \label{eq:Fourierderivative}
\end{equation}
Equivalently, the transform of a $\kappa^n$ harmonic contains
\begin{equation}
 2\pi\ii^n\e^{\ii n\phi_b}
 \int_0^\infty\dd k\,k(k/\M)^nJ_n(kb_T)c(x,k^2).
 \label{eq:Besseltransform}
\end{equation}
Both forms preserve the transverse harmonic while differentiating or Bessel transforming its radial coefficient. The small-$b_T$ operator expansion gives
\begin{equation}
 \widetilde{\mathcal O}^{a}_{Km}
 =\sum_j C^{a\leftarrow j}_{Km}\otimes
 \langle\mathcal O_j\rangle_{Km}+\text{power corrections},
 \label{eq:OPE}
\end{equation}
The coefficient functions depend on $b_T,\mu,\zeta$, and the convolution is over the longitudinal fraction. The collinear operators entering the expansion depend on the channel: for link-odd TMDs, the small-$b_T$ expansion can involve multiparton twist-three correlators, beyond the twist-two PDFs. A vanishing orbital-rank-zero PDF therefore does not eliminate the corresponding finite-$k_T$ TMD.

The mass in the momentum tensor of Eq.~\eqref{eq:Kn} is also a normalization convention. If $c_{[n]}^{(M)}$ multiplies a rank-$n$ tensor normalized with mass $M$, replacing $M_A$ by $M_0$ requires
\begin{equation}
 c_{[n]}^{(M_0)}=(M_0/\M)^nc_{[n]}^{(\M)}.
 \label{eq:massconversion}
\end{equation}
This rescaling leaves the coefficient--tensor product and every observable unchanged. Orbital rank specifies angular dependence, but by itself, it determines neither nuclear suppression, a unique partial wave, nor the number of correlated constituents. Its connection to a perturbative transverse-momentum tail requires channel-specific matching.

\subsection{Target calibration and nuclear structure}\label{sec:nuclear}

We first illustrate what calibrated target preparations can separate, then discuss the nuclear dynamics needed to calculate the target distributions. Take the longitudinal $\hat n=\hat z$ states of Eq.~\eqref{eq:rotatedrho}, with $S_L=S_{LL}=0$ and $S_{LLL}=\eta O$, where $\eta=\pm1$ and $0<O\leq1/2$. Their SIDIS cross sections are
\begin{equation}
 \begin{split}
 \sigma_{\lambda,\eta}=\mathcal N\bigl[
 \mathcal W_0^{\mathrm{SIDIS}}+\lambda\eta O\Dpol F^{\mathrm{L}}_{30}
 +\eta O\varepsilon\sin(2\phi_h)F^{\mathrm{T}[1]}_{30}\bigr],
 \end{split}
 \label{eq:octcross}
\end{equation}
Here, $\mathcal W_0^{\mathrm{SIDIS}}$ contains the unpolarized-target responses, and the structure functions are those of Table~\ref{tab:SF}. Weighting the four spin settings by the product of lepton helicity and target-preparation sign selects octupole helicity:
\begin{equation}
 \frac14\sum_{\lambda,\eta=\pm1}\lambda\eta\,
 \sigma_{\lambda,\eta}
 =\mathcal N O\Dpol\Conv[g^q_{30}D_1].
 \label{eq:octdouble}
\end{equation}
Averaging over beam helicity and taking the target difference instead selects $\mathcal N O\varepsilon\sin(2\phi_h)F^{\mathrm{T}[1]}_{30}$. These spin combinations isolate the corresponding structure functions, which still contain convolutions of flavor-dependent TMDs with fragmentation functions. The measurement model must include partial beam polarization, unequal luminosities, acceptance differences, and residual $S_L$. In particular, residual vector polarization adds the ordinary helicity response to Eq.~\eqref{eq:octdouble}.

General preparations and experimental coverage require a yield model. For preparation $a$ and kinematic bin $b$, write
\begin{equation}
 \begin{split}
 N_{ab}={}&\mathcal L_a\int_b\dd\mathcal P\,
 \mathcal A_a(\mathcal P)\mathcal N(\mathcal P)\\
 &\times\sum_\nu M_{a\nu}(\mathcal P;\rho_a,\lambda_a)
 F_\nu(\mathcal P)+N^{\mathrm{background}}_{ab}.
 \end{split}
 \label{eq:designfit}
\end{equation}
The matrix $M$ contains target amplitudes, angular functions, and lepton factors, and $\mathcal L$ and $\mathcal A$ denote luminosity and acceptance. Detector migration between bins and uncertainty in $\rho_a$ enter the same model. To separate responses that share a harmonic, the measured design matrix must contain enough independent response combinations, assessed through singular values or Fisher information. Extracting TMDs then requires the structure functions' flavor sums, fragmentation functions, evolution, and a specified factorization scheme, with complementary final states and kinematics. Finite-$b_T$ or Bessel-weighted observables can simplify the transverse convolution, while flavor separation and nonperturbative inputs remain necessary. Polynomial transverse moments require the UV treatment described above.

The nuclear dynamics can generate rank-three structure even when the active constituent has spin $1/2$, as its spin can correlate with orbital motion. Cloet \emph{et al.} provide a dynamical precedent, calculating polarized distributions for $^{7}\mathrm{Li}$ and $K=3$ spin-dependent $u$- and $d$-quark multipoles in $^{11}\mathrm{B}$ at the model input scale~\cite{Cloet2006}. To illustrate the spin--orbital coupling, let the Clebsch--Gordan isometry $V$ map the $J=3/2$ states into the $L=1\otimes s=1/2$ product basis. This specifies the angular coupling without a nuclear dynamical matrix element. Two states are
\begin{align}
 |\tfrac32,\tfrac32\rangle
 &=|1,1\rangle|\tfrac12,\tfrac12\rangle,\nonumber\\
 |\tfrac32,\tfrac12\rangle
 &=\sqrt{\tfrac23}|1,0\rangle|\tfrac12,\tfrac12\rangle + \sqrt{\tfrac13}|1,1\rangle|\tfrac12,-\tfrac12\rangle,
 \label{eq:LSstates}
\end{align}
and the remaining states follow by lowering. Projecting constituent spin, with and without an orbital-quadrupole weight, onto this coupled subspace gives
\begin{align}
 V^\dagger s_zV&=\tfrac13J_z,\nonumber\\
 V^\dagger[(3L_z^2-2)s_z]V&=O_{zzz}+\tfrac2{15}J_z.
 \label{eq:LSexample}
\end{align}
In descending $J_z$ order, the second operator has diagonal $(3,-5,5,-3)/6$. Comparing it with the octupole diagonal $(3,-9,9,-3)/10$ fixes both coefficients in Eq.~\eqref{eq:LSexample}. Constituent spin alone projects onto vector polarization, and weighting it by an orbital quadrupole also produces an octupole component. In a partonic calculation, the momentum-dependent matrix element supplies this weight. The example identifies an allowed angular-momentum mechanism, although a quantitative $^{7}$Li prediction requires its nuclear wave function and operator dynamics.

The target TMDs are matrix elements in the entire nucleus, including binding, spin recoupling, orbital motion, and gauge-link rescattering. A light-front impulse approximation provides a starting point for calculating them. Let $y_N=p_N^+/P_A^+$ be the active nucleon's fraction of nuclear plus momentum and $\bm p_{NT}$ its transverse momentum. The parton's fraction and transverse momentum relative to that nucleon are
\begin{equation}
 x_N=x/y_N,\qquad
 \bm k_{T,\mathrm{rel}}=\bk-(x/y_N)\bm p_{NT}.
 \label{eq:nuclearshift}
\end{equation}
Here, the parton fraction uses the active nucleon's momentum, which varies within the nucleus. Write the nucleon spin spectral density as $\mathcal S^{N,\rho}_{ss'}(y_N,\bm p_{NT})$, including removal-energy integration and the specified light-front spin rotations. The correlator is organized as
\begin{equation}
 \begin{split}
 \Phi_A[\rho](x,\bk)={}&\sum_{N=p,n}\sum_{ss'}
 \int_x^1\frac{\dd y_N}{y_N}\int\dd^2\bm p_{NT}\\
 &\times\mathcal S^{N,\rho}_{ss'}(y_N,\bm p_{NT})
 \Phi_N^{s's}(x_N,\bm k_{T,\mathrm{rel}})\\
 &+\delta\Phi_A,
 \end{split}
 \label{eq:nuclearconv}
\end{equation}
where the indices $s,s'$ refer to nucleon spin, and $\rho$ specifies the nuclear preparation. The correction $\delta\Phi_A$ accounts for rescattering, off-shell, and nonnucleonic contributions. We normalize the spectral function to the proton and neutron numbers:
\begin{equation}
 \int\dd y_N\,\dd^2\bm p_{NT}\Tr_s\mathcal S^{N,\rho}
 =\begin{cases}3,&N=p,\\4,&N=n.\end{cases}
\end{equation}
The spectral measure, wave function, spin basis, and binding treatment must be specified consistently with this normalization. Nuclear TMD studies include spin--orbit correlations and rescattering~\cite{Kovchegov2016}, but the validity of a large-nucleus approximation for \(A=7\) requires separate assessment.

Because $x$ is the parton's fraction of the total nuclear plus momentum, the pure-QCD PDFs satisfy
\begin{equation}
 \int_0^1\dd x\,x\left[
 \sum_q(f_1^{q/A}+f_1^{\bar q/A})+f_1^{g/A}\right]=1.
 \label{eq:nuclearmomentum}
\end{equation}
The total momentum operator is independent of the target's spin polarization, so the corresponding $x$-weighted sum of $f_{1LL}$ vanishes, even though individual quark and gluon tensor distributions may be nonzero. If electromagnetic partons are included, their distributions must also be included in both momentum sums. Together with the local-moment and positivity conditions, these sum rules constrain nuclear calculations but do not fix the detailed $x$- and transverse-momentum dependence, which is governed by the underlying nuclear and partonic dynamics.

\section{Conclusions and outlook}
\label{sec:conclusions}

Octupole polarization carries spin information beyond the vector and quadrupole moments of a spin-$3/2$ target. Our principal result is the complete leading-twist gluon transverse-momentum-dependent (TMD) basis for an arbitrary physical target density matrix. For each specified Wilson-line and color structure, the basis has 32 independent coefficients, fourteen in the octupole sector, of which three describe gluon number, four describe helicity, and seven describe linear polarization. Each quark flavor has the same coefficient count, consistent with established higher-spin classifications.

In heavy-quark-pair leptoproduction at small, nonzero pair imbalance, physical target states with opposite octupoles and identical lower moments isolate the rank-three contribution. Beam-helicity reversal, together with measurements at varied target orientations and pair-imbalance azimuths, separates all fourteen gluon octupole responses at leading-power Born order, provided the required target components and hard analyzing powers are nonzero. The fully transverse sector contains distinct first and fifth recoil harmonics for linear polarization, both at leading twist. Normalized moments give TMD ratios, while absolute normalization also requires an independently normalized unpolarized response.

The complete quark SIDIS catalog and DY octupole convolutions give the corresponding incoming-target responses. Time reversal implies that the three octupole quark number distributions reverse sign between the standard SIDIS and DY definitions, while the eleven helicity and transverse-spin distributions are unchanged. These relations remain valid under symmetry-preserving renormalization and evolution in a common scheme at matched scales. For gluons, complete reversal of a specified Wilson-line pair and color trace makes ten octupole coefficients odd and four even, while unrelated color-flow classes remain independent. These operator relations don't establish factorization or directly determine the signs of measured asymmetries. Testing them therefore requires each process’s hard scattering and fragmentation or second-hadron distributions.

The collinear limits recover the established spin-$3/2$ distribution sets. Local-moment constraints and momentum sum rules provide checks on nuclear calculations, as do positivity bounds when a positive spectral representation applies. Polarized $^{7}$Li offers an important nuclear application, as correlations between constituent spin and orbital motion can generate octupole structure even when the active constituent has spin $1/2$. The angular-momentum example identifies an allowed mechanism, while the distributions themselves require nuclear wave functions and operator dynamics.

The next steps are to calculate these nuclear distributions, develop and calibrate octupole target preparations, and assess experimental sensitivity with realistic angular coverage. Extending the heavy-pair analysis beyond Born order requires radiation and soft recoil to be included in a factorization formula for the chosen measurement. Linearity in the target density matrix preserves spin-rank projection, while the recoil response becomes a process-dependent convolution where factorization applies. These steps would allow the octupole correlations classified here to be tested quantitatively.

\section*{Data availability}
No experimental measurements were used in this study. The computational
code, exact certificates, numerical benchmark and figure data, and
reproduction instructions supporting the results are available in the
versioned Li7-TMDs repository~\cite{Li7TMDs}. The independent comparisons
and their analytical scope are summarized in
Appendix~\ref{app:validation}. All defining tensor, coefficient-projection,
and Born response formulas are given in this article.

\appendix

\section{Coefficient projections and tensor conventions}
\label{sec:inverse}

\subsection{Inverse projections}

Coefficient extraction proceeds in three steps: apply the partonic projections in Eqs.~\eqref{eq:qproject} and \eqref{eq:gproject}, isolate the target response with $\tau_{KM}$, then resolve its orbital tensors. The last step holds $x$, $|\bk|$, scales, and link class fixed.

For an isolated scalar response with $m>0$, let $\mathcal X_{Km}(\phi_k)$ be its value at $A_{Km,+}=1$. Writing $t=|\bk|/M_A$, the response is $t^m\mathscr Y_m(\phi_k)c_{Km}$, with $\mathscr Y_m$ equal to $\cos(m\phi_k)$ or $\sin(m\phi_k)$ as specified by Eq.~\eqref{eq:parityscalar}. Harmonic projection gives
\begin{equation}
 c_{Km}=\frac1{\pi t^m}\int_0^{2\pi}\dd\phi_k\,
 \mathscr Y_m(\phi_k)\mathcal X_{Km}(\phi_k)
 \label{eq:scalarinverse}
\end{equation}
for $m>0$. For $m=0$, take the azimuthal average of the unit-component response. Here, a unit component means a derivative with respect to a target moment, or an equivalent linear combination of physical preparations, and it need not correspond to a physical state with all other moments zero.

For quarks, undo the even-$K$ dual in the isolated transverse-spin projection and call the resulting complex component $V_+$. The contractions in Eqs.~\eqref{eq:T1}--\eqref{eq:T3} give
\begin{equation}
 \begin{split}
 V_+={}&A_{Km,+}(\kappa^*)^{m-1}h^{q[m-1]}_{Km}+\tfrac12 A_{Km,+}^*\kappa^{m+1}h^{q[m+1]}_{Km}
 \end{split}
 \label{eq:Vcomplex}
\end{equation}
for $m>0$. For $m=0$, $V_+=A_{K0}\kappa h^{q[1]}_{K0}$. The physical projection is $\calT_+=V_+$ for odd $K$ and $\calT_+=-\ii V_+$ for even $K$. The high-rank factor $1/2$ follows from $(\mathsf k_n)_+=2^{1-n}\kappa^n$ and the multiplicity of contracted STF indices. With $\bk$ along $x$, denote the responses at $A_{Km,+}=1$ and $A_{Km,+}=\ii$ by $v_1$ and $v_\ii$. Writing $h_-=h^{q[m-1]}_{Km}$ and $h_+=h^{q[m+1]}_{Km}$ gives
\begin{align}
 v_1&=t^{m-1}h_-+\tfrac12t^{m+1}h_+,\nonumber\\
 v_\ii/\ii&=t^{m-1}h_--\tfrac12t^{m+1}h_+.
\end{align}
Adding and subtracting the two responses gives
\begin{equation}
 h_- =\frac{v_1+v_\ii/\ii}{2t^{m-1}},\qquad
 h_+ =\frac{v_1-v_\ii/\ii}{t^{m+1}}.
 \label{eq:qinverse}
\end{equation}

For gluons, undo the odd-$K$ dual and denote the rank-two tensor's complex component by $\mathcal B_+$. The four transverse-rank cases are
\begin{align}
 \mathcal B_+(m=0)&=\tfrac12 A_{K0}\kappa^2h^{g[2]}_{K0},\nonumber\\
 \mathcal B_+(m=1)&=\tfrac12 A_{K1,+}\kappa h^{g[1]}_{K1}
 +\tfrac14 A_{K1,+}^*\kappa^3h^{g[3]}_{K1},\nonumber\\
 \mathcal B_+(m=2)&=A_{K2,+}h^{g[0]}_{K2}
 +\tfrac14 A_{K2,+}^*\kappa^4h^{g[4]}_{K2},\nonumber\\
 \mathcal B_+(m=3)&=A_{K3,+}\kappa^*h^{g[1]}_{K3}
 +\tfrac14 A_{K3,+}^*\kappa^5h^{g[5]}_{K3}.
 \label{eq:Bcomplex}
\end{align}
The physical tensor is $\calH_+=\mathcal B_+$ for even $K$ and $\calH_+=-\ii\mathcal B_+$ for odd $K$. For $m=1,2,3$, use the low-branch constants of Sec.~\ref{sec:jointobservables}, $c_m\equiv c_{m,|m-2|}$, with $(c_1,c_2,c_3)=(1/2,1,1)$, and set $n_-=|m-2|$. The same two unit-component responses $\beta_1,\beta_\ii$ at $\kappa=t>0$ then give
\begin{equation}
 h_- =\frac{\beta_1+\beta_\ii/\ii}{2c_mt^{n_-}},\qquad
 h_+ =\frac{2(\beta_1-\beta_\ii/\ii)}{t^{m+2}}.
 \label{eq:ginverse}
\end{equation}
Here, $h_-=h^{g[n_-]}_{Km}$ and $h_+=h^{g[m+2]}_{Km}$. Equations~\eqref{eq:scalarinverse}, \eqref{eq:qinverse}, and \eqref{eq:ginverse} project every coefficient at generic nonzero transverse momentum. If its prefactor vanishes at $k_T=0$, use the coefficient's radial continuation or limiting behavior instead of dividing at that point.

\begin{table*}[!t]
\caption{Notation for the complete quark and gluon coefficient sets. The scalar names apply to either species, with a superscript $q$ or $g$ understood. A pair $[n]:h$ assigns a name to the coefficient of orbital rank $n$ in the corresponding helicity-flip projection. All tensor normalizations and signs are fixed by the definitions in Sec.~\ref{sec:count}.}
\label{tab:names}
\centering
\begin{ruledtabular}
\begin{tabular}{cccll}
 $(K,m)$ & $f^a_{Km}$ & $g^a_{Km}$ & Quark $h^{q[n]}_{Km}$ & Gluon $h^{g[n]}_{Km}$\\
 $(0,0)$&$f_1$&---&$[1]:h_1^\perp$&$[2]:h_1^{\perp g}$\\
 $(1,0)$&---&$g_{1L}$&$[1]:h_{1L}^\perp$&$[2]:h_{1L}^{\perp g}$\\
 $(1,1)$&$f_{1T}^\perp$&$g_{1T}$&$[0]:h_1,\ [2]:h_{1T}^\perp$&$[1]:h_1^g,\ [3]:h_{1T}^{\perp g}$\\
 $(2,0)$&$f_{1LL}$&---&$[1]:h_{1LL}^\perp$&$[2]:h_{1LL}^{\perp g}$\\
 $(2,1)$&$f_{1LT}$&$g_{1LT}$&$[0]:h_{1LT},\ [2]:h_{1LT}^\perp$&$[1]:h_{1LT}^g,\ [3]:h_{1LT}^{\perp g}$\\
 $(2,2)$&$f_{1TT}$&$g_{1TT}$&$[1]:h_{1TT}^\perp,\ [3]:h_{1TT}^{\perp\perp}$&$[0]:h_{1TT}^g,\ [4]:h_{1TT}^{\perp\perp g}$\\
 $(3,0)$&---&$g_{1LLL}$&$[1]:h_{1LLL}^\perp$&$[2]:h_{1LLL}^{\perp g}$\\
 $(3,1)$&$f_{1LLT}^\perp$&$g_{1LLT}^\perp$&$[0]:h_{1LLT},\ [2]:h_{1LLT}^\perp$&$[1]:h_{1LLT}^g,\ [3]:h_{1LLT}^{\perp g}$\\
 $(3,2)$&$f_{1LTT}^\perp$&$g_{1LTT}^\perp$&$[1]:h_{1LTT}^\perp,\ [3]:h_{1LTT}^{\perp\perp}$&$[0]:h_{1LTT}^g,\ [4]:h_{1LTT}^{\perp\perp g}$\\
 $(3,3)$&$f_{1TTT}^\perp$&$g_{1TTT}^\perp$&$[2]:h_{1TTT}^\perp,\ [4]:h_{1TTT}^{\perp\perp}$&$[1]:h_{1TTT}^g,\ [5]:h_{1TTT}^{\perp\perp g}$
\end{tabular}
\end{ruledtabular}
\end{table*}

\subsection{Higher-rank tensors and distribution labels}

The dimensionless rank-four and rank-five momentum tensors can be written without a complex basis as
\begin{align}
 \mathsf k_4^{i_1\cdots i_4}
 ={}&\frac1{\M^4}\biggl[k^{i_1}\cdots k^{i_4}
 -\frac{\bk^2}{6}\sum_6\delta^{i_1i_2}k^{i_3}k^{i_4}
 \nonumber\\
 &\hspace{12mm}+\frac{(\bk^2)^2}{24}
 \sum_3\delta^{i_1i_2}\delta^{i_3i_4}\biggr],\nonumber\\
 \mathsf k_5^{i_1\cdots i_5}
 ={}&\frac1{\M^5}\biggl[k^{i_1}\cdots k^{i_5}
 -\frac{\bk^2}{8}\sum_{10}\delta^{i_1i_2}k^{i_3}k^{i_4}k^{i_5}
 \nonumber\\
 &\hspace{12mm}+\frac{(\bk^2)^2}{48}
 \sum_{15}\delta^{i_1i_2}\delta^{i_3i_4}k^{i_5}\biggr].
 \label{eq:K45}
\end{align}
The sums include each distinct index placement once, without averaging. Every tensor trace vanishes. At $\bk=(k,0)$, the all-$x$ components are $k^4/(8\M^4)$ and $k^5/(16\M^5)$, respectively, as in Eq.~\eqref{eq:Kcomplex}.

Table~\ref{tab:names} maps conventional polarization labels to this article's explicit tensor coefficients. Square brackets indicate orbital rank, not a transverse-momentum moment. Signs and normalizations should be compared using the full coefficient--tensor products, since rescaling $A_{Km}$ inversely rescales its coefficient. The scalar gluon $TT$ label denotes the minimal trace combination defined in Eq.~\eqref{eq:litcombination}; accordingly, the physical STF basis has no independent $h^{g[2]}_{22}$, $h^{g[2]}_{32}$, or $h^{g[3]}_{33}$.

\section{Relation to established parametrizations}
\label{app:literature}

\subsection{Conversion of the lower-spin gluon tensors}

This dictionary converts gluon conventions for the same matrix element. As in Sec.~\ref{sec:redundancy}, upright $\mathrm{B}$ labels the displayed $x/2$ parametrization of Refs.~\cite{Boer2016,Cotogno2017}. Initially match the overall correlator normalization, set the named components $S_L,S_T,S_{LL},S_{LT},S_{TT}$ equal in both bases, and use $M_A$ in both momentum tensors. This compares coefficient bases, without equating distributions of different target species.

\begin{table*}[!t]
\caption{Correspondence of octupole quark tensor channels with the spin-$3/2$ fragmentation classification of Ref.~\cite{Zhao2022}. The bracket $[n]$ is the orbital STF rank. Each entry denotes the FF occupying the same channel after the axes, metric, and projection normalizations are converted.}
\label{tab:FF_correspondence}
\centering
\begin{ruledtabular}
\begin{tabular}{clll}
 $m$ & Number channel & Helicity channel & Quark helicity-flip channels\\
 0 & --- & $g^q_{30}:G_{1LLL}$
 & $h^{q[1]}_{30}:H_{1LLL}^{\perp}$\\
 1 & $f^q_{31}:D_{1LLT}^{\perp}$ & $g^q_{31}:G_{1LLT}^{\perp}$
 & $h^{q[0]}_{31}:H_{1LLT},\quad h^{q[2]}_{31}:H_{1LLT}^{\perp}$\\
 2 & $f^q_{32}:D_{1LTT}^{\perp}$ & $g^q_{32}:G_{1LTT}^{\perp}$
 & $h^{q[1]}_{32}:H_{1LTT}^{\perp},\quad h^{q[3]}_{32}:H_{1LTT}^{\perp\perp}$\\
 3 & $f^q_{33}:D_{1TTT}^{\perp}$ & $g^q_{33}:G_{1TTT}^{\perp}$
 & $h^{q[2]}_{33}:H_{1TTT}^{\perp},\quad h^{q[4]}_{33}:H_{1TTT}^{\perp\perp}$
\end{tabular}
\end{ruledtabular}
\end{table*}

Lower contracted Lorentz transverse indices with $g_{Tij}=-\delta_{ij}$, retaining Cartesian values for the free components. Here, $k_T^{ij}$ is the STF tensor of Eq.~\eqref{eq:referenceSTF}, and $k_T^{i_1\cdots i_n}$ likewise denotes an STF momentum tensor. With $\epsilon_T^{xy}=+1$, this gives
\begin{align}
 S_{LT}^{\{i}k_T^{j\}}/M_A
 &=2\STF(S_{LT}^i\mathsf k_1^j)+\delta^{ij}\,\cC_1[S_{LT}],\nonumber\\
 {S_{TT}^{\{i}}_a k_T^{j\}a}/M_A^2
 &=-\delta^{ij}\,\cC_2[S_{TT}].
 \label{eq:literal_gluon_traces}
\end{align}
The braces symmetrize without averaging or subtracting the free-index trace. The STF momentum tensor must also be distinguished from a dyadic:
\begin{equation}
 \STF\!\left(S_{TT}^{\{i}{}_{\alpha}k_T^{j\}}k_T^\alpha\right)
 =-\bk^2 S_{TT}^{ij},
 \label{eq:dyadic_not_STF}
\end{equation}
which is generally nonzero. The reduction therefore uses the reference's explicit STF definition. In Eq.~\eqref{eq:literal_gluon_traces}, the first identity shifts a trace coefficient and retains an STF $LT$ term, while the second explains the redundant $TT$ label. The circular $TT$ contraction is
\begin{equation}
 {\epsilon_T^b}_c k_T^{ca}S_{TT,ab}/M_A^2
 =-\cS_2[S_{TT}].
 \label{eq:literal_gTT}
\end{equation}
The eighteen independent lower-rank coefficients convert as follows:
\begin{align}
 f^g_{00}&=f_1^{g,\mathrm{B}},& f^g_{11}&=f_{1T}^{\perp g,\mathrm{B}},\nonumber\\
 f^g_{20}&=f_{1LL}^{g,\mathrm{B}},&
 f^g_{21}&=f_{1LT}^{g,\mathrm{B}}+h_{1LT}^{g,\mathrm{B}},\nonumber\\
 f^g_{22}&=f_{1TT}^{g,\mathrm{B}}-h_{1TT}^{\perp g,\mathrm{B}},&&\nonumber\\
 g^g_{10}&=g_1^{g,\mathrm{B}},&g^g_{11}&=g_{1T}^{g,\mathrm{B}},\nonumber\\
 g^g_{21}&=g_{1LT}^{g,\mathrm{B}},&g^g_{22}&=-g_{1TT}^{g,\mathrm{B}}
 \label{eq:gluon_scalar_dictionary}
\end{align}
for the trace and circular projections, and
\begin{align}
 h^{g[2]}_{00}&=h_1^{\perp g,\mathrm{B}},&
 h^{g[2]}_{10}&=-h_{1L}^{\perp g,\mathrm{B}},\nonumber\\
 h^{g[1]}_{11}&=-h_1^{g,\mathrm{B}},&
 h^{g[3]}_{11}&=-h_{1T}^{\perp g,\mathrm{B}},\nonumber\\
 h^{g[2]}_{20}&=h_{1LL}^{\perp g,\mathrm{B}},&
 h^{g[1]}_{21}&=2h_{1LT}^{g,\mathrm{B}},\nonumber\\
 h^{g[3]}_{21}&=-h_{1LT}^{\perp g,\mathrm{B}},&
 h^{g[0]}_{22}&=h_{1TT}^{g,\mathrm{B}},\nonumber\\
 h^{g[4]}_{22}&=h_{1TT}^{\perp\perp g,\mathrm{B}}&&
 \label{eq:gluon_linear_dictionary}
\end{align}
for linear polarization. Lorentz index lowering and the clockwise transverse dual fix these signs, but the link-reversal classification is unchanged.

Convert operator and target normalizations separately from this dictionary. If operator definitions differ by a factor $C_\Gamma$, multiply every reference coefficient by $C_\Gamma$. For example, changing the $2/P^+$ fundamental-trace prefactor in Eq.~\eqref{eq:Gammaunsub} to $1/P^+$ halves the operator for the same fields and generator normalization. For target polarization, Refs.~\cite{Boer2016,Cotogno2017} use $T_{zz}^{\mathrm{B}}=2S_{LL}^{\mathrm{B}}/3$ and $T_{zi}^{\mathrm{B}}=S_{LT}^{i,\mathrm{B}}/2$, while Eq.~\eqref{eq:componentdefinitions} uses $T_{zz}=S_{LL}$. Matching Cartesian tensors instead of named components thus requires $S_{LL}^{\mathrm{B}}=3S_{LL}/2$: multiply the two $LL$ coefficients by $3/2$, with no analogous $LT$ or $TT$ rescaling. Specify what is matched so that polarization normalization remains separate from the TMD.

These combinations also appear in the published diagonal parton-helicity blocks~\cite{Cotogno2017}. The common shift $f_{1TT}^{g,\mathrm{B}}\mapsto f_{1TT}^{g,\mathrm{B}}+\delta$, $h_{1TT}^{\perp g,\mathrm{B}}\mapsto h_{1TT}^{\perp g,\mathrm{B}}+\delta$ leaves the correlator invariant. The dictionary removes this redundant coordinate before adding the rank-three sector.

\subsection{Quark fragmentation and incoming-target distributions}

Quark octupole distributions and fragmentation share the same spin representation and channel count. Table~\ref{tab:FF_correspondence} matches the tensor channels of Ref.~\cite{Zhao2022} by transverse target rank and orbital rank.

Convert the projected correlator to determine signs and factors. The fragmentation convention uses $\Delta^{[\Gamma]}=\Tr_D(\Delta\Gamma)$, while Eq.~\eqref{eq:qproject} uses a half trace. For the proper rotation $\mathscr R$ taking the reference hadron direction into $+z$, with Dirac representation $S(\mathscr R)$, define
\begin{equation}
 \Delta_{\parallel}=2S(\mathscr R)\Delta S(\mathscr R)^{-1}.
 \label{eq:FF_template}
\end{equation}
The half-trace projectors of Eq.~\eqref{eq:qproject} now preserve the reference normalization. Rotate target moments and transverse momenta together. For the rank-$K$ covariant tensor $\mathsf A_{\mathrm{Z}}$ of Zhao \emph{et al.}~\cite{Zhao2022}, with $\mathsf A=\mathsf S,\mathsf T,\mathsf R$ for $K=1,2,3$, the target-tetrad components are
\begin{equation}
 \begin{split}
 A^{\parallel}_{a_1\cdots a_K}
 ={}&(-1)^K e_{a_1\mu_1}\cdots e_{a_K\mu_K}\,\\
 &\times \mathscr R^{\mu_1}{}_{\nu_1}\cdots\mathscr R^{\mu_K}{}_{\nu_K}\,
 \mathsf A_{\mathrm{Z}}^{\nu_1\cdots\nu_K},
 \end{split}
 \label{eq:FF_rotation}
\end{equation}
where the spacelike tetrad $e_a$ is defined in Sec.~\ref{sec:kin} and $e_a\cdot e_b=-\delta_{ab}$ gives the factors $(-1)^K$. Use Eq.~\eqref{eq:componentdefinitions} for the longitudinal and transverse components, then Appendix~\ref{sec:inverse} to invert the coefficients. We note that rotating momentum and spin together preserves helicity, and changing only a sine sign with fixed target components does not implement that rotation. Index lowering provides a simple check, namely, $-k_T^{ij}S_j/M_A^2$ becomes $+\mathsf k_2^{ij}S^j$, whereas $S^{ij}k_{Tj}/M_A$ becomes $-S^{ij}\mathsf k_1^j$ in Euclidean components. Equations~\eqref{eq:FF_template} and \eqref{eq:FF_rotation} convert tensor conventions without analytically continuing an out-state matrix element into a target PDF.

The observable correspondence exchanges the distribution and
fragmentation roles. The final-state $TTT$ SIDIS response of
Ref.~\cite{ZhaoSIDIS2024} pairs a rank-one Boer--Mulders distribution
with rank-four polarized fragmentation. Equation~\eqref{eq:fifthharmonic}
pairs rank-four incoming-target transversity with rank-one Collins
fragmentation into an unpolarized hadron. Both carry one unit of quark
helicity flip. The contribution here is the distribution-side target
response and its extraction conventions, not the first octupole
SIDIS harmonic.

\section{Spin-matrix positivity}
\label{sec:positivity}

\subsection{Target and parton matrices}

The exact target condition $\rho\succeq0$ constrains its multipole moments jointly. Partonic positivity additionally requires a positive spectral representation, as in suitable unsubtracted correlators, cutoff models, or positivity-preserving input prescriptions. UV and rapidity subtractions, including fixed-order $\overline{\mathrm{MS}}$ distributions, need not preserve pointwise positivity. The following partonic bounds assume a positive Gram representation for the chosen link class and scheme~\cite{Cotogno2017,Fu2026,Handbook}.

For a fixed physical target state, the quark parton-spin matrix is
\begin{equation}
 D_q(\rho)=\frac12\begin{pmatrix}
 \calF^q+\calG^q&\calT^{q,x}+\ii\calT^{q,y}\\
 \calT^{q,x}-\ii\calT^{q,y}&\calF^q-\calG^q
 \end{pmatrix}.
 \label{eq:qD}
\end{equation}
Its eigenvalues, and those of the gluon matrix in Eq.~\eqref{eq:gD}, give
\begin{align}
 \calF^q&\geq\sqrt{(\calG^q)^2+(\calT^{q,x})^2+(\calT^{q,y})^2},
 \nonumber\\
 \calF^g&\geq\sqrt{(\calG^g)^2+(\calH^{g,xx})^2+(\calH^{g,xy})^2}.
 \label{eq:fixedrhobounds}
\end{align}
These bounds constrain the full projections in the selected target state, not individual multipole coefficients.

The joint parton--target matrix gives a stronger condition. Construct it by replacing moments with their spin operators: $S_i\to J_i$, $S_{LL}\to Q_{zz}$, $S_{LT}^i\to2Q_{zi}$, $S_{TT}^{ij}\to2\STF Q_{ij}$, and octupoles with the components of $O_{abc}$ normalized by Eq.~\eqref{eq:componentdefinitions}. Call the resulting operators $\widehat F$, $\widehat G$, $\widehat T_i$, and $\widehat H_{ij}$. Since $\calF=\Tr(\rho\widehat F)$, and similarly for the other projections, no factors from the inverse expansion of $\rho$ enter. The two $8\times8$ matrices are
\begin{align}
 \mathbb M_q&=\tfrac12\begin{pmatrix}
 \widehat F+\widehat G&\widehat T_+\\
 \widehat T_-&\widehat F-\widehat G
 \end{pmatrix},\nonumber\\
 \mathbb M_g&=\tfrac12\begin{pmatrix}
 \widehat F+\widehat H_{xx}&\widehat H_{xy}+\ii\widehat G\\
 \widehat H_{xy}-\ii\widehat G&\widehat F-\widehat H_{xx}
 \end{pmatrix},
 \label{eq:jointM}
\end{align}
where $\widehat T_\pm=\widehat T_x\pm\ii\widehat T_y$. Use the combined index $\alpha=(i,\Lambda)$, ordered as parton polarization then target spin. A positive spectral prescription gives
\begin{equation}
 \mathbb M_{\alpha\beta}
 =\sum_X\int\dd\mu_X\,
 \mathcal A_{X;\alpha}^*\mathcal A_{X;\beta},
 \qquad \dd\mu_X\geq0,
 \label{eq:Gram}
\end{equation}
and therefore $\mathbb M\succeq0$. Taking target expectations of its blocks gives the fixed-target parton matrix:
\begin{equation}
 \begin{split}
 D_{ij}(\rho)&=\sum_{\Lambda\Lambda'}\rho_{\Lambda'\Lambda}
                     \mathbb M_{i\Lambda,j\Lambda'}\\
             &=\Tr[\rho\,\mathbb M_{ij}].
 \end{split}
 \label{eq:joint_target_expectation}
\end{equation}
For $\rho=|c\rangle\langle c|$, the target contraction $c^\dagger\mathbb M_{ij}c$ retains populations and complex coherences without a partial transpose. Joint positivity makes the resulting parton matrix positive semidefinite for every target state. At finite transverse momentum, all eigenvalues, or equivalently all principal minors, must be nonnegative.

\subsection{Collinear bounds}

For either species in the collinear limit, write $f=f_1$, $f_Q=f_{1LL}$, $g=g_1$, and $g_O=g_{1LLL}$. The diagonal combinations of Eq.~\eqref{eq:populationresponses} are
\begin{align}
 U_{3/2}&=f+f_Q,& U_{1/2}&=f-f_Q,\nonumber\\
 G_{3/2}&=\tfrac32g+\tfrac3{10}g_O,&
 G_{1/2}&=\tfrac12g-\tfrac9{10}g_O.
 \label{eq:UG}
\end{align}
The corresponding positivity conditions are
\begin{equation}
 U_{3/2}\geq|G_{3/2}|,\qquad U_{1/2}\geq|G_{1/2}|.
 \label{eq:diagbound}
\end{equation}
For quarks, set $h=h_1^q$ and $h_O=h_{1LLT}^q$. The off-diagonal parton block contains the raising operator $hJ_++h_O(O_{zzx}+\ii O_{zzy})$, with
\begin{equation}
 O_{zzx}+\ii O_{zzy}=
 \begin{pmatrix}
 0&2\sqrt3/5&0&0\\0&0&-6/5&0\\
 0&0&0&2\sqrt3/5\\0&0&0&0
 \end{pmatrix}.
 \label{eq:Oraise}
\end{equation}
The outer and middle target-helicity amplitudes are $\sqrt3(h+2h_O/5)$ and $2h-6h_O/5$, respectively. Their minors give
\begin{align}
 3(h+\tfrac25h_O)^2&\leq(U_{3/2}+G_{3/2})(U_{1/2}-G_{1/2}),\nonumber\\
 |2h-\tfrac65h_O|&\leq U_{1/2}+G_{1/2}.
 \label{eq:qSoffer}
\end{align}
In the gluon circular basis, $(1,-\ii)/\sqrt2,(1,\ii)/\sqrt2$, Eq.~\eqref{eq:jointM} has helicity-flip block $h_{1TT}^gJ_+^2$, where $S_{TT}^{xx}+\ii S_{TT}^{xy}$ maps to $Q_{xx}-Q_{yy}+2\ii Q_{xy}=J_+^2$. The two nonzero entries, $2\sqrt3h_{1TT}^g$, give
\begin{equation}
 12(h_{1TT}^g)^2\leq(U_{3/2}+G_{3/2})(U_{1/2}+G_{1/2}).
 \label{eq:gSoffer}
\end{equation}
Within this spectral prescription, permuting rows and columns reduces the real collinear matrices to these $2\times2$ blocks and isolated diagonal entries. Hence, Eq.~\eqref{eq:diagbound}, together with Eq.~\eqref{eq:qSoffer} for quarks or Eq.~\eqref{eq:gSoffer} for gluons, is necessary and sufficient for positivity. The spin-normalization factors agree with the independent spin-$3/2$ PDF analysis~\cite{Fu2026}.

\section{Longitudinal polarization calibration}
\label{app:NMR}

The vector-polarization fraction commonly reported for a spin-$3/2$ target is
\begin{equation}
 P_z=\frac{\langle J_z\rangle}{J}=\frac23S_L.
 \label{eq:polpercent}
\end{equation}
Specifying the four normalized populations requires three longitudinal moments. The spin-temperature model,
\begin{equation}
 p_\Lambda=\frac{\exp(\beta\Lambda)}{Z(\beta)},\qquad
 Z(\beta)=2\cosh(3\beta/2)+2\cosh(\beta/2),
 \label{eq:spintemp}
\end{equation}
uses spin projections $\Lambda=\pm3/2,\pm1/2$ and relates all three moments through the spin-temperature parameter $\beta$. Selective radio-frequency manipulation can leave this one-parameter family. Then, independent quadrupole and octupole calibration requires additional population information.

For weak probing of resolved transitions in the high-field secular limit, the squared $J_+$ matrix elements give magnetic-dipole strengths $3,4,3$. Define the signed population differences
\begin{align}
 d_1&=p_{3/2}-p_{1/2},\nonumber\\
 d_2&=p_{1/2}-p_{-1/2},\nonumber\\
 d_3&=p_{-1/2}-p_{-3/2}.
 \label{eq:NMRd}
\end{align}
After correction for frequency-dependent response, their ideal line areas are
\begin{equation}
 I_1=3Cd_1,\qquad I_2=4Cd_2,\qquad I_3=3Cd_3,
 \label{eq:NMRareas}
\end{equation}
for a common calibration constant $C$ and fixed absorption-sign convention. Normalization gives $p_{-3/2}=(1-d_1-2d_2-3d_3)/4$, then the other populations follow by successively addding $d_3$, $d_2$, and $d_1$. Equation~\eqref{eq:popmoments} then yields
\begin{align}
 S_L&=\tfrac12(3d_1+4d_2+3d_3),\nonumber\\
 S_{LL}&=d_1-d_3,\nonumber\\
 S_{LLL}&=\tfrac3{10}(d_1-2d_2+d_3).
 \label{eq:NMRmultipoles}
\end{align}
The octupole uses a different line combination from the total absorption measuring the vector moment. Nonnegative reconstructed populations provide an independent consistency check. In solids, overlapping quadrupolar transitions, powder averaging, unequal excitation, circuit response, and state mixing must enter the calibrated forward NMR response before inversion. Populations measured along one axis determine $S_L,S_{LL},S_{LLL}$ there, while transverse components require controlled rotations or coherence tomography. Rotations mix the $2K+1$ components within $K$ (Sec.~\ref{sec:octobservables}) and cannot create octupole polarization from lower-rank moments.

\section{Independent validation and Born kinematics}
\label{app:validation}

The computational companion~\cite{Li7TMDs} provides implementations, exact certificates, numerical data, and reproduction instructions, with conventions, inputs, shared dependencies, and certificate provenance recorded. Exact spin and tensor checks use SymPy~\cite{SymPy2017}, and Born calculations use complex NumPy arrays~\cite{NumPy2020} and separate arbitrary-precision evaluations. The checks establish algebraic and numerical consistency under the main text's factorization, renormalization, and nuclear assumptions. Software checks and hashes document reproducibility, not independent physics results or scientific correctness.

\subsection{Spin, tensor, and coefficient-space constructions}

Ladder-algebra operators are compared with the independent Clebsch--Gordan construction of Eq.~\eqref{eq:spherical}. Trace conditions fix $5/4$ and $41/60$, and the trace metric fixes the inverse density weights. Reconstruction checks all sixteen Hermitian coordinates, including imaginary coherences, while rotated spectra and moments check positivity and rank selectivity. One full-rank set of unit preparation directions is
\begin{equation}
 \begin{gathered}
 (0,0,1),\quad(1,0,0),\quad(0,1,0),\quad\tfrac13(2,2,-1),\\
 \tfrac13(1,2,-2),\quad\tfrac13(2,1,-2),\quad
 \tfrac1{11}(2,6,-9).
 \end{gathered}
 \label{eq:sevendirections}
\end{equation}
In the companion's real orthonormal octupole basis, the response determinant is $-1400/35937$. This realizes Eq.~\eqref{eq:rotationinverse} with discrete settings, without optimizing for a particular acceptance or covariance.

Cartesian trace removal and complex helicity harmonics independently construct tensors through rank five for symbolic positive $M_A$, checking components, traces, duals, and mass rescalings. Projection of the reference gluon correlator checks the eighteen-coordinate dictionary and redundant trace direction, including the distinction between the STF contraction in Eq.~\eqref{eq:referenceSTF} and the dyadic's nonzero free-index STF part in Eq.~\eqref{eq:dyadic_not_STF}.

Each real $64\times32$ coefficient-to-matrix map has a nonzero maximal minor evaluated exactly at $(k_x,k_y,M_A)=(2,1,3)$. Its helicity entries are independently reconstructed, with normalization and row ordering recorded. The resulting lower bound complements the upper bound and spanning construction of Sec.~\ref{sec:count}. Covariance and homogeneous momentum powers relate maps at nonzero radii, while the origin is treated separately. Analytical projectors and independent inversion check every coefficient's signs and normalization, which rank alone cannot fix.

Equation~\eqref{eq:joint_target_expectation} fixes the spectral index ordering of Eq.~\eqref{eq:jointM}. An auxiliary realization used for coefficient-rank checks is related to it by
\begin{equation}
 \begin{split}
 \mathbb M_{\mathrm{source}}&=\tfrac12\mathsf T_p
                               (\mathbb M_{\mathrm{aux}}),\\
 (\mathsf T_p\mathbb M)_{i\Lambda,j\Lambda'}
     &=\mathbb M_{j\Lambda,i\Lambda'}.
 \end{split}
 \label{eq:auxiliary_coordinate_map}
\end{equation}
This invertible real-linear coordinate map preserves coefficient-map
rank, but it is not a unitary spin-basis change and need not preserve
an individual matrix's spectrum, rank, or positivity~\cite{Peres1996}.
Spectral examples, target expectations with complex coherences, and
positivity bounds therefore use the source ordering of Eq.~\eqref{eq:jointM}. Partial-transpose
counterexamples check the distinction.

\subsection{Process responses, limits, and separation}

Quark hard traces in two related Dirac representations are compared with independently specified reference results for all 32 SIDIS and fourteen octupole DY rows. Tests cover zero projections, target amplitudes, phases, flavor order, color factors, and differential normalization. Independent massless currents fix the DY longitudinal sign and dilepton angular integral, while charge conjugation, opposite-beam conventions, and link
reversal are checked separately. Cartesian convolutions, harmonic reductions, and Gaussian generating-function moments provide further comparisons. Weights odd in the second transverse momentum distinguish SIDIS subtraction from DY addition, which an unweighted radial convolution cannot do. Gaussian quadrature is exact for the specified polynomial tests, without assuming arbitrary TMD shapes.

Cartesian and helicity contractions independently give the gluon response. Physical density matrices generate synthetic rates independently of the reconstruction Fourier matrix. For $y_\nu=\lambda^{\epsilon_\nu}\mathscr Y_\nu$ in Table~\ref{tab:oct_fourier}, the normalized Gram matrix is exactly
\begin{equation}
 \begin{split}
 G_{\nu\mu}&=\frac12\sum_{\lambda=\pm1}\frac1{(2\pi)^2}
       \int\dd\varphi\,\dd\psi\,y_\nu y_\mu\\
 &=\diag(1,1/2,\ldots,1/2),
 \end{split}
 \label{eq:validation_angular_gram}
\end{equation}
with $\det G=2^{-13}$. Both azimuths run from $0$ to $2\pi$, and the azimuth-independent row is beam odd through $\lambda$. Recovery also requires nonzero cone factors, momentum powers, and analyzing powers (Sec.~\ref{sec:jointobservables}). Spin-setting rates check normalized moments, longitudinal responses, and continuous and eight-angle $TTT$ projections. Removing analyzers, preparation components, or angular information reveals null directions. Tests with known acceptance distinguish rank from numerical conditioning, without predicting calibration errors or yields.

Angular projection and straight-link symmetry give six quark and five gluon collinear distributions. Cartesian differentiation, Bessel transforms, and direct integration are compared on convergent inputs, while finite rotations and charge conjugation check local moments. Independent block reductions test Eqs.~\eqref{eq:qSoffer} and \eqref{eq:gSoffer} at interior, saturating, and violating points. UV matching, small-$x$ convergence, and the positive-spectral assumption of Appendix~\ref{sec:positivity} remain necessary. Finite-dimensional $PT$ checks implement the stated QCD field/link transformation, establishing neither factorization nor evolution kernels.

\subsection{Independent Born amplitudes and numerical domain}

To reconstruct the benchmark of Sec.~\ref{sec:gluonSF} from invariants, use the $\gamma^*g$ center-of-mass frame with $\hat s=(q+k)^2>4m_Q^2$ and $Q^2>0$:
\begin{align}
 k^\mu&=(\omega,0,0,\omega),&
 q^\mu&=(q^0,0,0,-\omega),\nonumber\\
 \omega&=\frac{\hat s+Q^2}{2\sqrt{\hat s}},&
 q^0&=\frac{\hat s-Q^2}{2\sqrt{\hat s}}.
 \label{eq:bornCM}
\end{align}
For $p=\sqrt{\hat s/4-m_Q^2}$ and $\hat p=(\sin\theta\cos\phi,\sin\theta\sin\phi,\cos\theta)$, the outgoing heavy-particle momenta are
\begin{equation}
 p_1^\mu=(\sqrt{\hat s}/2,p\hat p),\qquad
 p_2^\mu=(\sqrt{\hat s}/2,-p\hat p).
 \label{eq:bornfinal}
\end{equation}
A massless incoming electron of energy $E$ has
\begin{align}
 l^\mu&=(E,\ell_T,0,\ell_z),&l'^\mu&=l^\mu-q^\mu,\nonumber\\
 \ell_z&=\frac{-Q^2/2-Eq^0}{\omega},&
 \ell_T&=\sqrt{E^2-\ell_z^2}.
 \label{eq:bornlepton}
\end{align}
A nondegenerate lepton plane requires $E>\sqrt{\hat s}/2$, where equality gives the degenerate limit. These expressions satisfy the mass-shell conditions and four-momentum conservation. Conservation residuals have mass dimension one and invariant residuals dimension two, each of which is recorded with its corresponding scale.

An independent amplitude calculation checks massive-spinor normalization, Dirac equations, and completeness, then sums both heavy-quark-line amplitudes before squaring. Electron helicity is defined along its momentum. Direct current products distinguish $\sum j_\mu j_\nu^*$ from $\sum j_\mu^*j_\nu$ before comparison with Eq.~\eqref{eq:BornB} for both helicities and complex gluon polarizations. Full complex matrices agree, while independent color traces and spin averages fix the normalization left undetermined by Ward identities and normalized Stokes ratios.

The reference grid has 36 cases: $\theta=(0.4,0.8,1.7,2.7)$, $\phi=(0,0.4,1.2)$, and $P_\ell=(-1,0,1)$, where $P_\ell=\langle\lambda\rangle$ and zero means an incoherent helicity average. Amplitudes and traces are also compared at 161 uniformly spaced angles on $[0.05,\pi-0.05]$ with $\phi=0.4$ and all three polarizations, giving 483 evaluations without interpolation. Four independently rebuilt points evaluated at 50 and 80 decimal digits check arithmetic stability, without providing a third diagram derivation.

Full-matrix differences are of order $10^{-12}\,\mathrm{GeV}^2$ or smaller over the scan, with definitions, scales, and case values in Ref.~\cite{Li7TMDs}. Raw Hermiticity and finiteness are checked before diagonalization, and on-shell Ward residuals use nonvanishing amplitude and momentum scales. Negative controls omit a diagram or alter interference, normalization, or current ordering, and incorrect current ordering reverses the beam-odd response. Figure~\ref{fig:born} uses physical $\lambda=+1$ and Fig.~\ref{fig:rotation} shows exact target geometry, but neither models nuclear TMDs.

\bibliography{references}
\end{document}